\documentclass[12pt]{article}
\usepackage[utf8]{inputenc}
\usepackage{amsmath,setspace,geometry}
\usepackage{amsthm}
\usepackage{amsfonts}
\usepackage[shortlabels]{enumitem}
\usepackage{rotating}
\usepackage{pdflscape}
\usepackage{graphicx}
\usepackage{bbm}
\usepackage{comment}
\usepackage[dvipsnames]{xcolor}
\usepackage{hyperref}
\hypersetup{colorlinks=true, linkcolor= BrickRed, citecolor = BrickRed, filecolor = BrickRed, urlcolor = BrickRed, hypertexnames = true}
\usepackage[]{natbib} 
\bibpunct[:]{(}{)}{,}{a}{}{,}
\usepackage[english]{babel}
\usepackage{float}
\usepackage{caption}
\usepackage{subcaption}
\usepackage{booktabs}
\usepackage{pdfpages}
\usepackage{threeparttable}
\usepackage{lscape}
\usepackage{bm}

\usepackage{multirow}
\usepackage{adjustbox}
\begin{document}

\title{Elite Formation and Family Structure in Prewar Japan: Evidence from the Personnel Inquiry Records}
\author{Hiroshi Kumanomido\thanks{\href{mailto:}{hiroshi.kumanomido@econ.lmu.de}, Munich Graduate School of Economics, LMU Munich}, Suguru Otani\thanks{\href{mailto:}{suguru.otani@e.u-tokyo.ac.jp}, Market Design Center, Department of Economics, The University of Tokyo},
Yutaro Takayasu\thanks{\href{mailto:}{y.takayasu@r.hit-u.ac.jp}, Hitotsubashi Institute for Advanced Study, Hitotsubashi University\\
We thank Hidehiko Ichimura and Yasuyuki Sawada for generously sharing the dataset used in this study. The digitization of the dataset was supported by JSPS KAKENHI Grant Numbers JP18330040 and JP22K21341. We are also grateful to Mari Tanaka, and Chiaki Moriguchi for their valuable advice. This work was supported by JST ERATO Grant Number JPMJER2301, Japan. We take sole responsibility for errors in this paper. All data and code are available upon request.
}
}

 \date{
First version: July 9, 2025\\
Current version: \today
}
\maketitle

\begin{abstract}
    This paper introduces a newly constructed individual-level dataset of prewar Japanese elites using the ``Who's Who'' directories published in 1903--1939. Covering approximately the top 0.1\% of the population, the dataset contains rich information on social group, education, occupation, and family structure. By reconstructing intergenerational links and family networks, we provide descriptive evidence on elite formation and persistence across geography, social groups, and education during transitions from a feudal system to a modern system. We also use family records to document elite marriage patterns and family-based mobility, showing stable age assortative matching, widening husband--wife age gaps, and associations between marriage-age structure, adoption, and elite persistence. The dataset provides a foundational empirical resource for studying intergenerational and intergroup mobility, and institutional development during Japan's transition to a modern society. \\
\textbf{Keywords}: Elite formation, Intergenerational mobility, Social mobility, Institutional development \\
\textbf{JEL code}: 
N35, 
J62, 
D63

\end{abstract}


\section{Introduction}

Modernization, through the introduction of new education systems, bureaucratic recruitment, and modern business firms, can weaken barriers based on hereditary status and create new paths to leadership positions. Historical studies have shown that the introduction of such institutions can reshape long-run economic outcomes \citep{acemoglu2002reversal, banerjee2005history, becker2016empire, markevich2018economic}. Yet, it does not necessarily eliminate older mechanisms of elite persistence. Pre-existing elite networks, marriage ties, family succession, and inherited norms may continue to shape access to opportunities and the transmission of advantage across generations \citep{munshi2006traditional, ager2021intergenerational}. Understanding elite formation therefore requires examining not only how new institutions broadened access to elite positions, but also how inherited social structures continued to shape the persistence of elite status.

Prewar Japan provides a useful setting for studying this coexistence. After the Meiji Restoration of 1868, the government built new institutions for a modern state and economy. These reforms weakened the direct link between hereditary status and elite positions. Alongside these changes, however, modernization did not immediately dissolve locally rooted networks and longstanding family practices. This paper examines how social systems formed under the feudal order continued to shape elite formation after the transition toward a modern economy. We focus on two aspects. The first is the relationship between geographical mobility and locally rooted elite formation. The second is marriage and family succession practices inherited from the feudal order and maintained under the Meiji Civil Code. In particular, we examine age matching between husbands and wives at marriage and the use of adopted sons as a strategy of transferring family businesses and assets.

To examine these questions, we construct a historical dataset based on the Japanese Personnel Inquiry Records (PIR), which listed individuals who held leadership positions in the public and business sectors in prewar Japan. The PIR has a structure similar to biographical directories such as the ``Who's Who'' directories published in the United States, the United Kingdom, and other countries. It contains detailed biographical information, including an individual's name, birth year, birthplace, address, social group (hereditary nobles (\textit{Kazoku}), samurai, or commoners), educational background, occupational trajectories throughout his lifetime, and family members. Furthermore, we reconstruct family links between listed individuals and their children. This allows us to examine not only the characteristics and geographical mobility of listed individuals, but also their family structure and the intergenerational transmission of elite status.

First, we examine geographical mobility as one aspect of elite formation. Using information on birthplace and residence, we show that two distinct patterns coexisted in prewar Japan. Some individuals moved from their birth prefectures to major metropolitan areas and became elites, while many others remained in their birth prefectures and built their careers within local communities. These results show the coexistence of elite formation through migration to metropolitan centers and locally rooted elite formation.

Second, we examine marriage patterns as another aspect of elite formation. Even after the transition to a modern state, the Meiji Civil Code gave legal form to household practices that had existed under the earlier family system. To examine marriage patterns in this context, we compare spouses' ages at marriage, which is one of the most important factors in marriage markets \citep{choo2006estimating}, across marriage cohorts from the 1870s to the 1920s. We find that age assortative matching remained strong and stable throughout our study period. We also find that the age gap between husbands and wives widened over time. This widening husband--wife age gap suggests that older features of family formation persisted and may have become more pronounced as Japan transitioned to a modern economy.

Third, we further examine how spouses' age structure at marriage is associated with elite family formation. A wife's age at marriage and the age gap between spouses may be related to the number of biological children through the length of the reproductive period. Under the household system formalized by the Meiji Civil Code, elite families also needed to secure an heir to inherit the family business and assets to the next generation. When a family did not have a biological son, adopting was an alternative to secure an heir. We show that a later age at marriage for wives and a larger husband--wife age gap are associated with fewer biological children and a higher probability of having an adopted son.

This paper makes three contributions. First, we construct an individual-level dataset from five editions of the Japanese Personnel Inquiry Records published between 1903 and 1939. Existing studies have used the PIR to examine institutional changes in education and elite recruitment, as well as the intergenerational transmission of elite status \citep{moriguchi2024meritocracy, ichimura_last_2024, takayasu2024intergenerational, kumanomido_elite_2026}. We compile information on individuals' birthplaces, residences, social groups, education, occupations, and family members. We also link the same individuals across editions and identify sons of listed elites who subsequently appear in the PIR. These linked records allow us to examine geographical mobility, marriage and fertility, adoption, and the intergenerational persistence of elite status using a common empirical framework. The dataset also contributes to the growing use of linked historical microdata to study long-run social mobility \citep{ferrie1996new, abramitzky2021automated, feigenbaum2025examining}.

Second, we contribute to the literature on the geography of opportunity and locally rooted elite formation. Previous research documents large spatial differences in upward mobility and shows that exposure to different places can affect later-life outcomes \citep{chetty_opportunity_2018, chetty_impacts_2018}. Other studies show that elite and family status can persist even after major institutional and economic disruptions \citep{ager2021intergenerational, barone2021intergenerational}. Prewar Japan provides a setting in which metropolitan opportunities expanded while locally rooted elite structures remained important. 

Third, we connect the historical-demography literature on marriage and fertility with research on family succession and business continuity. Historical research emphasizes that age at marriage was an important constraint on completed family size \citep{hayami1987fossa, saito1988edo, guinnane2011}. Related demographic work examines the relationship between spousal age differences and reproductive outcomes \citep{rotering2019age}. Research on Japan, in turn, shows that the adoption of a son provided an alternative way to secure an heir when a biological heir was unavailable \citep{kurosu_adoption_1995, moriguchi_comparative_2010, kurosu2013adoption, mehrotra_adoptive_2013, kumon_adoption_2025, kumanomido_elite_2026}. We connect these strands of literature by examining how marriage-age structure is associated with both biological fertility and adoption.

This study proceeds as follows. Section \ref{sec:background} describes the historical background and explains the emergence of demand for biographical directories in prewar Japan. Section \ref{sec:data} introduces the data, and Section \ref{sec:descriptive_stats} presents descriptive evidence on listed elites and their family structure. Section \ref{sec:result} examines elite formation, focusing on geographical mobility, marriage patterns, and intergenerational links. Section \ref{sec:conclusion} concludes.

\section{Historical Background and the Emergence of Biographical Registries}\label{sec:background}

During the Tokugawa period (1603--1868), Japan was governed by the Tokugawa family under a rigid feudal hierarchy. Political authority was delegated to local lords (\textit{Daimyo}) and their subordinate samurai who were hereditarily engaged in military and administrative roles. Legal and institutional barriers prohibited social mobility (e.g., educational, occupational, residential, and marriage opportunities) between samurai and commoners (farmers, artisans, and merchants).

However, the new Meiji government was established in 1868 in response to growing pressure from domestic and Western powers. The new government abolished hereditary privileges and restructured the former classes into three groups: hereditary nobles (\textit{Kazoku}), samurai, and commoners. A series of stipend reforms beginning in 1869 reduced the financial basis of samurai privilege, and the Conscription Law of 1873 introduced universal military service, weakening the dominance of samurai over military roles.\footnote{The stipend reforms gradually reduced and then commuted hereditary payments to former samurai families. The Conscription Law of 1873 introduced universal military service and reduced the institutional connection between samurai status and military service.} Although Japan's institutional path differed from that of Western industrializing countries, these reforms shared a broader feature of modern state formation: the weakening of hereditary status as a formal basis of social and political authority.

At the same time, the Meiji government created new pathways to elite positions. The Education Order of 1872 established a national framework for modern schooling, and the Civil Service Examination Law of 1893 strengthened the role of competitive examinations in bureaucratic recruitment. Professional qualifications for physicians, lawyers, teachers, and other occupations were also increasingly tied to schooling, licensing, and examinations. These reforms did not eliminate inherited advantage, but they changed the institutional channels through which individuals could enter elite positions.

These institutional changes also changed how elites were identified and evaluated. During the Tokugawa period, social position was closely tied to hereditary status, and family background and local reputation were often known within local communities. After the Meiji Restoration, however, elite positions were no longer defined by hereditary status alone. Education, bureaucratic service, professional qualifications, business careers, and public offices also became important aspects of social position. As a result, information on education, career history, occupation, public service, income, honors, and family connections became useful for evaluating an individual's social status beyond family background and local reputation. In this context, formal biographical registries such as the Japanese Personnel Inquiry Records (PIR) became a practical source of information for marriage, business, and other social networks.

\section{Data}\label{sec:data}

This section introduces our data source and explains how we construct the dataset used in the analysis. We first describe the historical background and coverage of our dataset, then explain the structure of our individual-edition-level data and family records. Finally, we describe how we classify key variables and link individuals and their family members across editions to study elite formation during our study period.

\subsection{Japanese Personnel Inquiry Records}

This paper focuses on individuals who achieved leadership positions in politics and business in prewar Japan. To capture individuals in such positions, we collected a historical dataset based on the Japanese Personnel Inquiry Records (PIR). We compiled the PIR published in 1903, 1915, 1928, 1934, and 1939.\footnote{We digitized 5 editions of PIR out of 15 editions, which were published before WWII.} The number of listed individuals is 2,892 in 1903 (0.006\% of the population), 13,759 in 1915 (0.026\%), 24,931 in 1928 (0.040\%), 25,846 in 1934 (0.038\%), and 54,497 in 1939 (0.076\%). This dataset is comparable to the ``Who's Who'' directory in developed countries, as it is a selective list of distinguished individuals, such as high-income earners, CEOs, bureaucrats, politicians, military servants, and professional elites (judges, lawyers, scholars, and physicians).\footnote{\cite{doxey2022democratization} uses Wikidata to collect biographical information for distinguished individuals in the United States, such as judges, politicians, and business managers. \cite{serafinelli_creativity_2022} also collected data on notable individuals born in Europe between the eleventh and the nineteenth centuries from Wikidata.}

One important purpose of the PIR was to provide information for identifying and evaluating potential marriage partners, business partners, and other socially relevant connections. This purpose reflected the conditions of the period covered by our study, when changes in industrial structure and turnover among political and business leaders made the composition of the elite more fluid. As a result, reliable and up-to-date information on who held elite positions, as well as on their careers and family backgrounds, became important. The PIR responded to this demand by providing detailed biographical information, including birth year, birthplace, current residence, occupational trajectories, educational background, and family members. We describe the more detailed information for PIR in Appendix Section \ref{sec:PIRDataSource}.

\begin{figure}[htbp]
    \begin{center}
    \includegraphics[width = 0.7\textwidth]{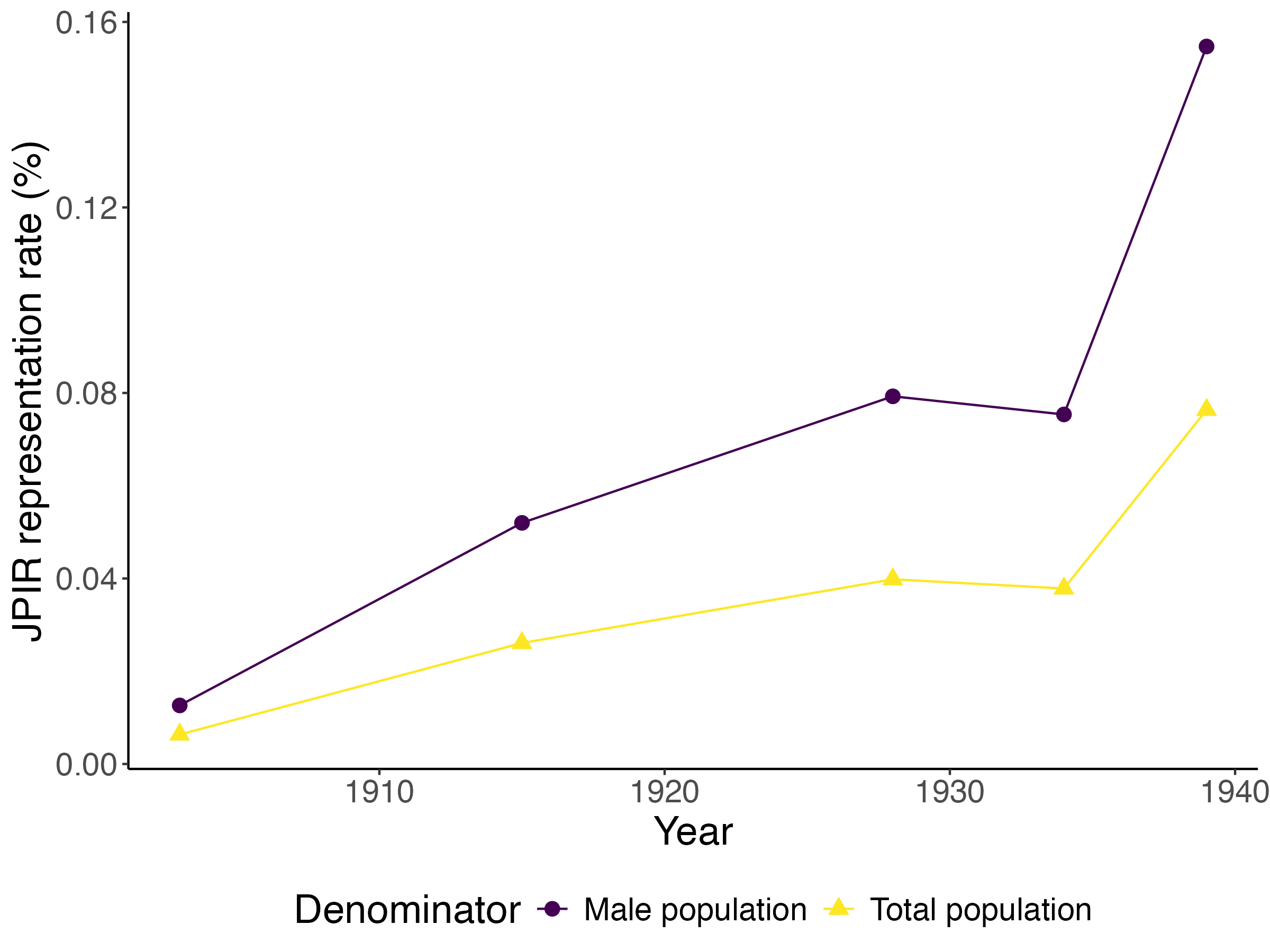}
    \end{center}
    \caption{Share of elites listed in each PIR edition}
    \label{fg:listing_ratio} 
    \footnotesize
    \textit{Note}: This figure shows the representation rate of individuals listed in the PIR relative to the population for each edition year (1903, 1915, 1928, 1934, and 1939). We use the estimated population by the Statistics Bureau of Japan. The purple line shows the number of listed individuals divided by the total male population, and the yellow one is divided by the total population. The number of elites is 2,892 in the 1903 PIR (0.006\% of the total population; 0.013\% of the male population), 13,759 in the 1915 PIR (0.026\%; 0.052\%), 24,931 in the 1928 PIR (0.040\%; 0.079\%), 25,846 in the 1934 PIR (0.038\%; 0.075\%), and 54,497 in the 1939 PIR (0.076\%; 0.155\%).
\end{figure}

We begin our data description by discussing the inclusion criteria used in the PIR. Over time, the PIR gradually broadened its coverage of elites. While early editions primarily focused on individuals from hereditary nobles or public servants, later editions expanded to include individuals engaged in the private sector, such as professional elites and CEOs.

Figure \ref{fg:listing_ratio} shows the proportion of individuals listed in the PIR relative to Japan's total population and male population in each edition. The PIR covered an increasingly large share of the population over time, accounting for between approximately 0.006\% and 0.076\% of the total population, or between 0.013\% and 0.155\% of the male population.

\subsection{Data Structure}\label{sec:data_structure}

Using biographical records of elites, we construct an individual-level dataset that includes detailed information on family members (their parents and children). To briefly explain the structure of our dataset, we distinguish elites' information between two components: (1) information on elites and their families within each edition, and (2) linkages of individuals and their family members across editions.

\paragraph{Individual-edition Level Data}

We use five editions of the PIR, published in 1903, 1915, 1928, 1934, and 1939. Let $\mathcal{T}=\{1,4,8,10,12\}$ denote the corresponding set of editions, and let $\mathcal{I}_{t}$ denote the set of individuals listed in edition $t\in \mathcal{T}$. The unit of observation in the raw data is an individual $i\in \mathcal{I}_{t}$ in edition $t$.

For each individual-edition observation, the data contain individual characteristics and family information. We summarize this information as
\begin{align*}
    D_{it} = \{X_{it},m_{i},f_{i},w_{it},s_{it},a_{it},d_{it}\},
\end{align*}
where $X_{it}$ is the set of individual characteristics recorded for individual $i$ in edition $t$, including time-invariant characteristics, such as name, birth year, birthplace, social group, educational background, as well as time-varying characteristics, such as current residence, occupational history.

The other elements of \( D_{it} \) include family member information recorded in the PIR. We denote the records for the mother and father of individual \( i \) by \( m_{i} \) and \( f_{i} \), respectively, and the record for the wife of individual \( i \) in edition \( t \) by \( w_{it} \). We also denote the sets of records for biological sons, adopted sons, and daughters of individual \( i \) in edition \( t \) by \( s_{it} \), \( a_{it} \), and \( d_{it} \), respectively. These records include the family-member information available in the PIR, such as names, birth years, social group, and relationships to the listed individual. For the marriage analysis, we use the wife's birth year and the birth year of the eldest biological child to approximate each couple's marriage year as one year before the eldest child's birth year; this construction allows us to measure spouses' ages at marriage but necessarily excludes couples without a datable biological child.

\paragraph{Family Structure within Each Edition}

Using the family-member records defined above, we describe the within-edition family structure for each edition $t\in \mathcal{T}$. The parent set is defined as
\begin{equation*}
\mathcal{P}_t = \{ m_i, f_i \mid i \in \mathcal{I}_t \},
\end{equation*}
and the child set as
\begin{equation*}
\mathcal{C}_t = \{ c \mid c \in s_{it} \cup a_{it} \cup d_{it},\ i \in \mathcal{I}_t \}.
\end{equation*}
Thus, within each edition, the data consist of listed individuals and the family-member records associated with them. The recorded family relationships link each listed individual to his parents and children. Specifically, individual \( i \) is linked to parents \( m_i \)  and \( f_i \), and to children in \( s_{it} \), \( a_{it} \), and \( d_{it} \).

\paragraph{Linking Family Records across Editions}
We use the family information recorded in each PIR edition to link individuals and family members across editions. First, we identify the same listed individuals across editions using name and birth year. Based on this matching, we examine whether individuals \( i \in \mathcal{I}_t \) are listed in later editions \( t' > t \), and use this information to measure persistence and turnover among listed individuals.

Second, we link sons recorded in one edition to listed individuals in later editions. Each edition provides a snapshot of parent-child relationships, but these relationships can be extended across editions when a son of a listed individual later appears as a listed elite. Specifically, if individual \( i \in \mathcal{I}_t \) lists a biological or adopted son \( c \in s_{it} \cup a_{it} \) in edition \( t \), and this son is matched to a listed individual \( j \in \mathcal{I}_{t'} \) in a later edition \( t' > t \) using identifiers such as name and birth year, we construct an inter-edition family link from \( i \) to \( j \). Formally,
\begin{align*}
    \text{if } c \in s_{it} \cup a_{it} \text{ and } c \equiv j \in \mathcal{I}_{t'}, \text{ then } (i, j) \in E_{t,t'}.
\end{align*}
This procedure allows us to extend the within-edition family information beyond a single snapshot and to study whether sons of listed individuals subsequently appear as listed individuals. Furthermore, we show the intergenerational transmission of elite status by examining the extent to which father-son pairs are both listed across the PIR in Section \ref{result:intergeneration}.

\subsection{Variable Construction and Classification}\label{data:definition}

\begin{table}[!htbp]
\caption{Availability of Biographical Variables across Who's Who Editions}
\label{tb:variable_list} 
\begin{center}
\begin{tabular}[t]{lcccccccccc}
\toprule
Edition & YOB & \shortstack{Social\\Group} & Occ. & Educ. & Residence & Tax & Child & Father & Wife & Relation. \\
\midrule
Edition 1 & Y & Y & Y & Y & Y &  & Y & Y &  & Y\\ 
Edition 4 & Y & Y & Y & Y & Y &  & Y & Y &  & Y\\ 
Edition 8 & Y & Y & Y & Y & Y &  & Y & Y & Y & Y\\ 
Edition 10 & Y & Y & Y & Y & Y & Y &  &  &  & \\ 
Edition 12 & Y & Y & Y & Y & Y & Y &  &  &  & Y\\ 
\hline
\end{tabular}
\end{center}
\footnotesize
\textit{Note:} This table summarizes the availability of key variables across editions. Variables include each listed individual's year of birth (YOB), social group, occupation (``Occ.''), final education institutions (``Educ.''), residential address (both birth and current prefecture), names and birth years of their children, names of their fathers, names and birth years of their wife, and familial relationships to their fathers (e.g., firstborn sons, adopted sons). ``Y'' indicates that the variable is available in the corresponding edition; blank cells indicate that the information is not included in that edition.
\end{table}

To provide an overview of the information recorded in our data source, Table \ref{tb:variable_list} summarizes the key variables. While core attributes such as birth year, address (birth prefecture and current residential address), education (final education institutions), occupation, and occupation title are consistently available across all editions, the availability of family-related variables--such as information on tax payments, children, parents, spouse, and household relationships--varies by edition.

In the following, we provide a detailed explanation of how we construct individual-edition-level covariates $X_{it}$, including social group, occupation, education, residential address, income tax, and recipients of medals awarded for distinguished achievements based on text-based information extracted from our dataset.\footnote{\citet{moriguchi2024meritocracy} identifies the following elite subgroups: (1) individuals who earn more than top 0.05\% and 0.01\% income percentiles according to the national income distribution, (2) recipients of medals awarded for distinguished achievements, (3) corporate executives (individuals who hold an executive position in a corporation and pay any positive amount of income or business tax), (4) top politicians and bureaucrats (individuals who serve as members of the Imperial Diet, cabinet members, or high-ranking officials in the central government), (5) professors or associate professors at Imperial Universities.}

\paragraph{Social Group}

We classify each individual into one of three social groups: nobles (\textit{Kazoku} in Japanese), former samurai, and commoners. The PIR explicitly records noble and samurai backgrounds, while individuals without such information are coded as commoners.

\paragraph{Occupation}

We classify occupational information along two dimensions: industry affiliation and occupational title. Industry categories include finance, manufacturing and construction, professional and academic services, wholesale and retail trade, information services, military, public administration, and transportation. Occupational titles are grouped into categories such as executives, military officials, scholars and engineers, judges and lawyers, physicians, and teachers. Since the PIR records occupational trajectories rather than a single current occupation, an individual may be assigned to multiple industries or occupations if several affiliations appear in his career record.

\paragraph{Education}

We classify educational background into four categories: ``Imperial University'', ``Other University'', ``Private School'', and ``No Higher Education''. These categories are identified using school names and related keywords in the PIR. Individuals without information on higher educational institutions are coded as having no higher education.\footnote{The PIR records higher educational institutions but generally does not report secondary education or lower levels of schooling.} 

\paragraph{Geography}

We standardize birthplaces and residential addresses by mapping historical place names to modern prefectures. The classification uses prefectural names, historical regional names, and domain names from the Tokugawa period. We assign each birthplace and current residence to one of Japan's 47 prefectures or to an additional category for foreign regions, including Taiwan, Korea, and parts of China.

\paragraph{Income Tax Payments \& State Honors}

For the 1934 and 1939 editions, we use income tax payments to identify individuals at the upper end of the income distribution. We compare the number of listed taxpayers in each income bracket with the Tax Bureau Statistical Yearbook, following \citet{moriguchi2024meritocracy}. We define top income earners as individuals in the top 0.1\% income bracket based on income tax records. This income-based threshold provides a consistent benchmark for identifying economic elites across the 1934 and 1939 editions.

In addition to income-based measures, we use information on state honors and decorations to capture institutional and symbolic forms of elite recognition. Following \citet{moriguchi2024meritocracy}, we focus on decorations ranked from 1st to 5th class, which indicate formal state recognition of distinguished service in public, military, business, or academic fields.

These two indicators also help assess the selectivity of the PIR. The share of medal recipients remains around 20\% across editions, except for the first edition. Because the total number of listed individuals increased over time, this stable share implies that later editions included a larger absolute number of individuals with formal state recognition. The share of top income earners was approximately 53\% in the 1934 edition and 21\% in the 1939 edition, suggesting that the later PIR editions captured a substantial share of the national economic elite, while the expanded coverage in 1939 included a broader set of listed individuals (see Table \ref{tb:summary_statistics_individual_version_covariates}).

\section{Descriptive Evidence of Elites and Family Structure}\label{sec:descriptive_stats}

Before examining elite formation, we first present descriptive evidence on two dimensions of the PIR. First, we summarize the characteristics of listed individuals, including both time-invariant and time-varying characteristics $X_{it}$ across PIR editions $t$. Second, we describe information on their children, which provides evidence on the family structure of listed individuals and examines whether sons from listed individuals are listed in later editions. We exclude individuals who do not have information on birth year or whose reported age at the time of listing is below 17.

\subsection{Individuals' Characteristics}

We first provide descriptive evidence on how the characteristics of listed individuals vary across PIR editions. We focus on four aspects: (1) the age of listed individuals and the timing of the birth of their first child, (2) social group backgrounds, (3) industry affiliations and occupational titles, and (4) educational backgrounds.

\paragraph{Distribution of Elites' Age}

\begin{figure}[htbp]
    \begin{center}
    \subfloat[Age at Being Listed]{
    \includegraphics[width = 0.45\textwidth]{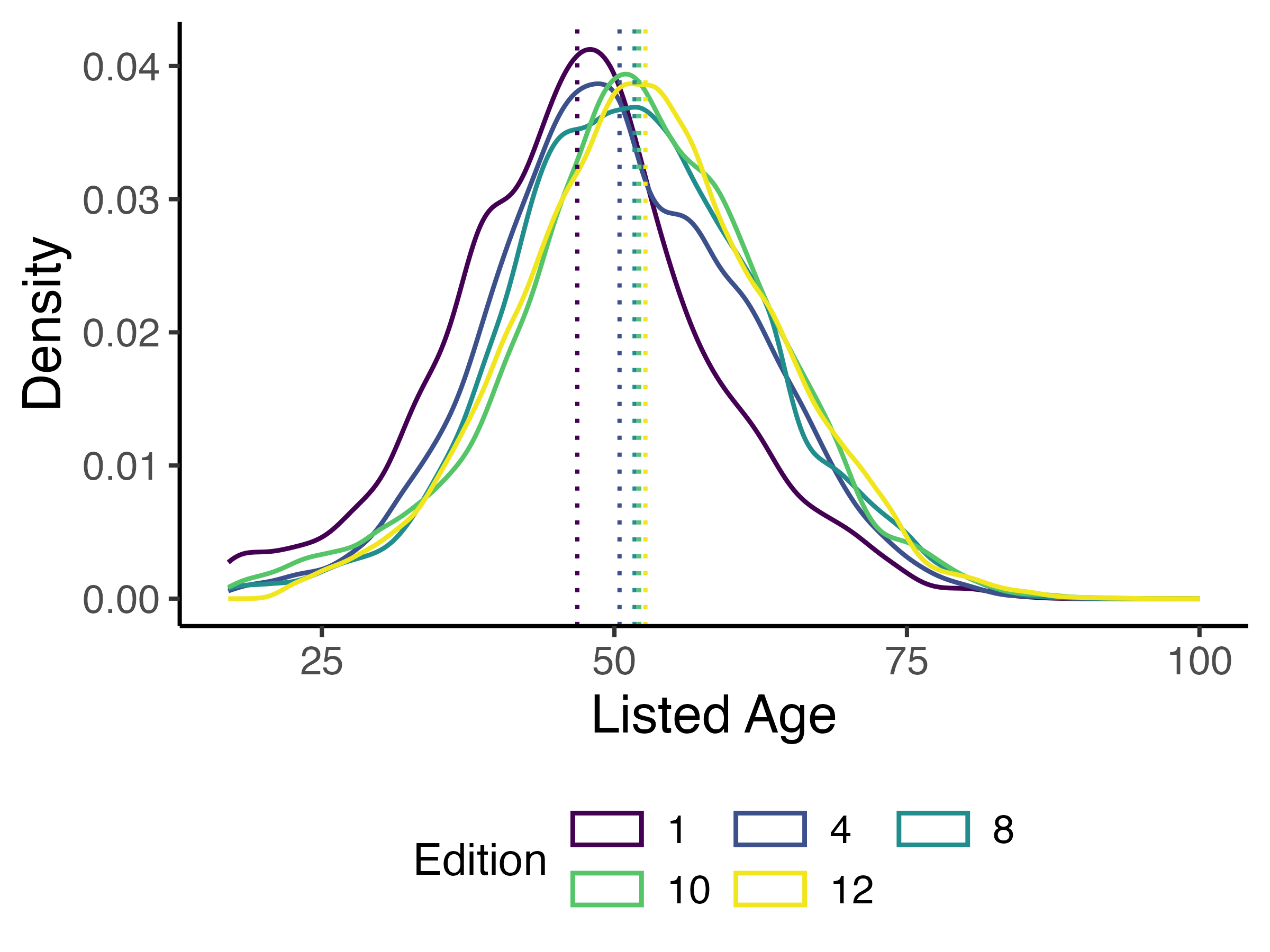}
    }
    \subfloat[Age of First Birth]{
    \includegraphics[width = 0.45\textwidth]{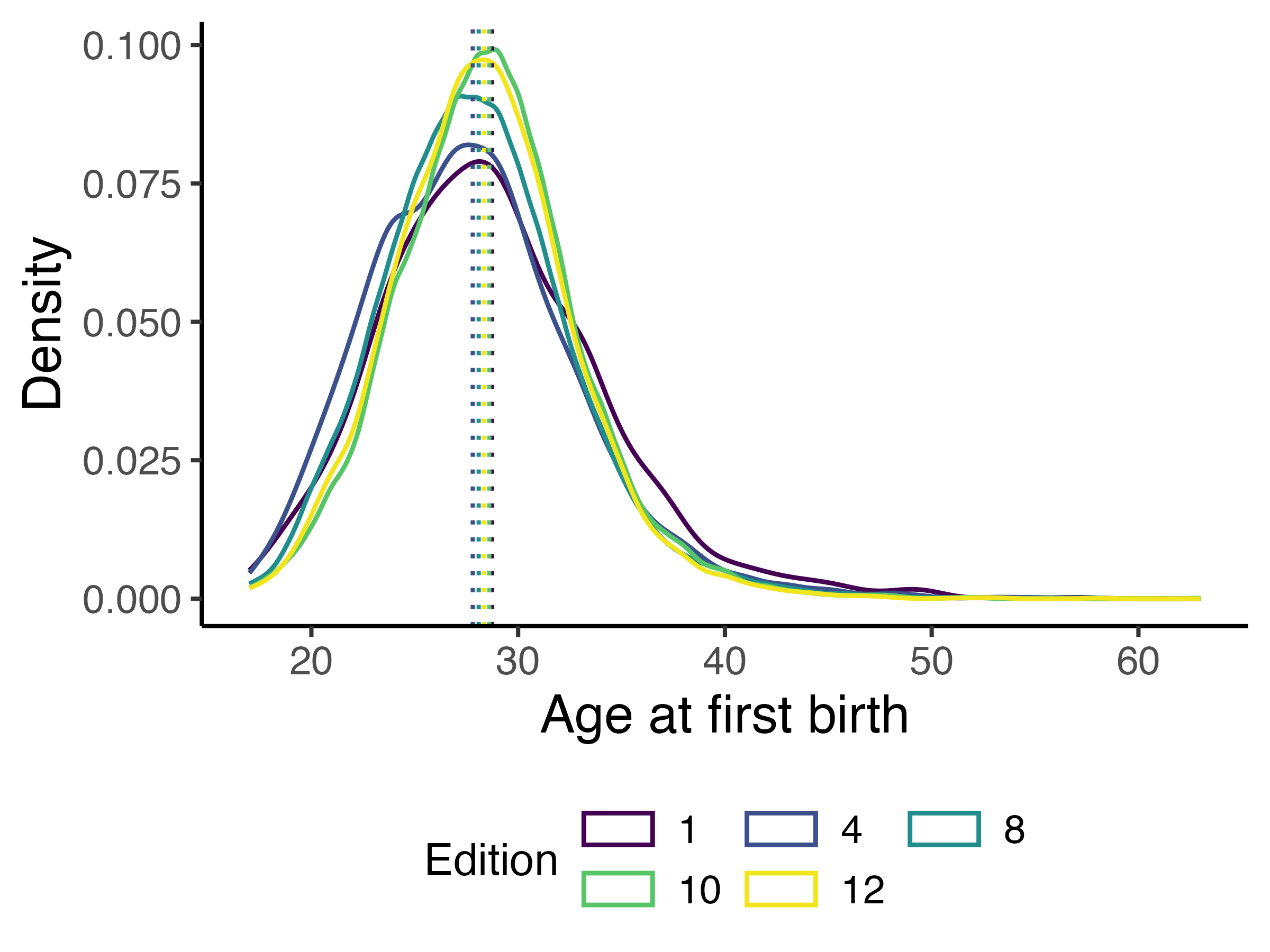}
    }
    \end{center}
    \caption{Distribution of Elites' Ages}
    \label{fg:age_firts_birth_distribution}
    \footnotesize
    \textit{Note:} These figures show the distribution of ages at the time individuals are first listed in the PIR (panel (a)) and the distribution of elites' ages at the time of their first recorded childbirth (panel (b)) with dotted vertical lines marking edition means. In panel (a), we exclude individuals who do not have information on birth year or whose reported age at the time of listing is below 17. In panel (b), we further exclude elites who do not have any biological children.
\end{figure}

Figure \ref{fg:age_firts_birth_distribution} provides detail on listed individuals' ages. Panel (a) shows the distribution of ages at the time individuals are first listed in the PIR.\footnote{The age of listed individuals is calculated as the publication year of each edition minus birth year.} The average ages range from 46.8 (edition 1) to 52.7 (edition 12). Although the distributions for editions 8 and 10 are slightly skewed to the right, they remain stable across editions. Panel (b) presents the distribution of elites' ages at the time of their first recorded childbirth. Average ages are similar across editions, with the lowest in edition 4 (27.8) and the highest in edition 1 (28.7). As the sample size increases, the standard deviations decrease from 5.50 (edition 1) to 4.37 (edition 12).

\paragraph{Social Groups}

\begin{figure}[htbp]
    \begin{center}
    \includegraphics[width = 0.75\textwidth]{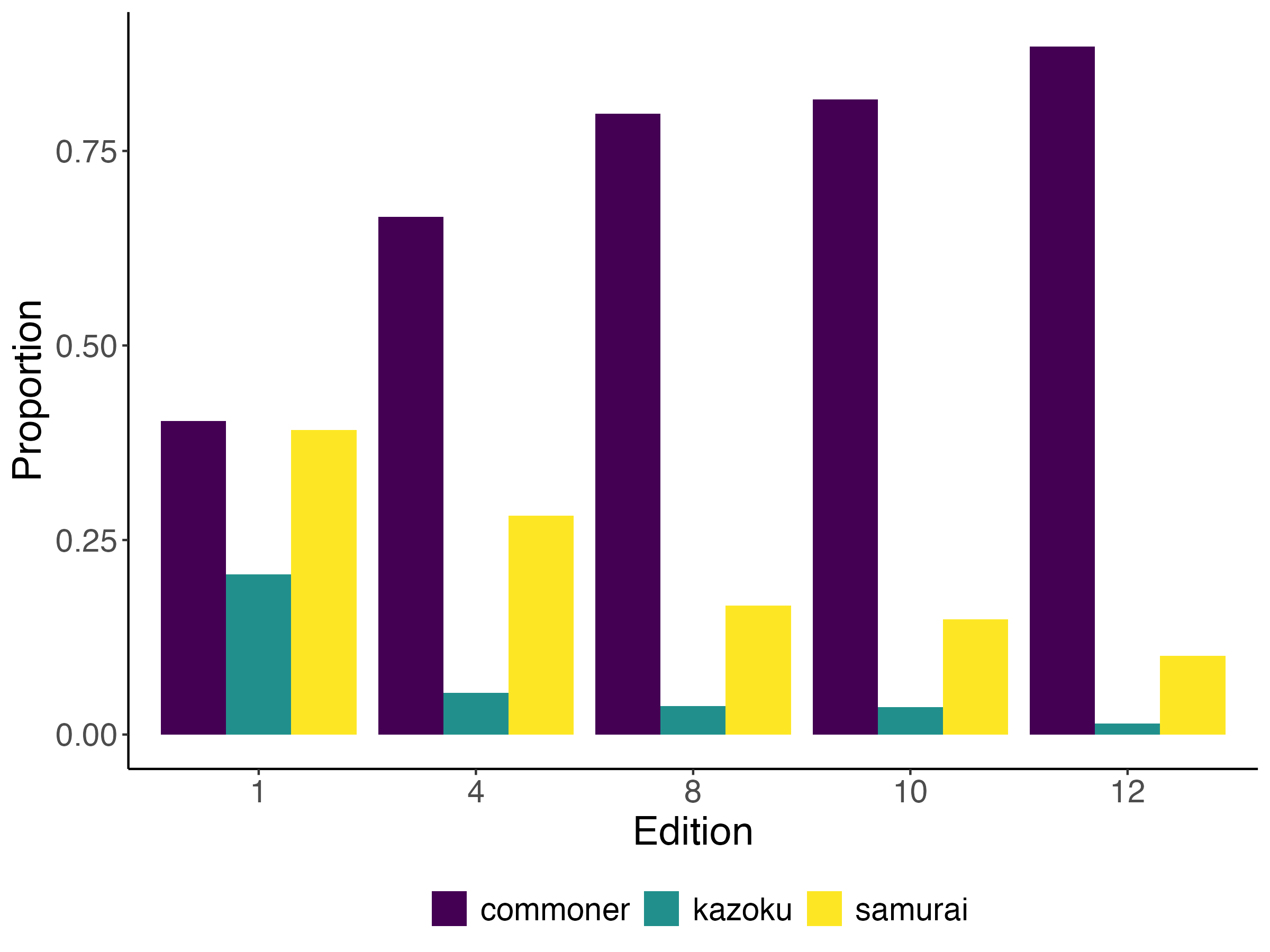}
    \end{center}
    \caption{Transition in Social Group by Editions}
    \label{fg:bargraph_social_group}
    \footnotesize
    \textit{Note:} This figure presents the share of elites by social group across editions. We show the proportion of social groups, defined in the early Meiji period: commoners (merchants, artisans, and farmers in the Tokugawa period), \textit{Kazoku} (hereditary nobles, including court nobles and local lords in the Tokugawa period), and samurai.
\end{figure}

Next, we show the differences in the share of social groups among elites. Figure \ref{fg:bargraph_social_group} shows the share of social groups \textit{Kazoku} (hereditary nobles), samurai, and commoners across PIR editions. In 1903, samurai represented 39\% of the listed elites, but their share declined sharply to 10\% by 1939. In contrast, commoners accounted for 40\% in 1903, and their share rose to 90\% by 1939. Although samurai still remained overrepresented in elite positions even in 1939, these results indicate a substantial rise of commoners among elite positions in prewar Japan.\footnote{The shares of \textit{Kazoku}, samurai, and commoners were 0.1\%, 5\%, and 95\% of the population, respectively. Due to the gradual relaxation of the PIR inclusion criteria, the share of \textit{Kazoku} declined drastically from 20\% to 1\% over the same period.}

\paragraph{Industry \& Occupation Titles}

We describe elites' industrial affiliations and occupational titles using all occupational records observed for each individual. The late nineteenth and early twentieth centuries saw the expansion of modern corporate organizations in Japan, especially after the enactment of the Commercial Code in 1899 and its revision in 1911. These institutional developments created a range of modern corporate positions, including directors, executive managers, auditors, and other managerial titles. To capture these changes, we classify industrial affiliations using the ISIC and occupational titles using the ISCO.

\begin{figure}[htbp]
    \begin{center}
    \subfloat[Industry]{
    \includegraphics[width = 0.7\textwidth]{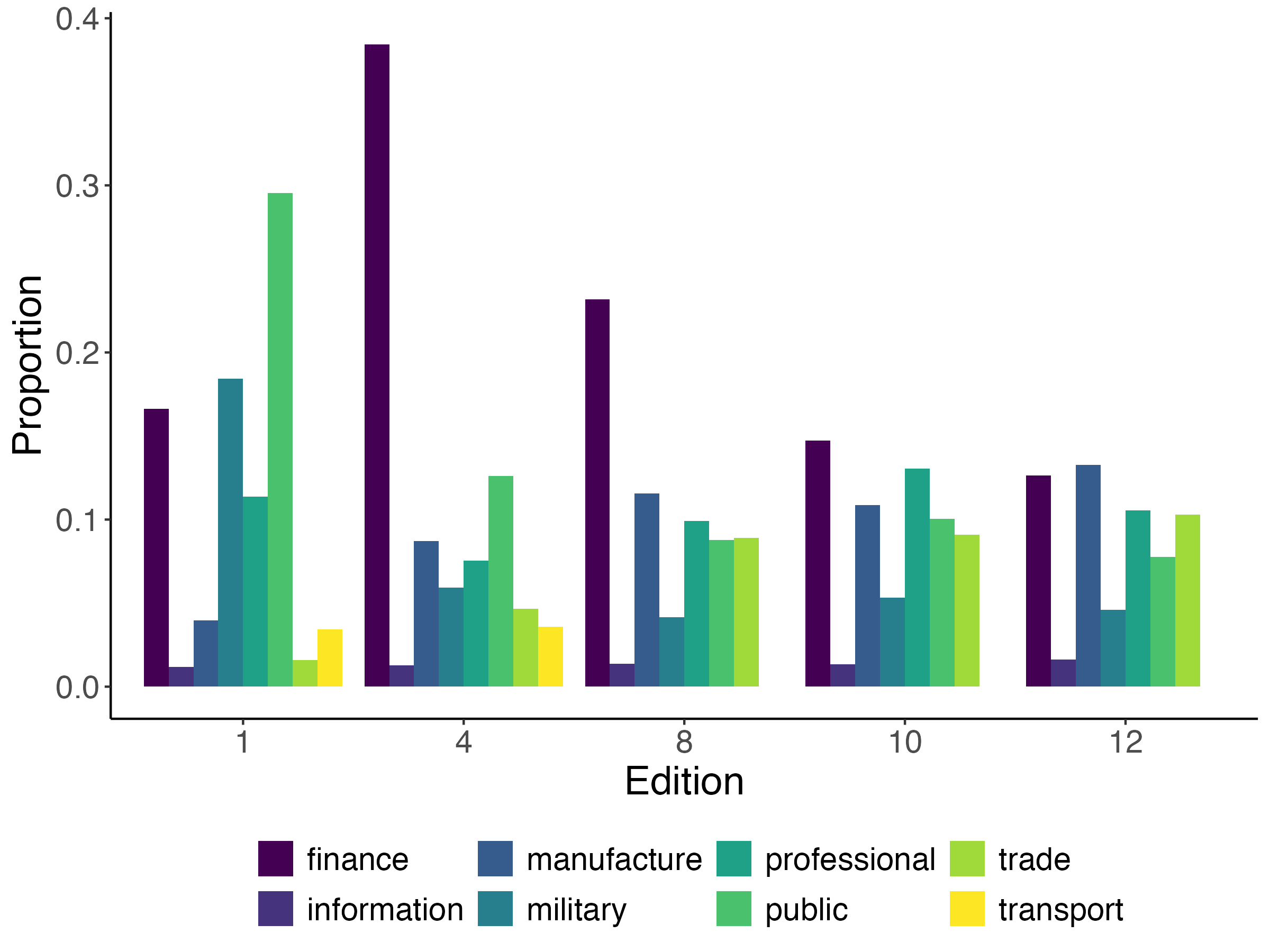}
    }\\
    \vspace{0.3in}
    \subfloat[Occupational Titles]{
    \includegraphics[width = 0.7\textwidth]{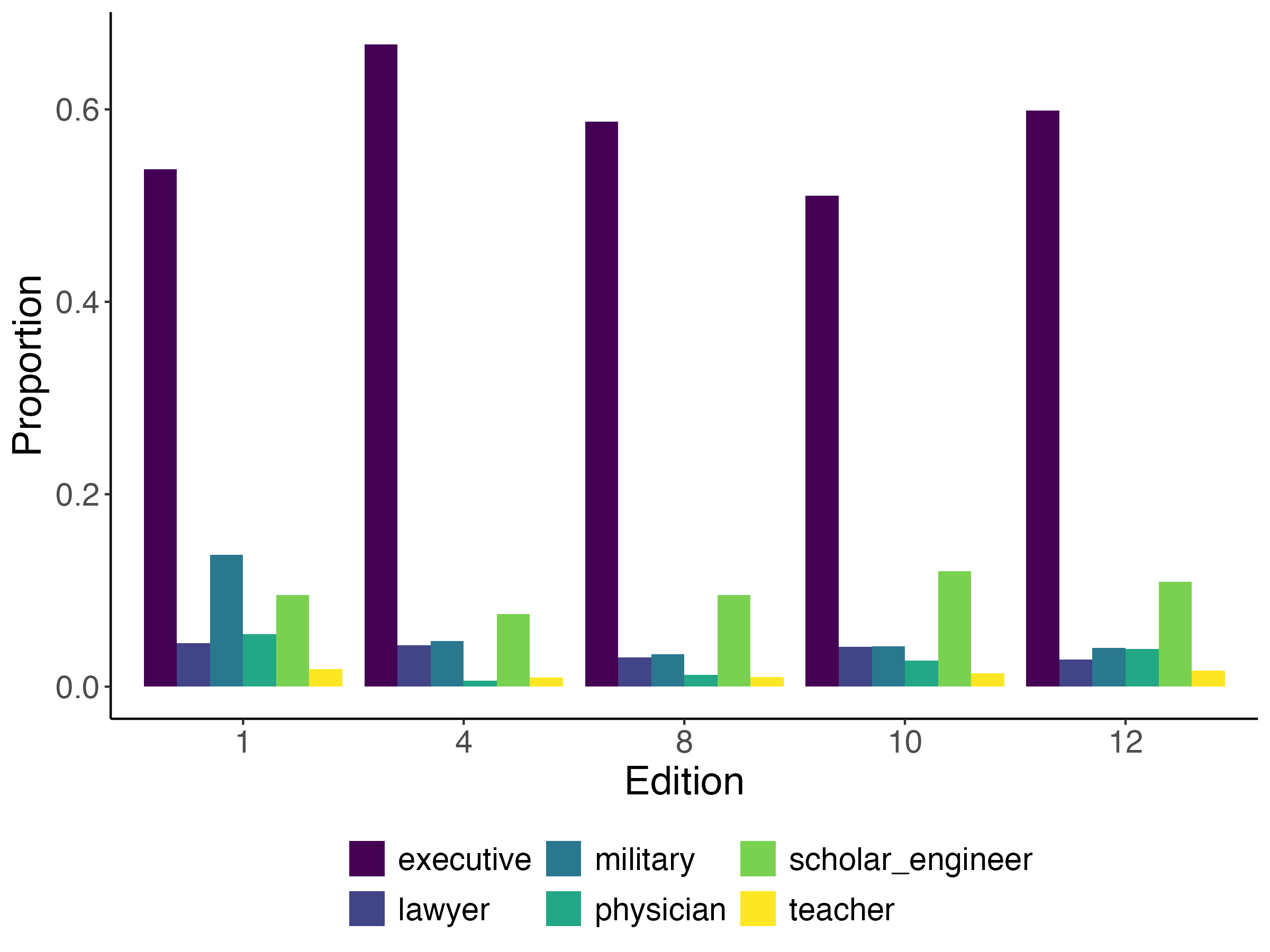}
    }
    \end{center}
    \caption{Transition in Occupation Share of Elites by Editions}
    \label{fg:bargraph_occupation}
    \footnotesize
    \textit{Note:} These figures present the share of elites by industrial affiliations (panel (a)) and occupations (panel (b)) across editions. Panel (a) shows the share of industry categories. Panel (b) shows their occupational titles. We exclude individuals who do not have any industry information within their occupational records in panel (a). We also exclude individuals who do not have any occupational-title information within their occupational records in panel (b).
\end{figure}

Panel (a) of Figure~\ref{fg:bargraph_occupation} reports the share of elites affiliated with each industry. Individuals without industry information are excluded. Because individuals may have multiple occupational records over their lifetime, industry categories are not mutually exclusive. The figure shows that the share of elites affiliated with financial institutions was about 38\% in 1915, before declining to about 23\% in 1928 and 13\% in 1939. Other major sectors, including manufacturing, retail services, and the public sector, display more stable shares after 1903. The industrial composition of listed elites may reflect both changes in Japan's industrial structure and changes in the coverage of the PIR across editions.

Panel (b) of Figure~\ref{fg:bargraph_occupation} reports the distribution of occupational titles. Individuals without occupational-title information are excluded, and title categories are again not mutually exclusive. Executives and senior managers consistently account for a large share of listed elites, roughly 51--67\%, although the industrial affiliations shown in panel (a) varied across editions. The stable share of executives and senior managers suggests that corporate and organizational leaders remained a central group among listed elites throughout our study period.

\paragraph{Education}

\begin{figure}[htbp]
    \begin{center}
    \includegraphics[width = 0.75\textwidth]{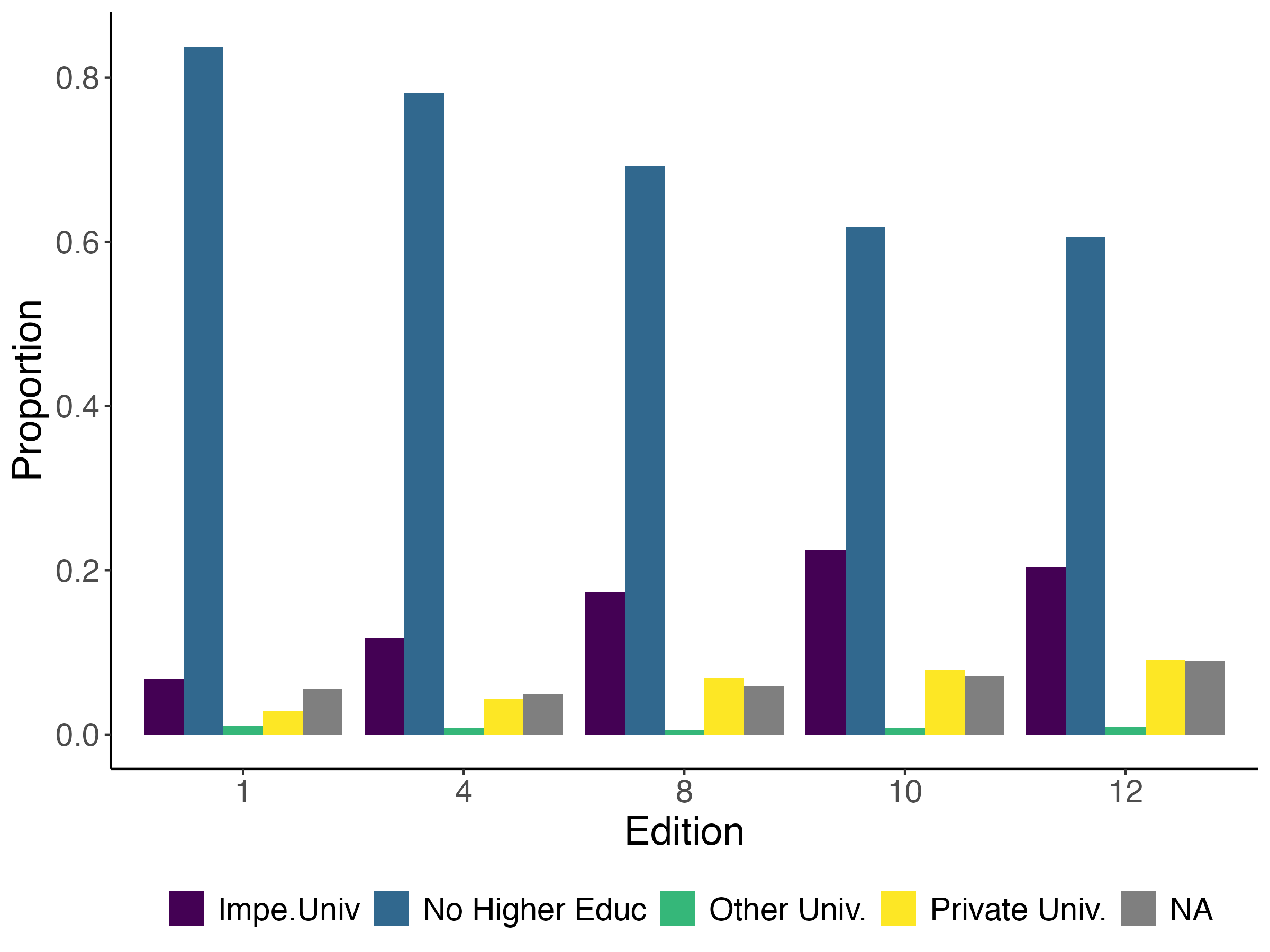}
    \end{center}
    \caption{Transition in the Educational Share of Elites by Editions}
    \label{fg:bargraph_school_cat}
    \footnotesize
    \textit{Note:} This figure shows the educational background of listed individuals. ``Impe.Univ'' includes imperial universities and selective high schools. ``Other Univ.'' includes foreign universities and language schools. ``Private Univ.'' refers to private institutions that served as alternatives to imperial universities or vocational schools. Many of them were considered preparatory schools for qualification exams, especially for aspiring lawyers. ``No Higher Educ'' includes individuals who did not proceed to any higher education institution, typically ending their formal education at the secondary level or below. ``NA'' denotes records not assigned to these categories.
\end{figure}

Figure \ref{fg:bargraph_school_cat} shows the educational attainment of elites across PIR editions. We classify educational background into ``Impe.Univ'', ``Other Univ.'', ``Private Univ.'', ``No Higher Educ'', and unclassified records. The results show a substantial increase in the share of Imperial University graduates over time, particularly between 1928 and 1934. Although the share of individuals who did not pursue higher education declined during this period, it remained above 60\% in every edition. These results indicate that graduating from an Imperial University was not a prerequisite for attaining elite positions \citep{ichimura_last_2024}.

\clearpage

\subsection{Family Structure and Intergenerational Transmissions}

In this section, we provide descriptive evidence on the family structure of listed individuals. We then examine the intergenerational transmission of elite status from listed individuals to their sons by identifying sons who appear in subsequent editions of the PIR.

\paragraph{Family Composition}

\begin{figure}[htbp]
    \begin{center}
    \subfloat[Distribution of Number of Children per Elite by Edition]{
    \includegraphics[width = 0.70\textwidth]{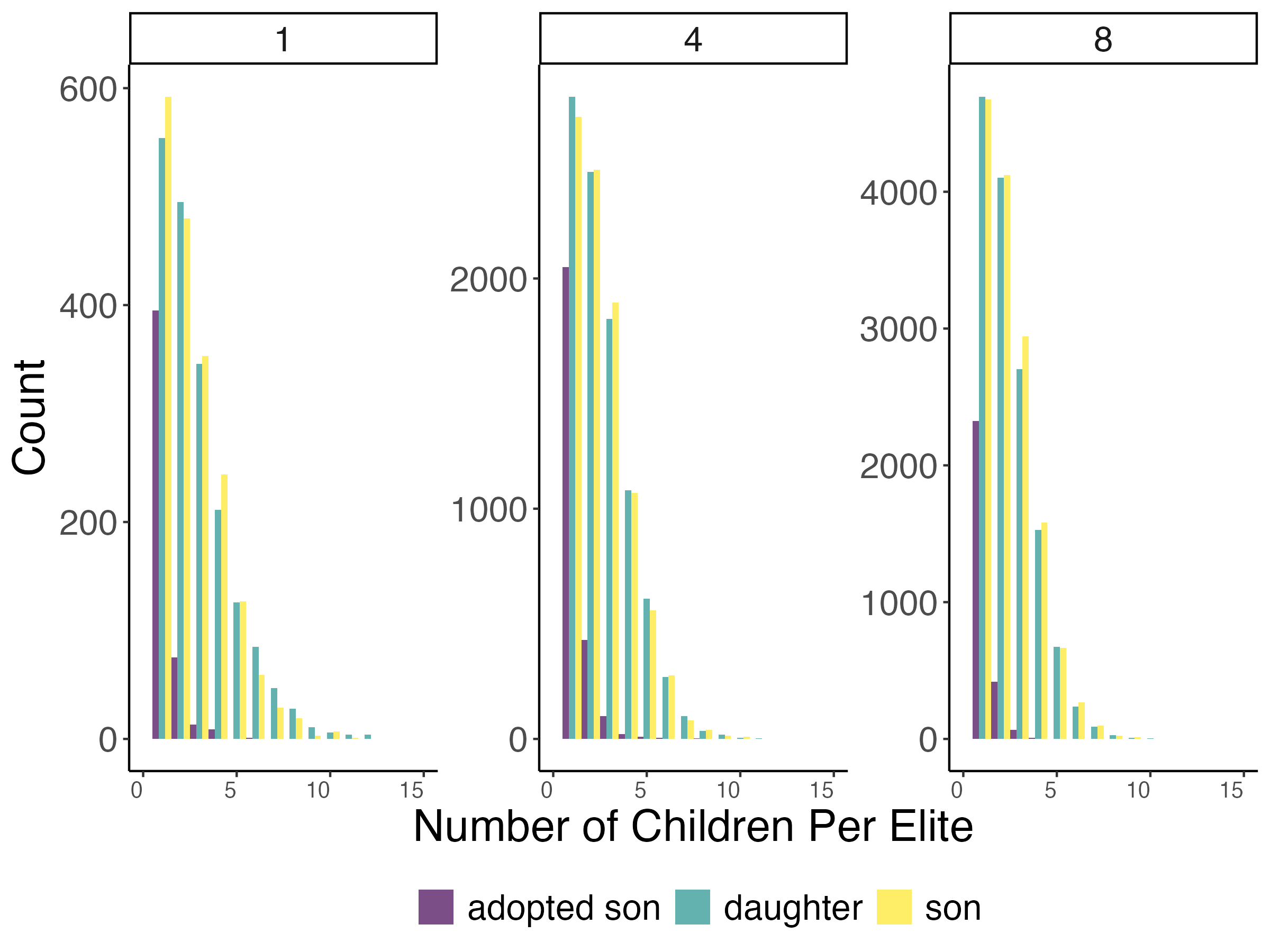}
    }\\
    \vspace{0.3in}
    \subfloat[Distribution of Child Birth Order by Edition]{
    \includegraphics[width = 0.45\textwidth]{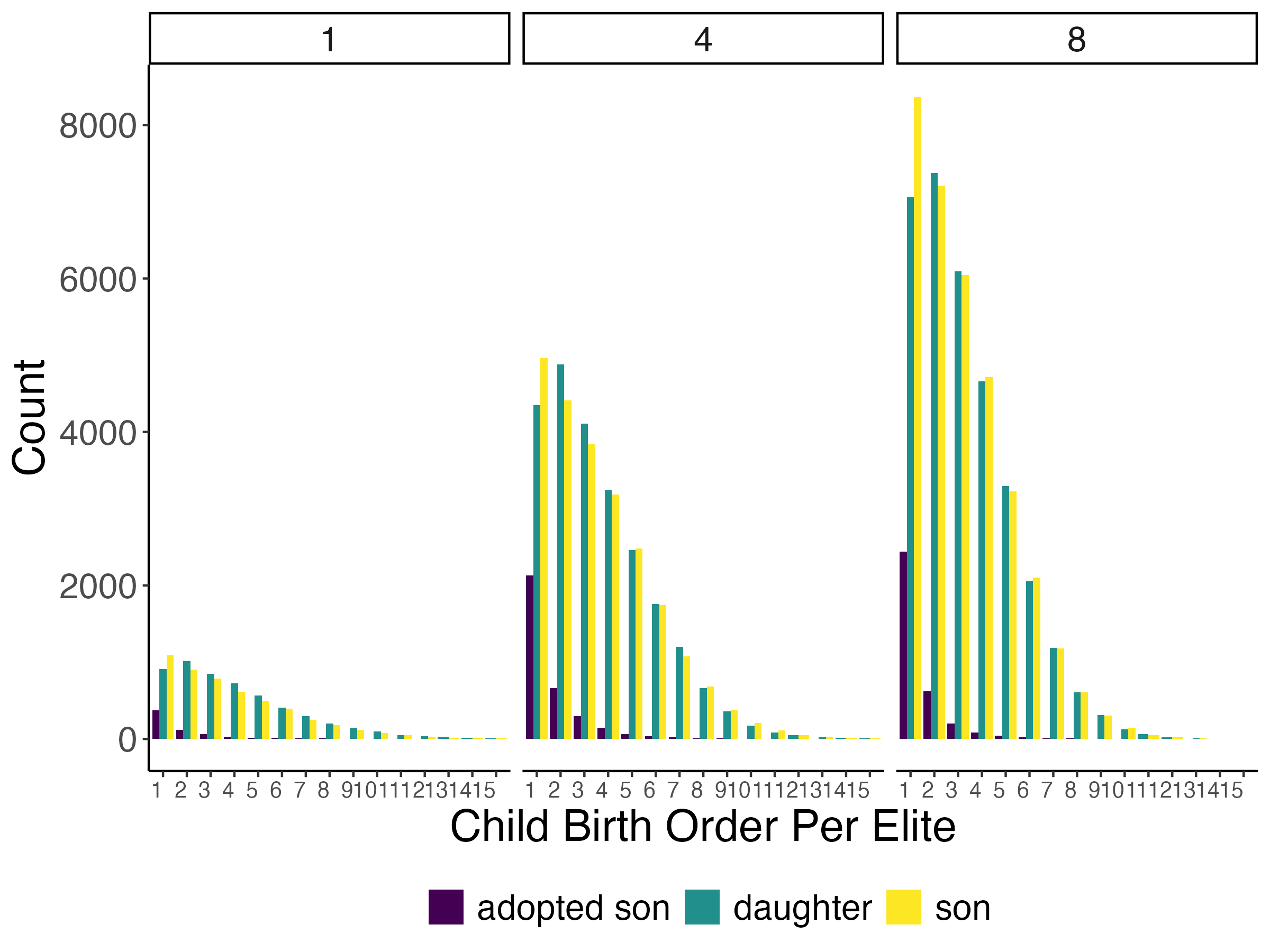}\includegraphics[width = 0.45\textwidth]{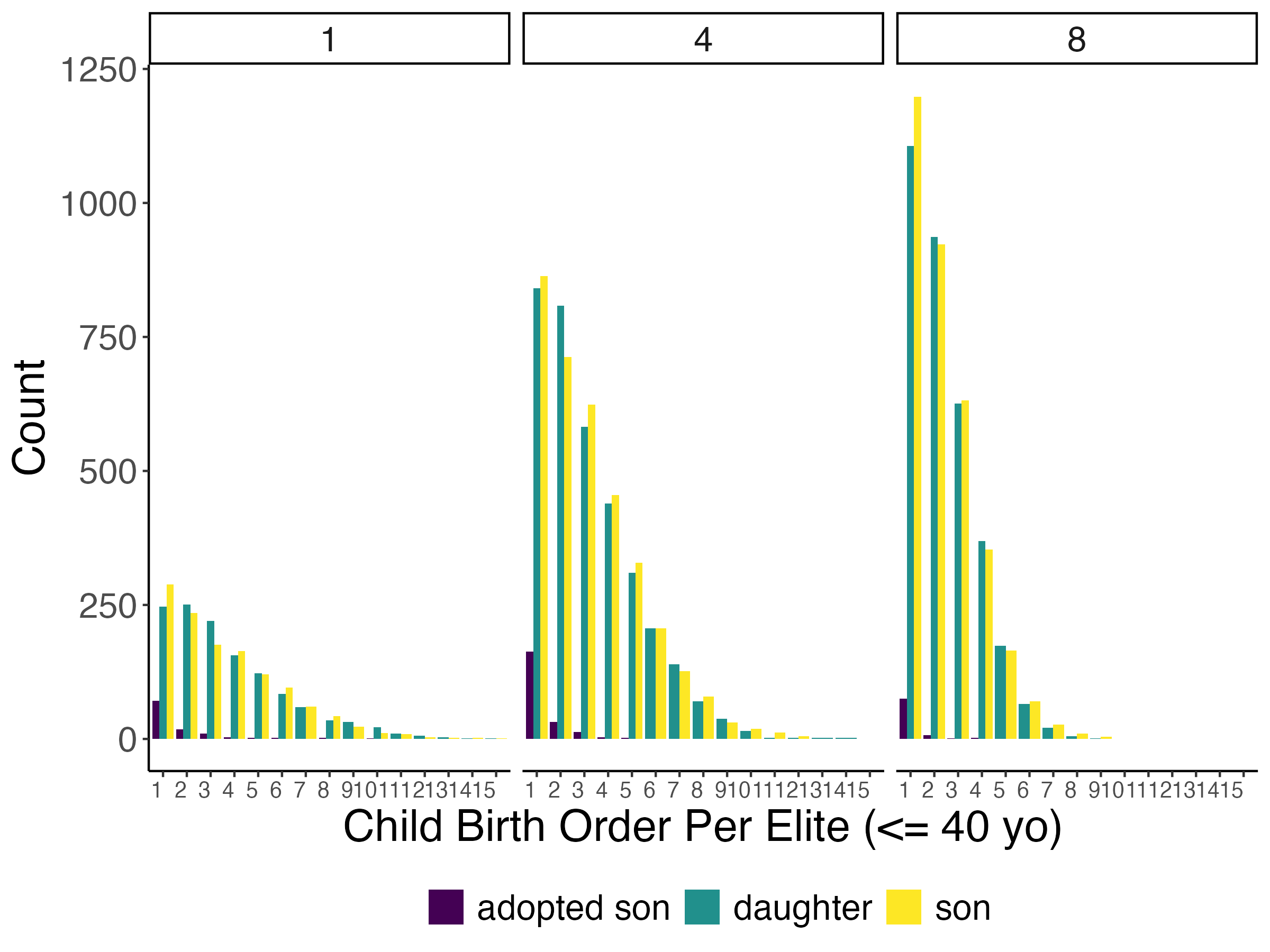}
    }
    \end{center}
    \caption{Family Composition Change}
    \label{fg:histogram_number_of_children_type_per_birthorder}
    \footnotesize
    \textit{Note:} These figures show the distribution of children recorded for listed elites across PIR editions. Panel (a) shows the distribution of the number of adopted sons, biological daughters, and biological sons per elite, calculated separately by child type. Panel (b, left) presents the distribution by overall birth order within the family. We calculate the order of all children in a family based on age, then count the number of adopted sons, daughters, and sons at each order. Panel (b, right) applies the same calculation as panel (b, left), but restricts the sample to elite fathers who were first listed in the PIR before the age of 40, to account for the possibility that daughters in older households may have already married into other families by the time of listing. Tables \ref{tb:summary_statistics_daughter_version_covariates}, \ref{tb:summary_statistics_son_version_covariates}, and \ref{tb:summary_statistics_adopted_son_version_covariates} show the summary statistics of adopted sons, daughters, and sons.
\end{figure}

Figure \ref{fg:histogram_number_of_children_type_per_birthorder} examines changes in family composition across PIR editions. Panel (a) shows that, among families with adopted sons, most had one adopted son. The number of families falls sharply from one adopted son to two adopted sons in all editions. This pattern is consistent with the practice of adopting a single male heir from outside the biological lineage \citep{mehrotra_adoptive_2013, kumanomido_elite_2026}.

Panel (b, left) shows child type (adopted sons, daughters, and biological sons) by overall birth order within the family. Adopted sons are often placed at the oldest birth order, while daughters are underrepresented among firstborn children. This pattern suggests that adopted sons were often brought into elite families as husbands of daughters and as heirs to the family line.

Panel (b, right) restricts the sample to elite fathers who were first listed before age 40. This restriction helps address the concern that older daughters may have already married out of the household before the family was recorded in the PIR. Even in this younger sample, daughters remain underrepresented relative to biological sons. This descriptive evidence on elites' children suggests that the underrepresentation of daughters among firstborn children may partly reflect the custom of non-reporting of daughters who had married out of the family, but it is unlikely to be fully explained by this factor alone.

\clearpage

\paragraph{Intergenerational Transmission of Elite Status}\label{result:intergeneration}

\begin{table}[!htbp]
  \begin{center}
      \caption{Intergenerational Transmission of Elite Status Across PIR Editions}
      \label{tb:intergenerational_elite_num_across_version}
      \subfloat[Number of Elites' Sons]{
\begin{tabular}[t]{r|rrrrr}
\toprule
Parent PIR & Child PIR 1 & Child PIR 4 & Child PIR 8 & Child PIR 10 & Child PIR 12\\
\midrule
1 & 15 & 155 & 285 & 313 & 464\\
4 & 0 & 47 & 798 & 821 & 1942\\
8 & 0 & 0 & 104 & 366 & 1505\\
\bottomrule
\end{tabular}
} \vspace{0.3in} \\
      \subfloat[Proportion of Elites' Sons in Each PIR Edition]{
\begin{tabular}[t]{r|rrrrr}
\toprule
Parent PIR & Child PIR 1 & Child PIR 4 & Child PIR 8 & Child PIR 10 & Child PIR 12\\
\midrule
1 & 0.005 & 0.011 & 0.011 & 0.012 & 0.009\\
4 & 0.000 & 0.003 & 0.032 & 0.032 & 0.036\\
8 & 0.000 & 0.000 & 0.004 & 0.014 & 0.028\\
\bottomrule
\end{tabular}
}        
  \end{center}\footnotesize
  \textit{Note}: These tables show the intergenerational transmission of elite status across editions. Panel (a) reports the number of elites' sons (including both biological and adopted sons) who appear in each PIR edition (columns). Each cell indicates the number of sons who appear in the PIR edition shown in the column, and whose fathers were listed in the PIR edition shown in the row. Panel (b) reports the share of such sons relative to the total number of elites in each PIR edition (column-wise).
\end{table} 

Lastly, we describe the intergenerational transmission of elite status from elite fathers to their sons. We define elites as individuals listed in any edition of the PIR and include both biological and adopted sons.

Panel (a) of Table \ref{tb:intergenerational_elite_num_across_version} reports the number of elite sons listed in each PIR edition. Rows indicate the edition in which their fathers were listed, and columns indicate the edition in which the sons were listed. Because the total number of listed individuals increased substantially over time, panel (b) reports the corresponding column-wise shares, using the total number of listed elites in each edition as the denominator.

The table shows that sons of elites listed in early editions appear most frequently in later editions. Among fathers listed in 1903, the share of sons peaks in 1934, while among fathers listed in 1915, it peaks in 1939. Although sons of listed elites account for only about 0.1\% of the relevant male birth cohorts, they account for about 3\% of listed individuals in PIR edition 12. As a result, sons of listed individuals were substantially overrepresented among elites, suggesting strong intergenerational persistence of elite status.

\section{Elite Formation in Early Twentieth-Century Japan}\label{sec:result}

In this section, we examine three aspects of elite formation in early twentieth-century Japan: geographical mobility, assortative matching in marriage, and the husband--wife age gap at marriage in elite family formation. Together, these analyses treat elite formation as a process shaped not only by where elites built their careers, but also by how elite households organized marriage and succession. First, we study where listed elites were born and where they lived, using prefecture-level migration flows and gravity-style estimates. Second, we examine age assortative matching among elite marriages by reconstructing spouses' age at marriage from family information in the PIR. Third, we study how the husband--wife age gap at marriage is associated with completed family size and the use of adoption as an heir-substitution margin.

\subsection{Geographical Mobility}

Geographical mobility helps us understand elite formation in early twentieth-century Japan. During this period, the broader process of modernization, including the expansion of modern education, broadened access to bureaucratic institutions, and the introduction of new technologies from abroad, may have created new opportunities for individuals to achieve elite positions. At the same time, local family networks and regionally rooted business or political connections may have helped maintain inherited advantages \citep{nakamura2010local, tanimoto2007wealthy}. Examining where elites were born and where they lived therefore allows us to characterize two broad patterns of elite formation: locally rooted elite formation and elite formation through movement toward metropolitan centers.

First, we show the spatial distribution of listed individuals by prefecture. Figures \ref{fg:listing_num_pref_ver12} and \ref{fg:listing_ratio_pref_ver12} present the geographical variation in the number of elites and the share of elites relative to the population across PIR editions. We use prefectural population data from 1920 for editions 1 and 4, and from 1930 for editions 8, 10, and 12.\footnote{Population census in Japan was conducted every five years from 1920.} Both the number and share of elites were consistently higher in major metropolitan areas such as Tokyo, Kyoto, and Osaka. There are two possible explanations for this pattern. First, as documented by \citet{chetty_where_2014, chetty_impacts_2018-1, chetty_impacts_2018, chetty_opportunity_2018}, individuals who grow up in urban areas may experience greater social and intergenerational mobility. This suggests that some elites from non-elite backgrounds may have achieved elite status through upward mobility opportunities in urban environments. Second, some elites who were born in rural prefectures may have migrated to urban centers in search of better educational and economic opportunities.

\begin{figure}[htbp]
    \begin{center}
    \subfloat[Edition 1 (1903)]{
    \includegraphics[width = 0.45\textwidth]{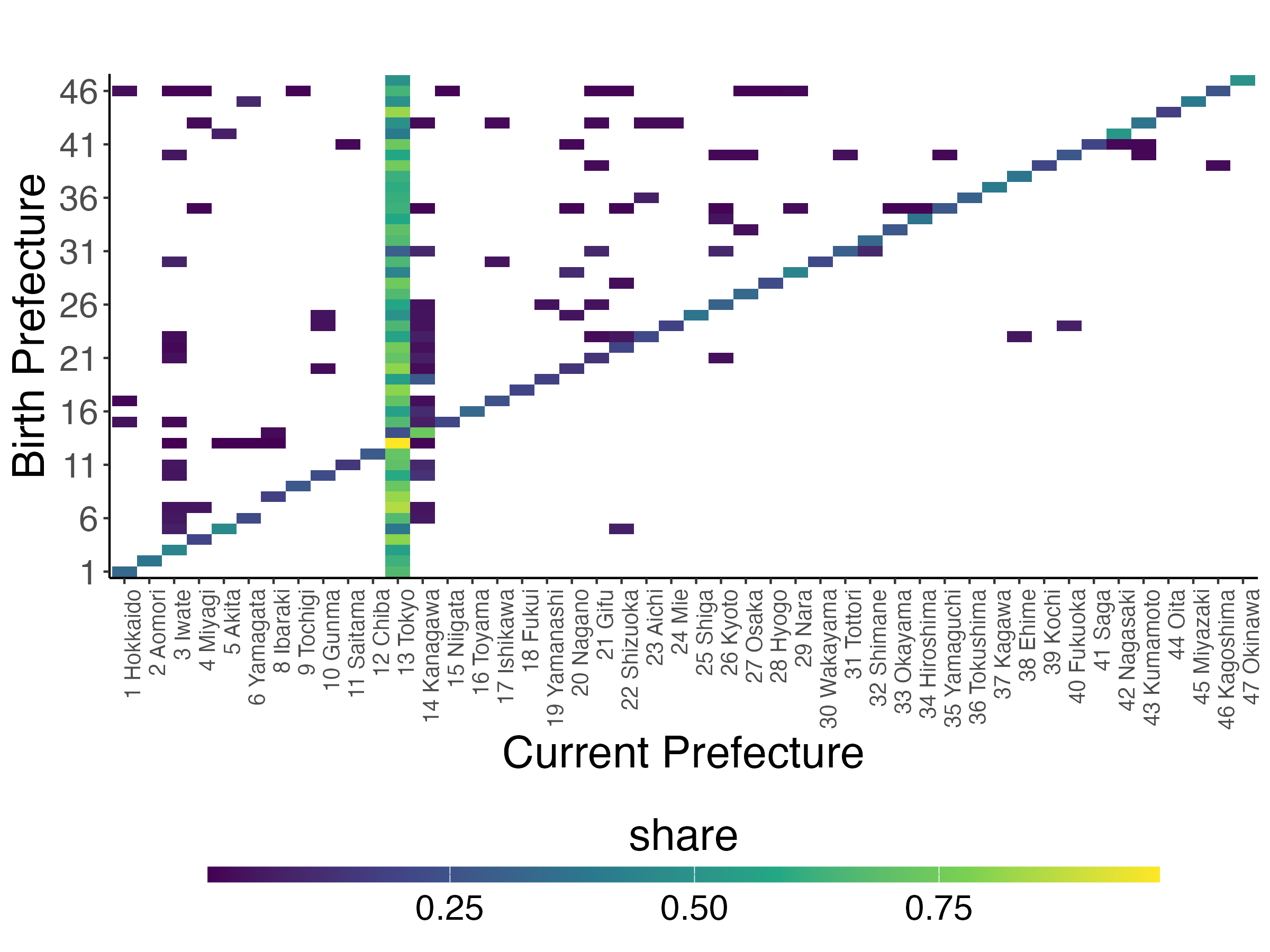}
    }
    \subfloat[Edition 4 (1915)]{
    \includegraphics[width = 0.45\textwidth]{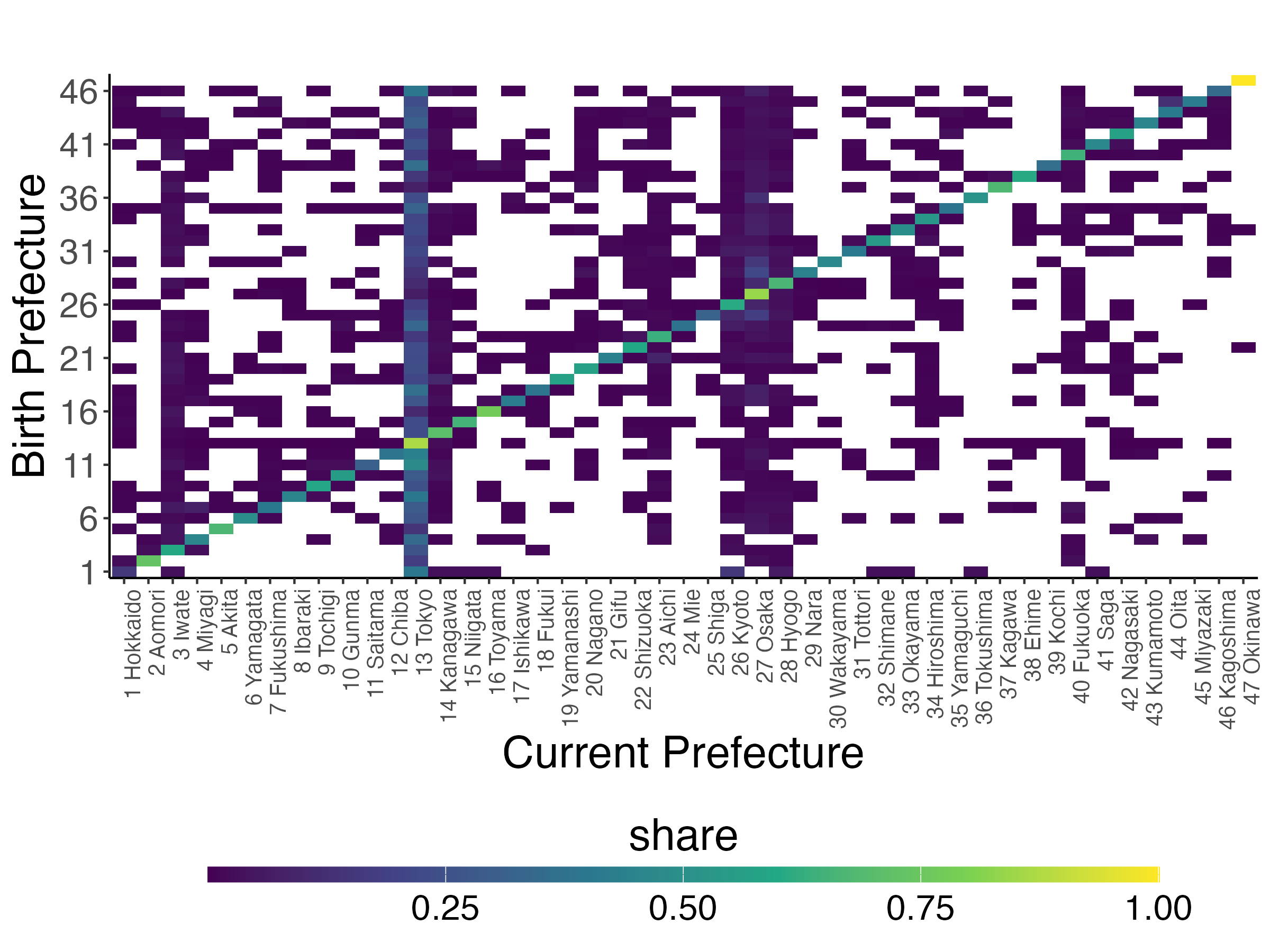}
    }\\
    \subfloat[Edition 8 (1928)]{
    \includegraphics[width = 0.45\textwidth]{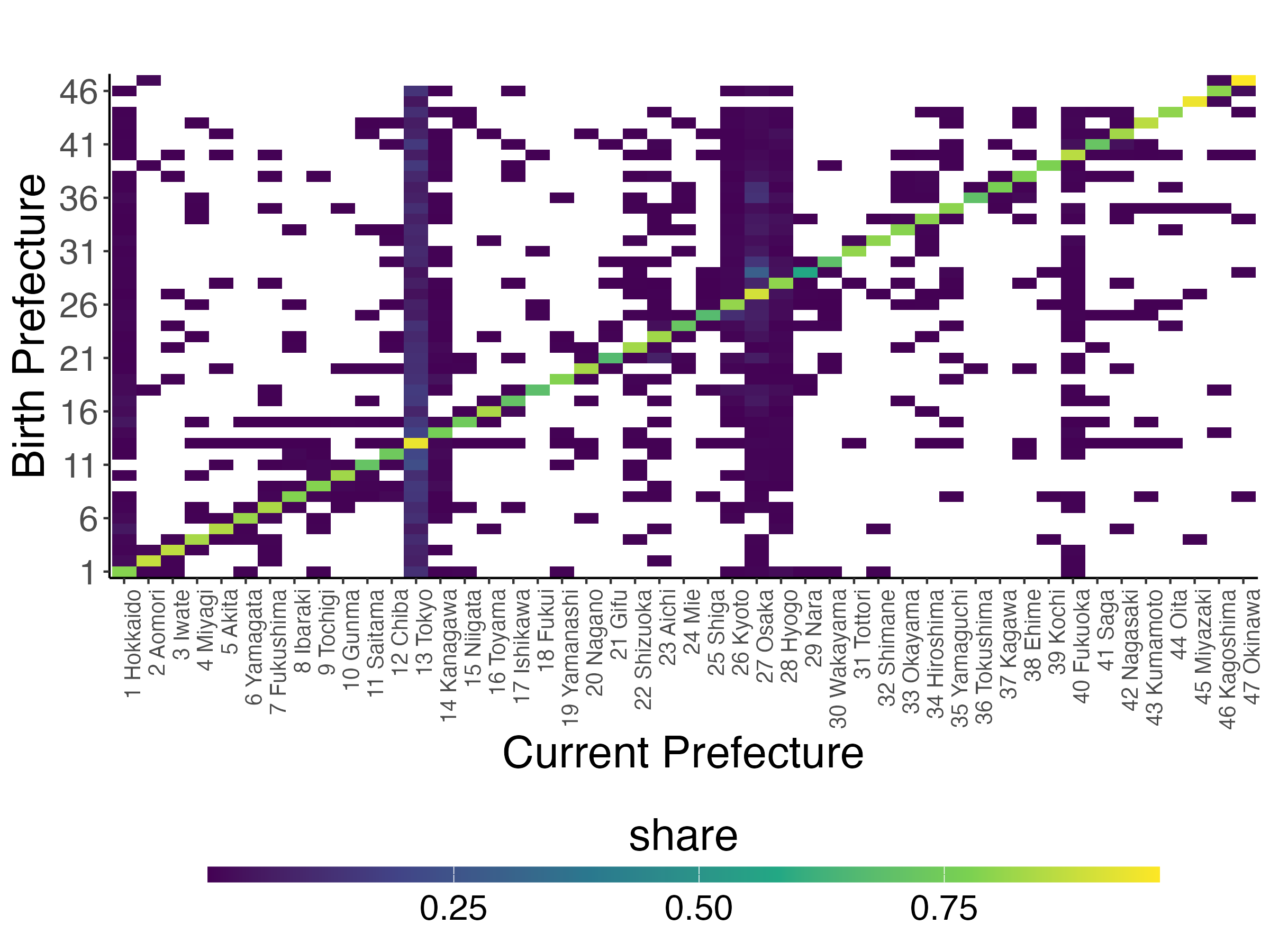}
    }
    \subfloat[Edition 10 (1934)]{
    \includegraphics[width = 0.45\textwidth]{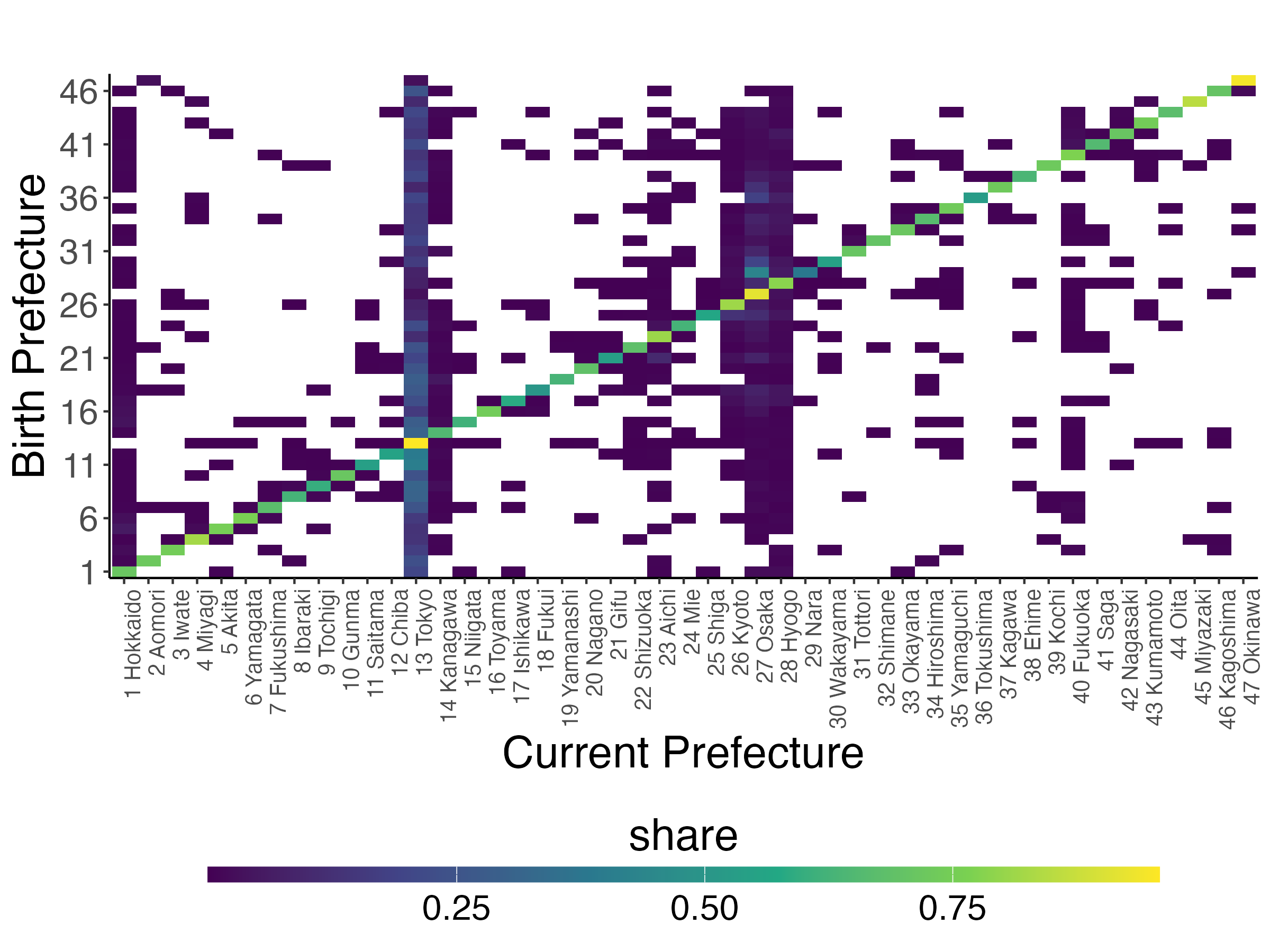}
    }\\
    \subfloat[Edition 12 (1939)]{
    \includegraphics[width = 0.45\textwidth]{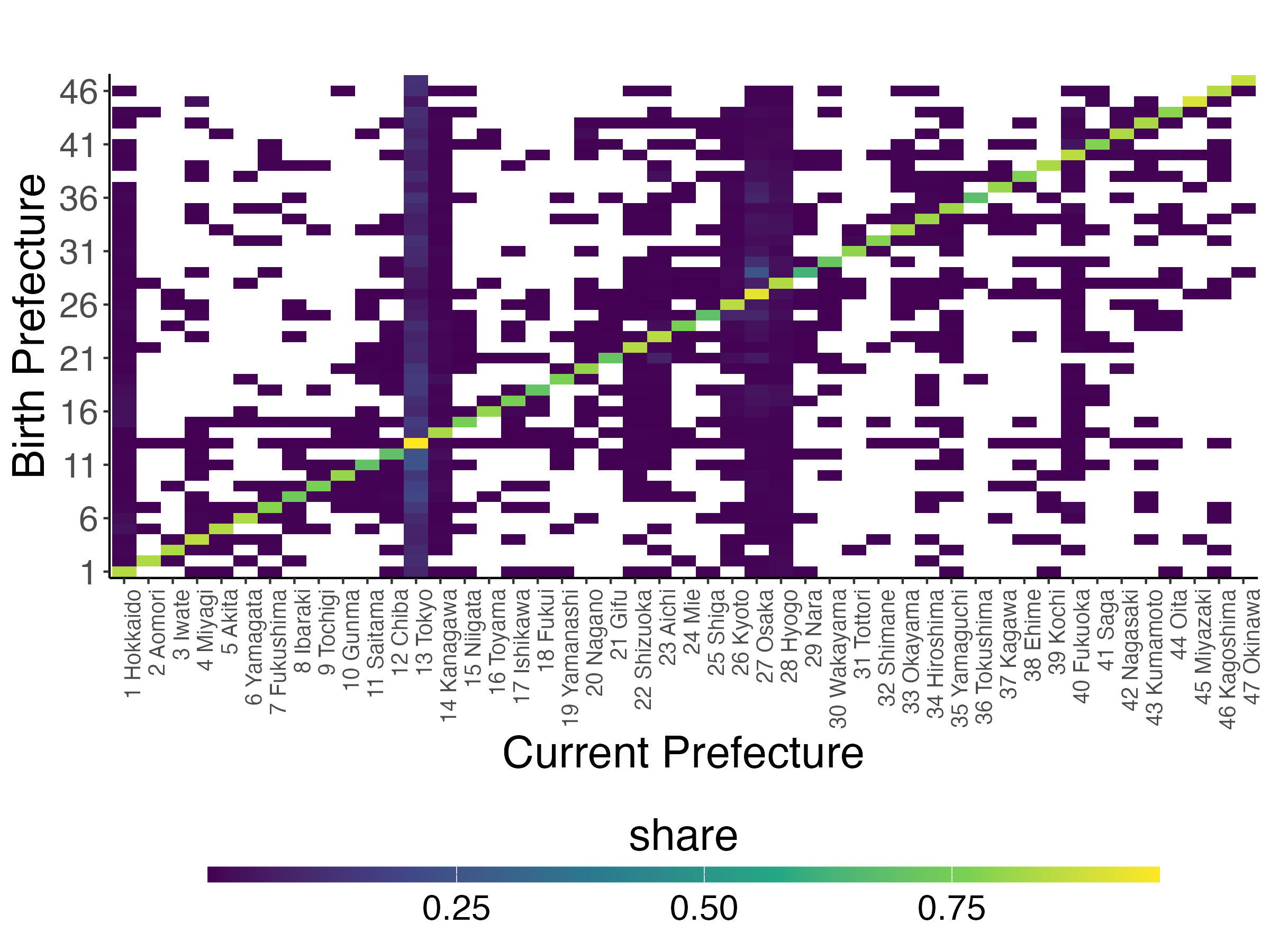}
    }
    \end{center}
    \caption{Elite Mobility Matrix by 47 $\times$ 47 Prefectures}
    \label{fg:matrix_pref_flow12}
    \footnotesize
    \textit{Note:} The figures show the flow of spatial mobility from birth prefecture to current residential prefecture by edition. Each cell represents the share of individuals who moved from their birth prefecture to the destination prefecture based on information about their birthplace and current residential address at the time of being listed. Individuals with missing residential information or who live in foreign countries at the time of listing are excluded.
\end{figure}

To examine the spatial mobility of elites, we calculate the flows from birth prefectures to current residential prefectures across PIR editions. Figure \ref{fg:matrix_pref_flow12} presents the results. From 1915 to 1939, the share of individuals who remained in their birth prefecture was highest in most regions. Also, urbanized prefectures such as Kanagawa, Kyoto, Osaka, Hyogo, and Fukuoka exhibit significant inflow, indicating increasing migration toward metropolitan areas. 

These results suggest two aspects of elites' spatial mobility during this period. First, many elites remained in their birth prefecture throughout their careers. They possibly maintained strong local networks and continued their careers in their birth prefectures. Second, access to economic and social opportunities in major urban areas might be one of the key factors behind the increasing number of elites in metropolitan areas. Overall, the results show both the persistence of local elite structures and the emergence of new metropolitan elites.

To quantify these mobility patterns, we estimate a gravity-style model. For each edition, we construct a prefecture-pair panel by counting the number of listed elites born in prefecture $o$ and residing in prefecture $r$. This gives a balanced panel of $47 \times 47$ origin--destination pairs for each edition. Geographical distance is measured by the centroid distance between prefectures. For within-prefecture observations, we define internal distance as $Distance_{oo} = 0.67 \sqrt{\frac{Area_o}{\pi}}$, which approximates the average distance within a prefecture.

We estimate the following Poisson pseudo-maximum likelihood specification:
\begin{equation}\label{eq:ppml_gravity}
Flow_{ort} = \exp \left(\beta_t Distance_{or} + FE_{ot} + FE_{rt} \right),
\end{equation}
where $Flow_{ort}$ is the number of listed elites born in prefecture $o$ and residing in prefecture $r$ in edition $t$. The terms $FE_{ot}$ and $FE_{rt}$ denote origin-by-edition and destination-by-edition fixed effects. These fixed effects absorb time-varying prefectural characteristics, such as population size, urbanization, industrial development, and other local trends.

\begin{figure}[htbp]
    \begin{center}
    \includegraphics[width = 0.7\textwidth]{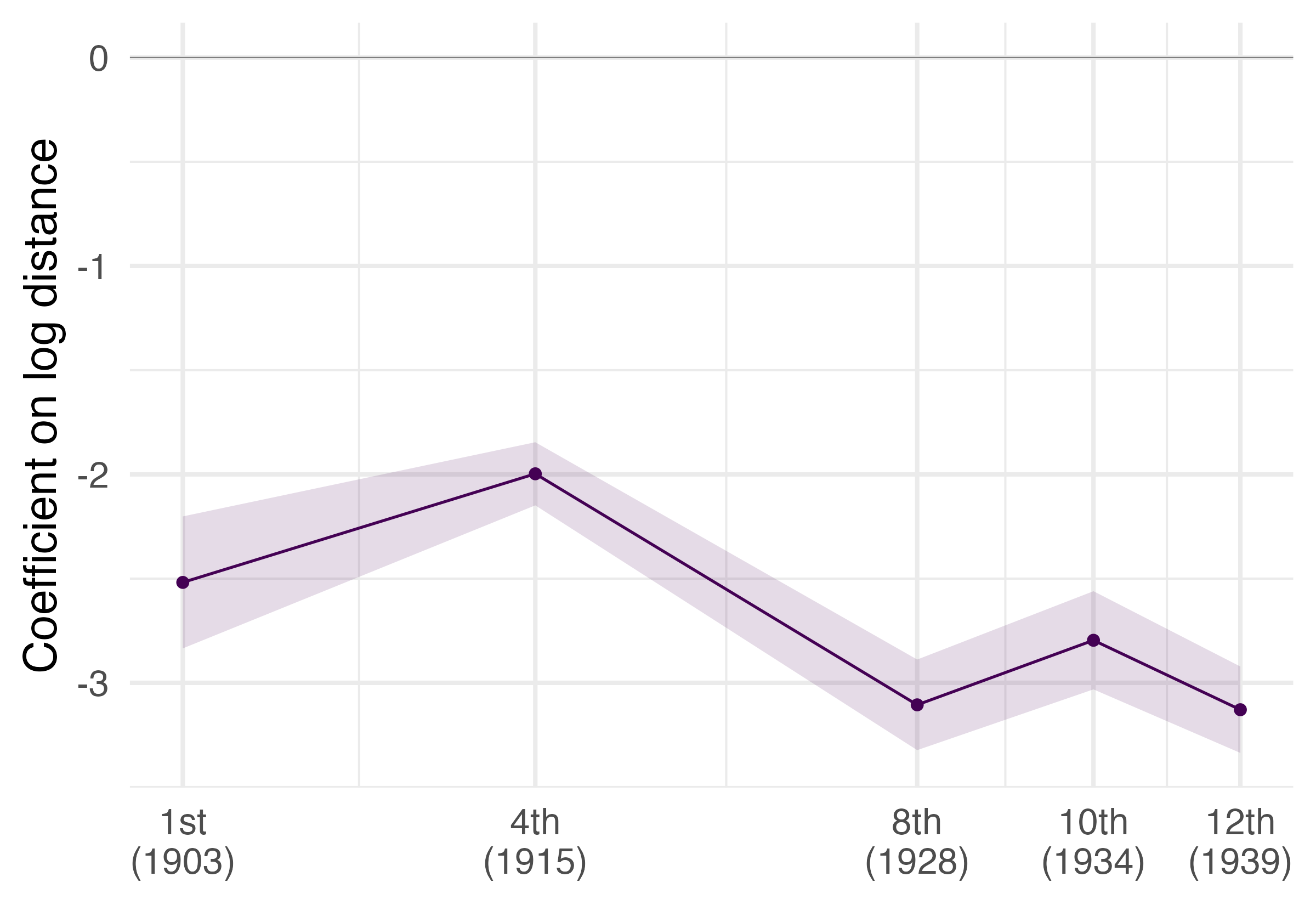}
    \end{center}
    \caption{Coefficient Plots on Distance in PPML Gravity}\label{fig:gravity_ot_rt_fe}
    \footnotesize
    \textit{Note:} This figure reports the estimated coefficient on geographical distance from Poisson pseudo-maximum likelihood gravity regressions. The dependent variable is the number of listed elites between origin and destination prefectures in each PIR edition. All specifications include origin-by-edition and destination-by-edition fixed effects. Vertical bars indicate 95\% confidence intervals. Standard errors are clustered at the origin-destination pair level.
\end{figure}

Figure~\ref{fig:gravity_ot_rt_fe} reports the estimated distance coefficients by edition. The coefficients are negative and statistically significant in all editions, indicating that distance constrained elite mobility. The magnitude of the coefficient becomes larger over time. This does not necessarily mean that migration costs increased. During this period, transportation and communication networks improved, which should have reduced the cost of mobility. A more plausible interpretation is that later PIR editions included more locally influential individuals. As the coverage of the PIR expanded, the sample increasingly captured elites who remained active within their birth prefectures.

We conduct two additional checks (Figure~\ref{fig:gravity_ot_rt_fe_sub}). First, we exclude observations related to major metropolitan prefectures, including Tokyo, Osaka, Kyoto, Aichi, and Kanagawa. The results are similar, suggesting that the stronger distance effect is not driven only by metropolitan concentration. Second, we exclude within-prefecture observations. In this case, the distance coefficient becomes much smaller, indicating that the growing representation of local elites is an important part of the aggregate pattern.

\begin{figure}[htbp]
    \begin{center}
    \includegraphics[width = 0.7\textwidth]{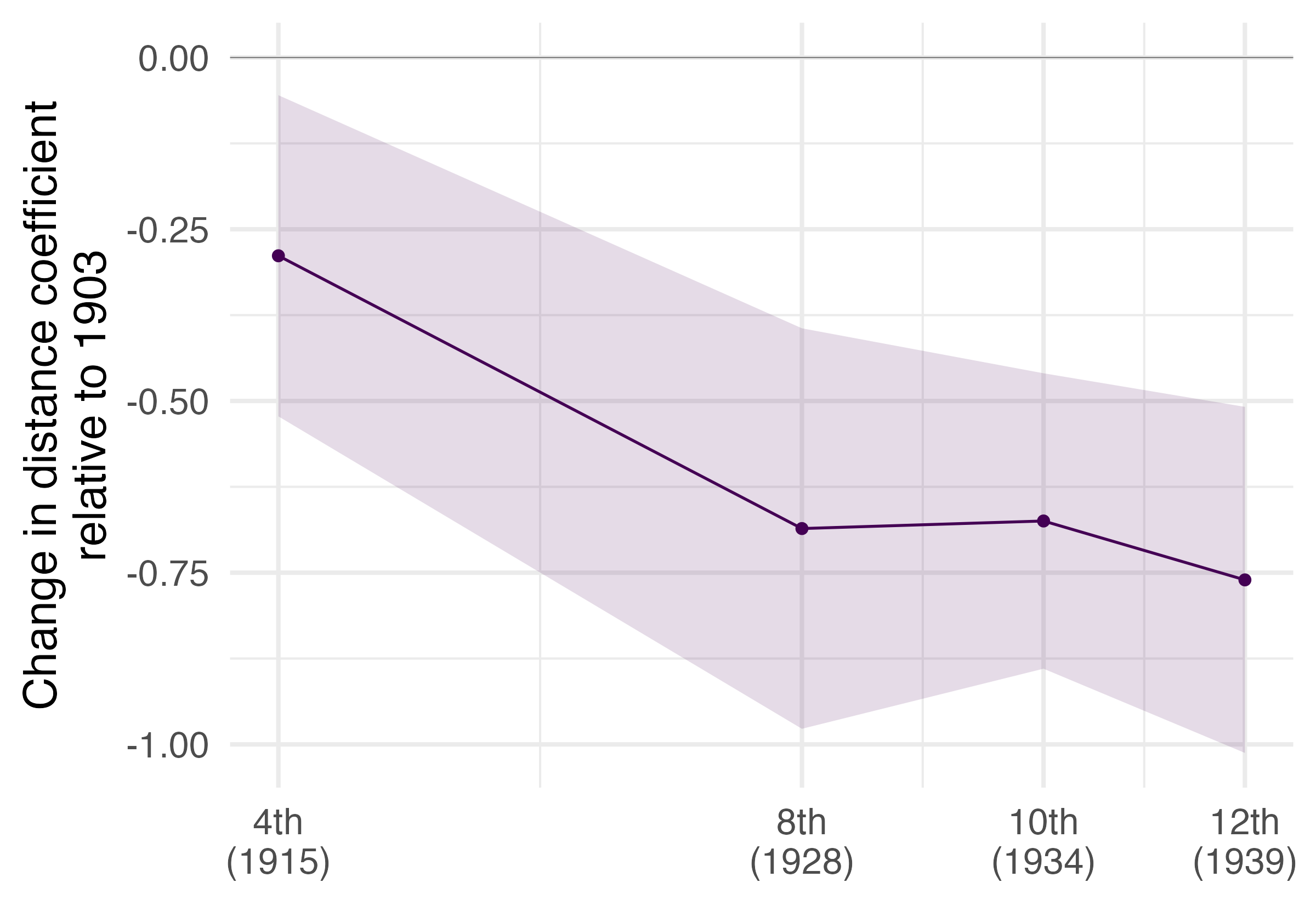}
    \end{center}
    \caption{Coefficient Plots on Distance in PPML Gravity with Origin-Destination Fixed Effect}\label{fig:gravity_or_t_fe}
    \footnotesize
    \textit{Note:} This figure reports estimates from a specification including origin--destination pair fixed effects and edition fixed effects. Coefficients are normalized relative to Edition 1 (1903) and capture changes in the distance gradient over time. Vertical bars indicate 95\% confidence intervals. Standard errors are clustered at the origin-destination pair level.
\end{figure}

We then estimate an alternative specification with origin--destination pair fixed effects. This specification controls for time-invariant bilateral characteristics, such as historical trade links, cultural proximity, migration networks, and geographical barriers. Figure~\ref{fig:gravity_or_t_fe} presents the results. Relative to edition 1, the magnitudes of distance coefficients become larger in later editions. Taken together, the baseline and pair-fixed-effect results suggest that the apparent strengthening of the distance effect is largely driven by changes in sample composition. Later PIR editions appear to include more local elites, rather than showing a simple increase in the importance of distance.

Finally, we compare the mobility of elites with that of the general male population. Using the 1920 and 1930 population censuses, we calculate the share of males born in each prefecture who lived outside their birth prefecture. We compare these rates with the corresponding outflow rates among listed individuals. The 1920 census is matched to edition 4, and the 1930 census is matched to edition 8.

\begin{figure}[htbp]
    \begin{center}
    \includegraphics[width = 0.9\textwidth]{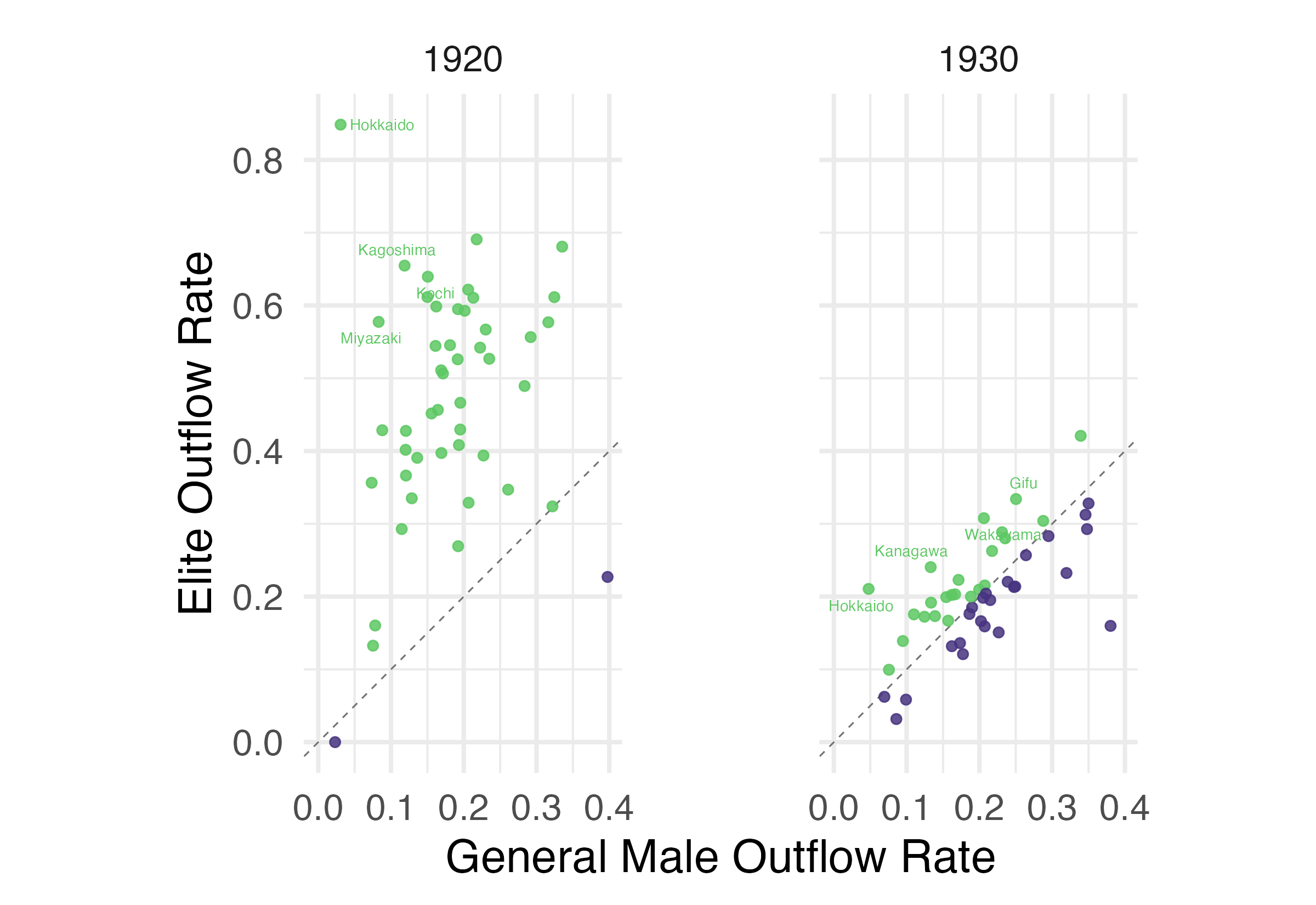}
    \end{center}
    \caption{Comparison of Outflow Rate between Census and PIR-listed Elites}\label{fig:elite_general_mobility}
    \footnotesize
    \textit{Note:} This figure compares prefecture-level out-migration rates of PIR-listed elites and the general male population. The horizontal axis shows the share of males residing outside their birth prefecture based on population census data, while the vertical axis shows the corresponding share among PIR-listed elites. The left panel matches the 1920 Population Census with Edition 4 (1915), and the right panel matches the 1930 Population Census with Edition 8 (1928). The dashed 45-degree line indicates equal mobility between elites and the general population. Prefectures above the line exhibit higher mobility among observed elites.
\end{figure}

Figure \ref{fig:elite_general_mobility} shows that elites were more mobile than the general male population in edition 4. However, this difference should be interpreted with caution. Early PIR editions covered only a small fraction of the population, so the observed sample may disproportionately consist of highly successful individuals whose careers required large-scale migration. Consequently, selection into the PIR may mechanically generate higher measured mobility. In edition 8, the difference between elites and the general population becomes much smaller, with many prefectures close to the 45-degree line. Given the substantial expansion of PIR coverage between editions, this convergence is consistent with a reduction in selection bias rather than a decline in elite mobility itself.

Overall, these results suggest that elite formation in this period was characterized by two different patterns. Some elites moved from their birth prefectures to urban centers, while many others remained active within their local communities. Thus, the geographical pattern observed in the PIR should be interpreted not simply as a change in mobility costs, but as the coexistence of metropolitan elite formation and locally rooted elite formation.

\subsection{Age Assortative Matching}\label{sec:age_assortativeness}

The previous subsection shows that elite formation was shaped by both migration to urban centers and the persistence of local elite structures. We now turn to marriage, another important aspect of elite formation. Age at marriage provides a simple way to examine matching patterns in the elite marriage market. Historical studies show that men generally married at older ages than women in late nineteenth-century Japan, although marriage ages varied substantially across regions and occupational groups \citep{hayami1987fossa, saito1988edo}. We examine whether elite men and women tended to marry partners of similar ages and whether the age gap between husbands and wives changed across marriage cohorts.

Measuring age sorting requires us to distinguish age at listing from age at marriage. Then, we approximate each couple's \emph{age at marriage} by the birth year of the couple's eldest biological child minus one. Pooling all editions, we assign each marriage to a ten-year \emph{marriage cohort} and retain cohorts with at least $1{,}000$ marriages: the 1870s through the 1920s, with $25{,}459$ marriages in total. For each cohort, we group husbands and wives into the same age bins. We then construct an age matching matrix $M=(\mu_{ij})$, where each cell $\mu_{ij}$ counts the number of marriages between husbands in age bin $i$ and wives in age bin $j$. The diagonal cells of this matrix correspond to marriages between husbands and wives in similar age bins. Therefore, when more marriages are observed along the diagonal, we interpret this as stronger age assortative matching.

Comparing assortativeness across cohorts is not a matter of simply counting same-aged couples, because the marginal age distributions themselves change over time. As \citet{chiappori2025changes} emphasize, the central difficulty is to separate a genuine change in the matching \emph{structure} from the mechanical effect of shifting marginals. The natural reference point is statistical independence: if a husband's and a wife's ages were unrelated, the matching matrix would simply be the outer product of the marginal age distributions, $\mu_{ij}^{0} = \mu_i \mu_j / \lvert M \rvert$, where $\mu_i = \sum_k \mu_{ik}$, $\mu_j = \sum_k \mu_{kj}$, and $\lvert M \rvert = \sum_{i,j}\mu_{ij}$. Assortative matching is the departure of the observed $M$ from this benchmark, and recent work argues axiomatically that a credible measure of it should be (i) invariant to relabeling or rescaling the types, (ii) increasing when mass shifts toward like-with-like pairings, and---decisively for a cross-cohort comparison---(iii) \emph{marginal independent}, i.e., unchanged by shifts in marginal age distributions that leave the association structure intact. We deliberately rely on model-free measures rather than estimating a structural matching model such as the transferable-utility model of \citet{choo2006estimating}, for two reasons. First, model-free indices are computed from the matched couples alone and require no data on the unmatched (single) population on either side of the market; this is essential here because the PIR records only marriages and provides no sample of unmarried individuals, in particular none for women, so the masses of singles that a structural surplus requires are simply unobserved. Second, they impose no parametric assumptions on preferences or on the distribution of idiosyncratic match-specific tastes, whereas recovering a structural surplus requires such assumptions (for example, type-I extreme-value shocks). We therefore report two transparent, model-free indices and read them against each other.

The first measure is the aggregate likelihood ratio (ALR), the ratio of the observed same-age (diagonal) mass to the mass expected under independence,
\begin{equation}\label{eq:alr}
  \mathrm{ALR}(M)
  \;=\;
  \frac{\sum_i \mu_{ii}}{\sum_i \mu_{ii}^{0}}
  \;=\;
  \frac{\sum_i \mu_{ii}}{\sum_i \mu_i \mu_i / \lvert M \rvert}.
\end{equation}
By construction, $\mathrm{ALR}=1$ under independence, while $\mathrm{ALR}>1$ indicates that spouses of the same age marry more often than would be expected from the marginal age distributions. The ALR therefore provides an intuitive summary of diagonal concentration. It is, however, anchored to the strict diagonal and fails property (iii): it is not invariant to the marginals and can move even when the underlying sorting is unchanged. We therefore treat the ALR as a descriptive summary rather than a structural measure of sorting.\footnote{\citet{imamura2025note} modify an axiomatization of the aggregate likelihood ratio examined in \cite{chiappori2025changes}.}

The second measure addresses this limitation by reducing the matching table to two age groups. For a threshold $\tau$, we classify each spouse as young $(<\tau)$ or old $(\geq\tau)$ and collapse $M$ into a $2\times2$ table. Association in this table is measured by the odds ratio
\begin{equation}\label{eq:tor}
  \mathrm{OR}(\tau) \;=\; \frac{\mu_{\mathrm{YY}}\,\mu_{\mathrm{OO}}}{\mu_{\mathrm{YO}}\,\mu_{\mathrm{OY}}},
\end{equation}
An odds ratio greater than one, or equivalently a positive log odds ratio, indicates positive age sorting. Because the odds ratio is a cross-product of cells, multiplying any row or column by a positive constant leaves it unchanged: it satisfies marginal independence by construction. \citet{chiappori2025changes} show that, in two-type markets, the odds ratio is the \emph{unique} assortativeness ordering satisfying the basic axioms together with marginal independence, and that it has a structural interpretation as the average per-couple gain from assortative matching in a transferable-utility marriage model \citep{choo2006estimating}. It is therefore our preferred measure of age sorting.

Why report both? For more than two types, \citet{chiappori2025changes} prove an impossibility result: no single index can be at once monotone in the marginals and robust to how the underlying variable is discretized, so there is no fully satisfactory \emph{global} multitype measure. We thus follow their prescription of evaluating the marginal-invariant $2\times2$ odds ratio at a range of thresholds, while reporting the full-matrix ALR as a complementary---but marginal-confounded---summary. The contrast between the two is itself diagnostic: when the ALR trends across cohorts but the odds ratio does not, the trend reflects the shifting age marginals rather than a change in sorting.

\begin{figure}[htbp]
    \begin{center}
    \subfloat[Marriage counts by cohort (editions pooled)]{
    \includegraphics[width = 0.6\textwidth]{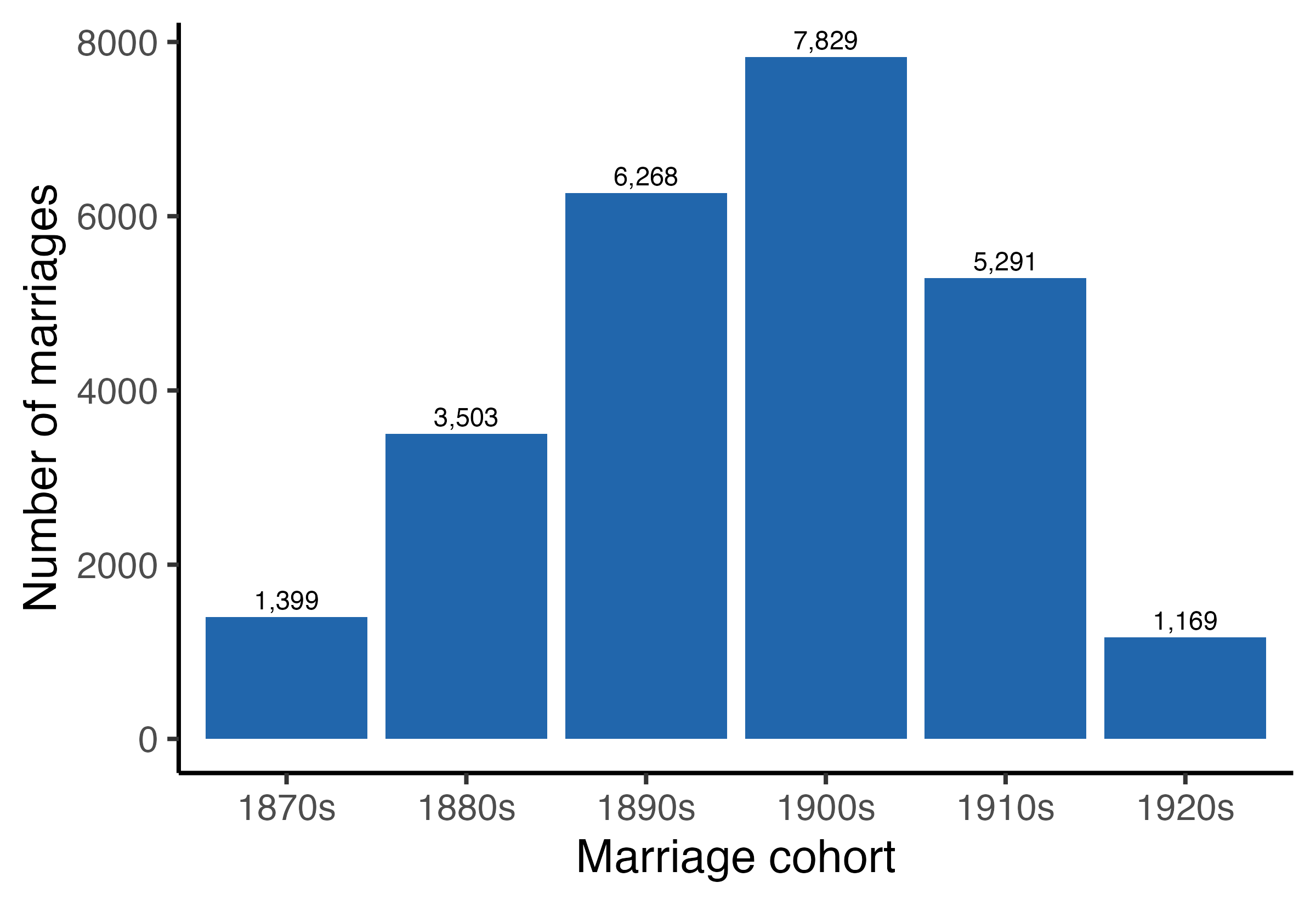}
    }\\
    \vspace{0.6in}
    \subfloat[Matched male age at marriage]{
    \includegraphics[width = 0.48\textwidth]{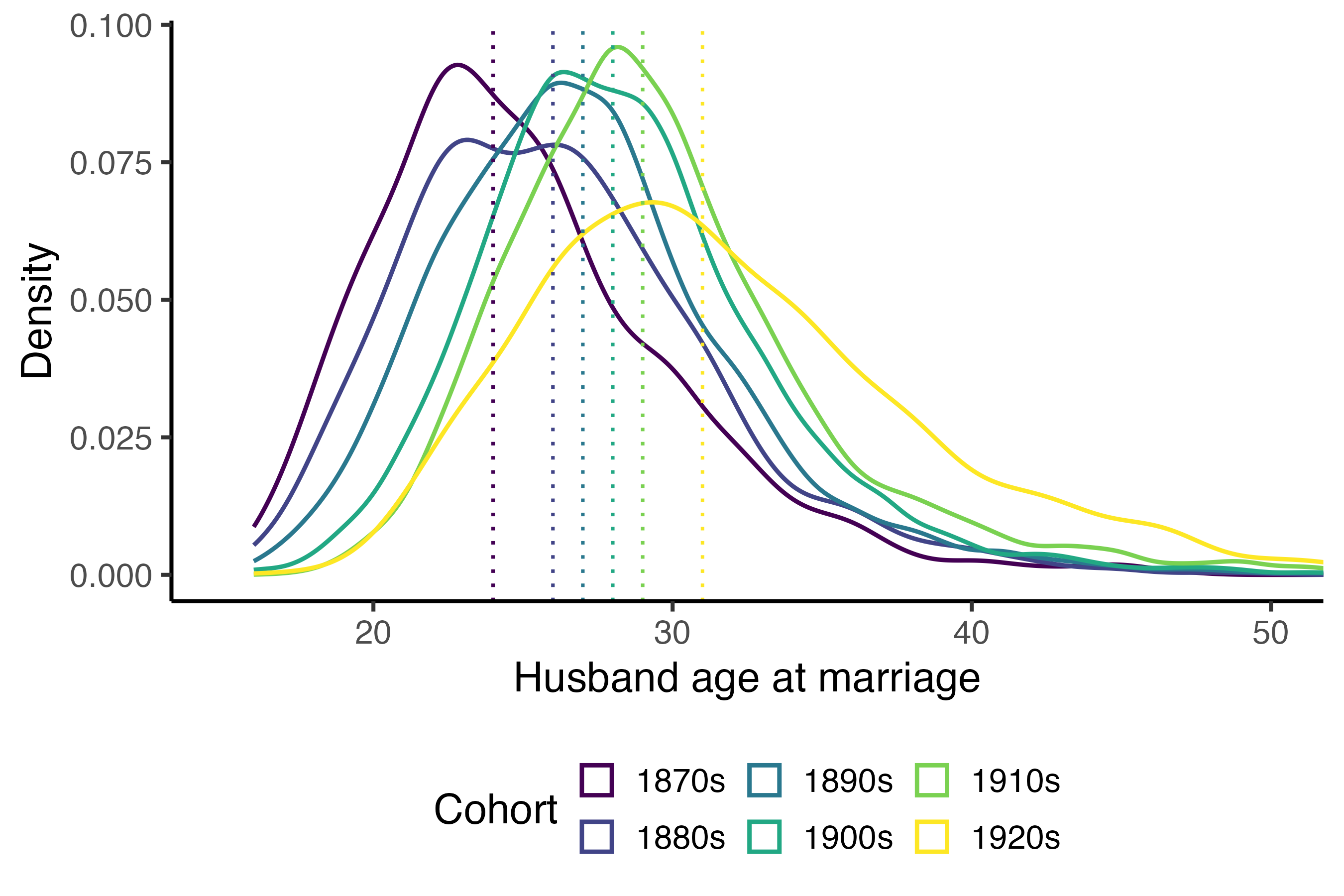}
    }
    \subfloat[Matched female age at marriage]{
    \includegraphics[width = 0.48\textwidth]{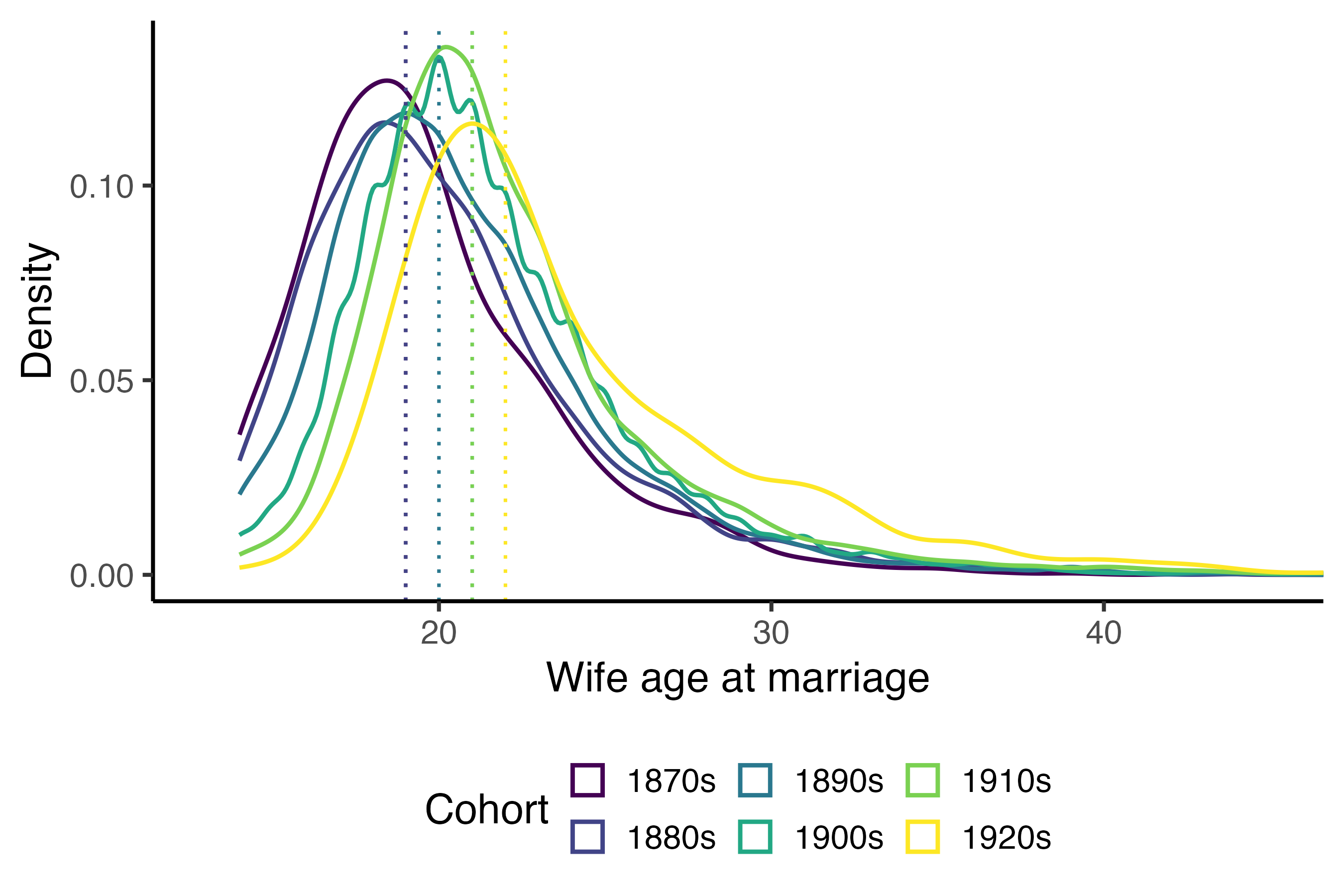}
    }
    \end{center}
    \caption{Cohort-Level Marriage Counts and Age-at-Marriage Marginal Distributions}\label{fg:cohort_counts_marginals}
    \footnotesize
    \textit{Note:} Marriages are pooled across PIR editions (one row per marriage) and assigned to ten-year marriage cohorts using the approximate marriage year (eldest biological child's birth year minus one). Panel (a) shows the number of marriages per cohort; only cohorts with at least $1{,}000$ marriages (1870s--1920s) are retained. Panels (b) and (c) show kernel densities of husbands' and wives' age at marriage by cohort, with dotted vertical lines marking cohort medians. Couples with implausible ages (husband outside $[16,55]$ or wife outside $[14,50]$) are excluded.
\end{figure}

Figure \ref{fg:cohort_counts_marginals} reports the number of marriages in each cohort in panel (a) and the marginal distributions of husbands' and wives' age at marriage (panels b and c). Marriage is concentrated at relatively young ages: the median age at marriage is 27 for husbands and 21 for wives. Both distributions nevertheless shift gradually to the right across cohorts. The mean age at marriage rises from approximately 25 to 32 for husbands and from 20 to 24 for wives, while the average husband--wife age gap widens from roughly five to eight years. The widening husband--wife age gap reflects changes in the age distributions of husbands and wives and does not imply a change in age sorting.

\begin{figure}[htbp]
    \begin{center}
    \subfloat[1870s]{
    \includegraphics[width = 0.32\textwidth]{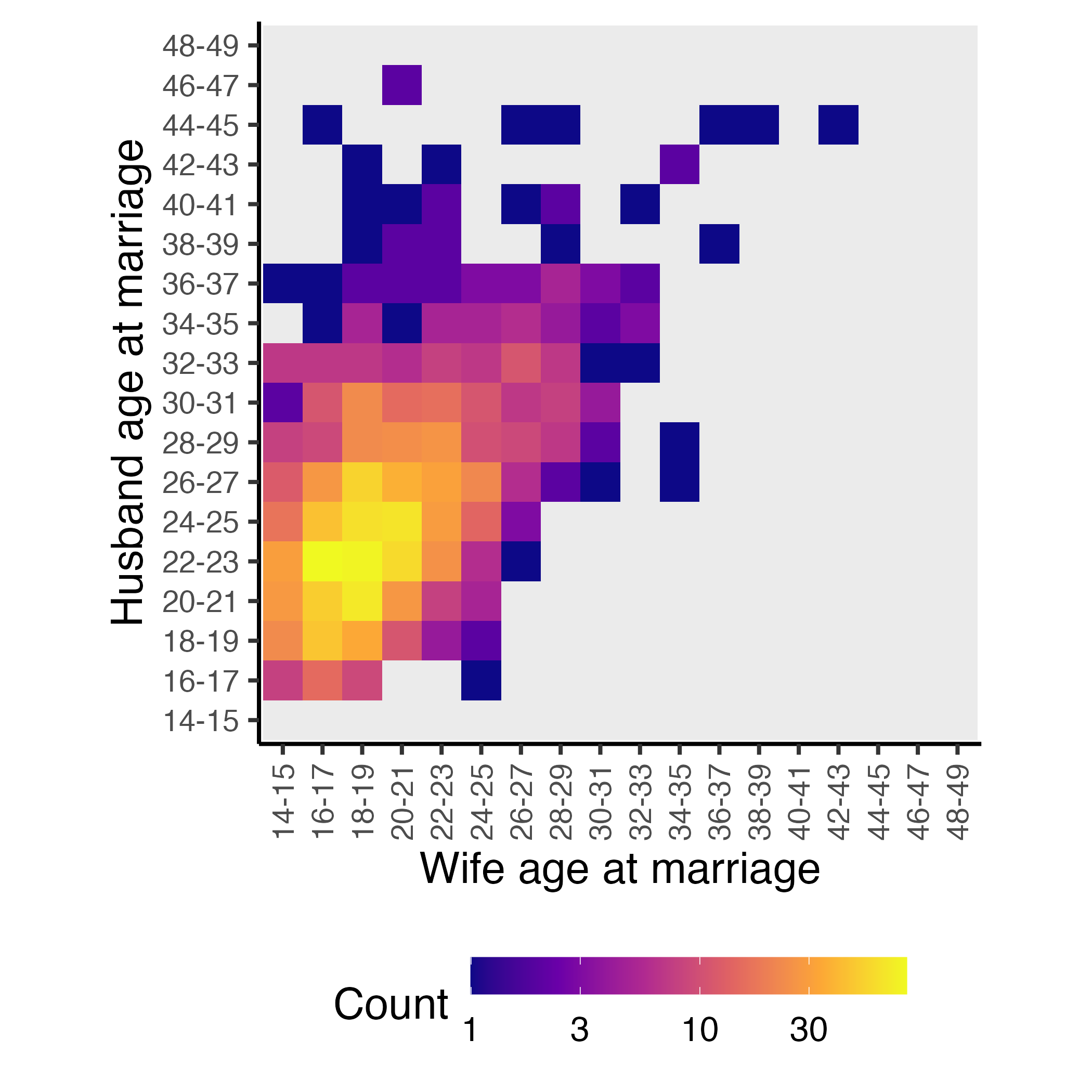}
    }
    \subfloat[1880s]{
    \includegraphics[width = 0.32\textwidth]{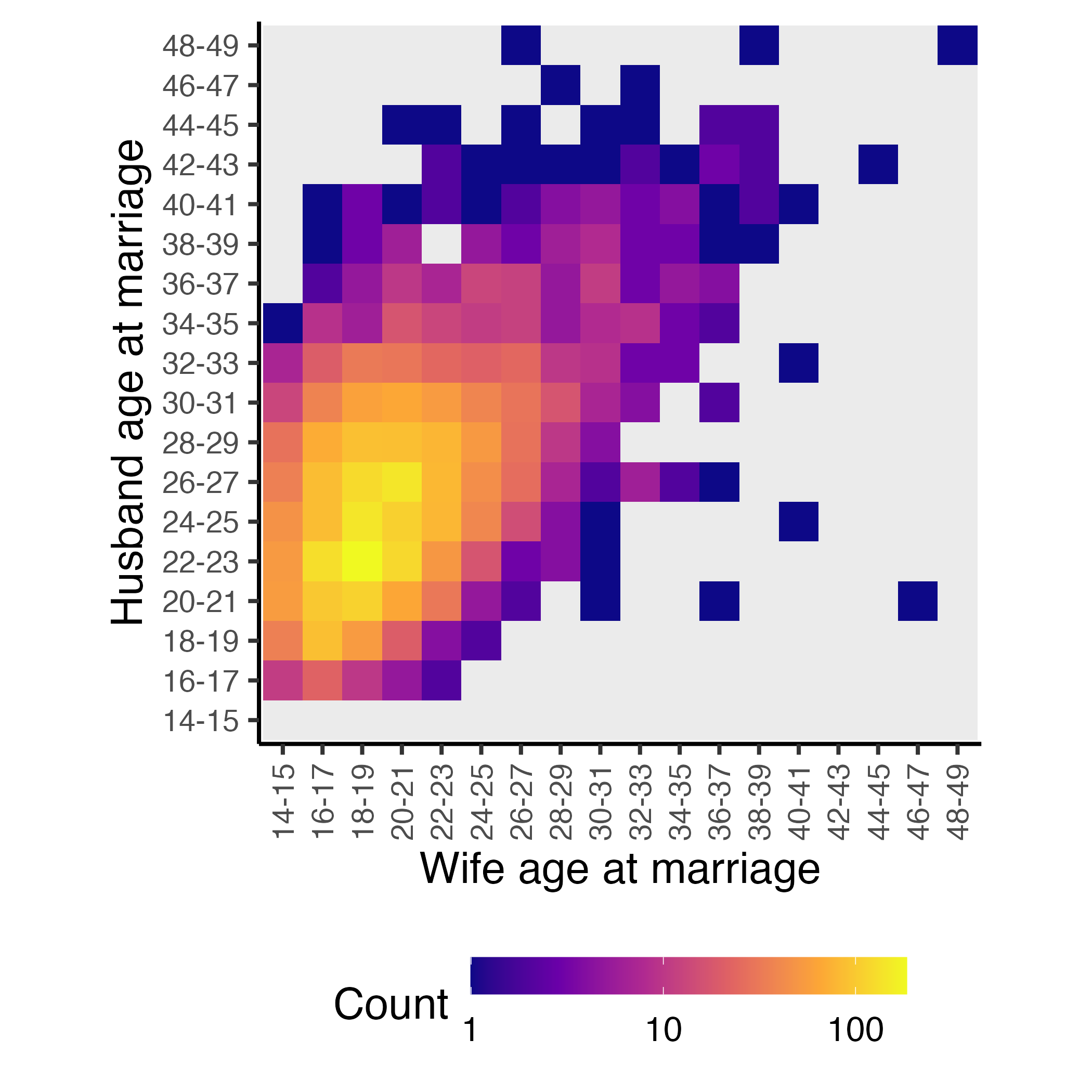}
    }\\\vspace{0.2in}
    \subfloat[1890s]{
    \includegraphics[width = 0.32\textwidth]{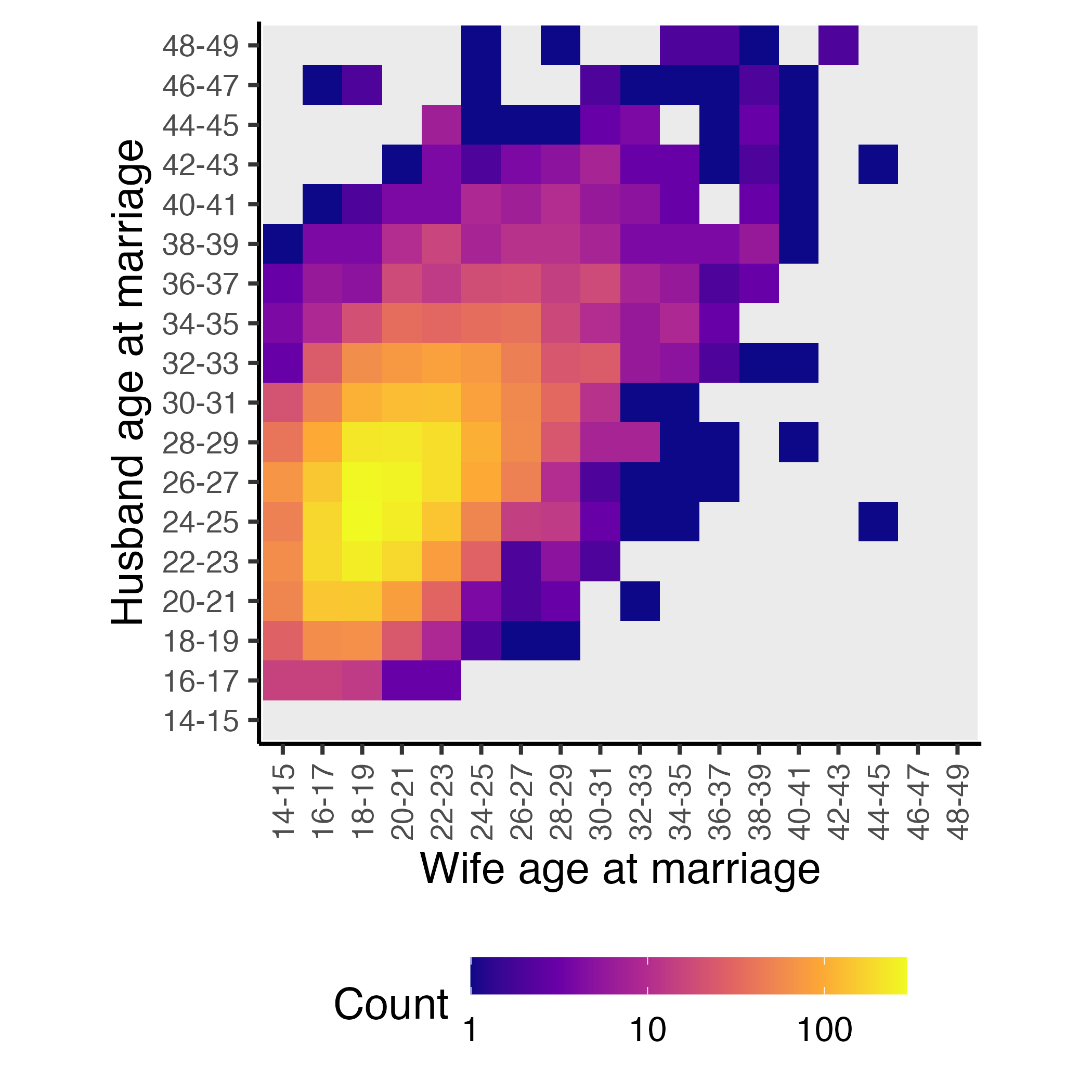}
    }
    \subfloat[1900s]{
    \includegraphics[width = 0.32\textwidth]{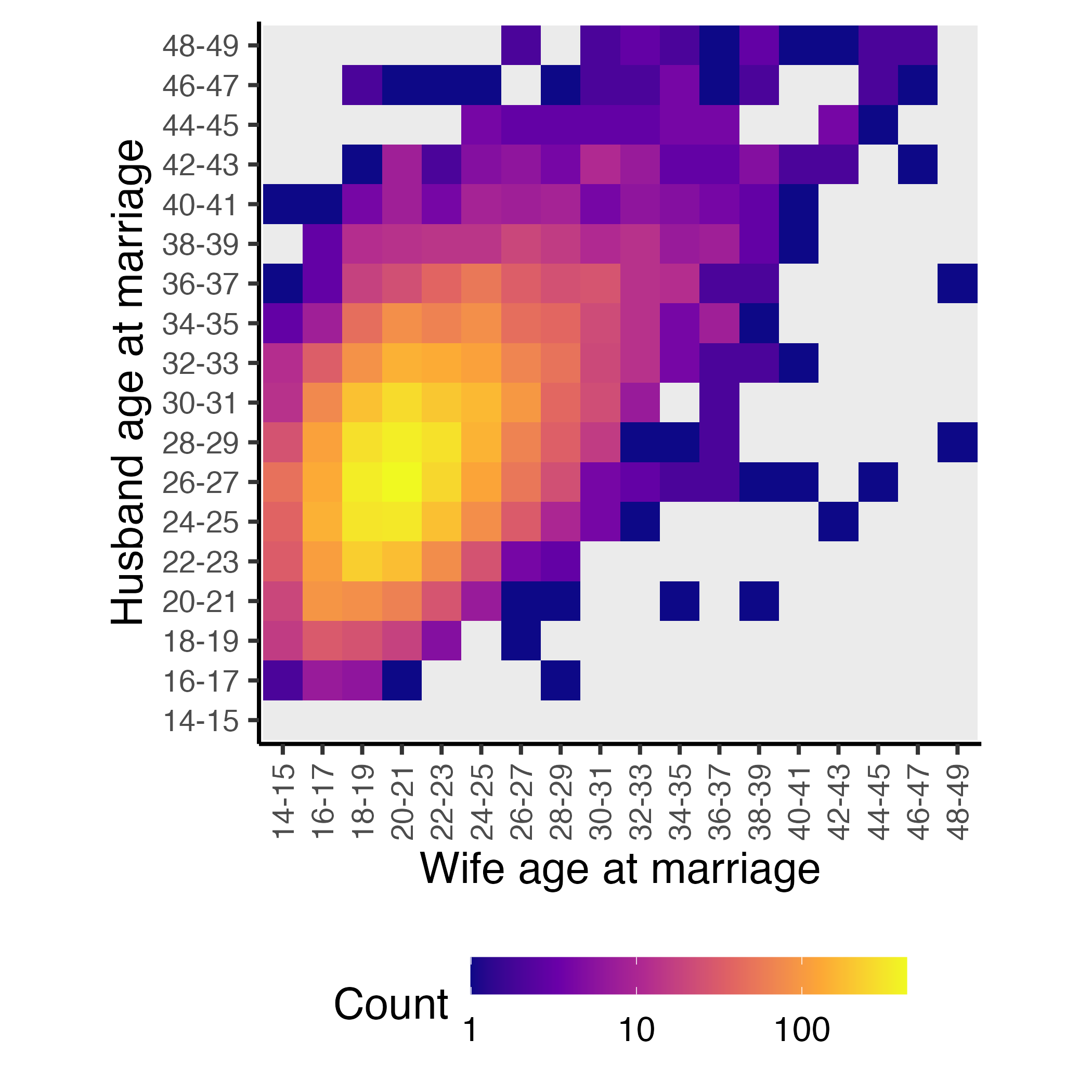}
    }\\\vspace{0.2in}
    \subfloat[1910s]{
    \includegraphics[width = 0.32\textwidth]{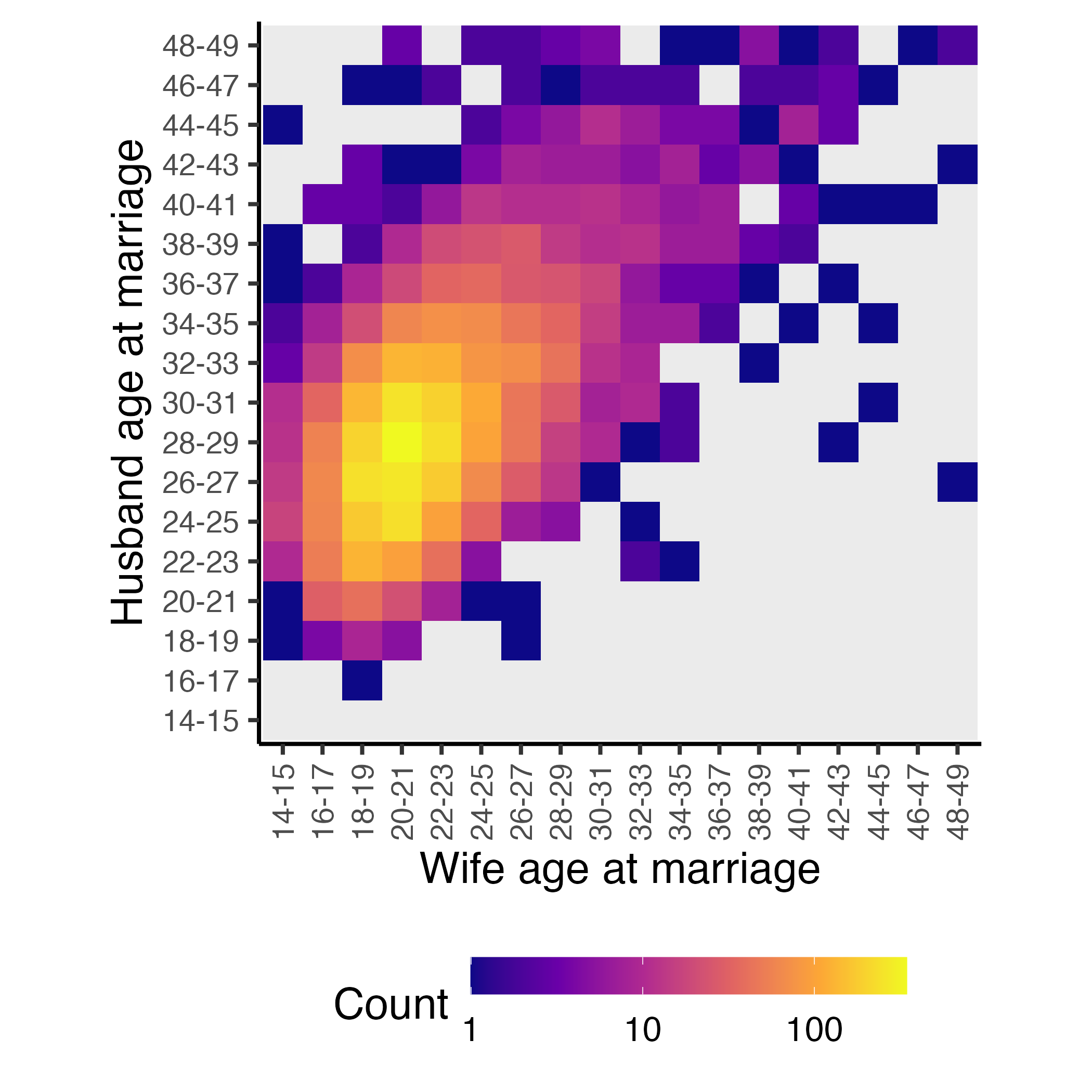}
    }
    \subfloat[1920s]{
    \includegraphics[width = 0.32\textwidth]{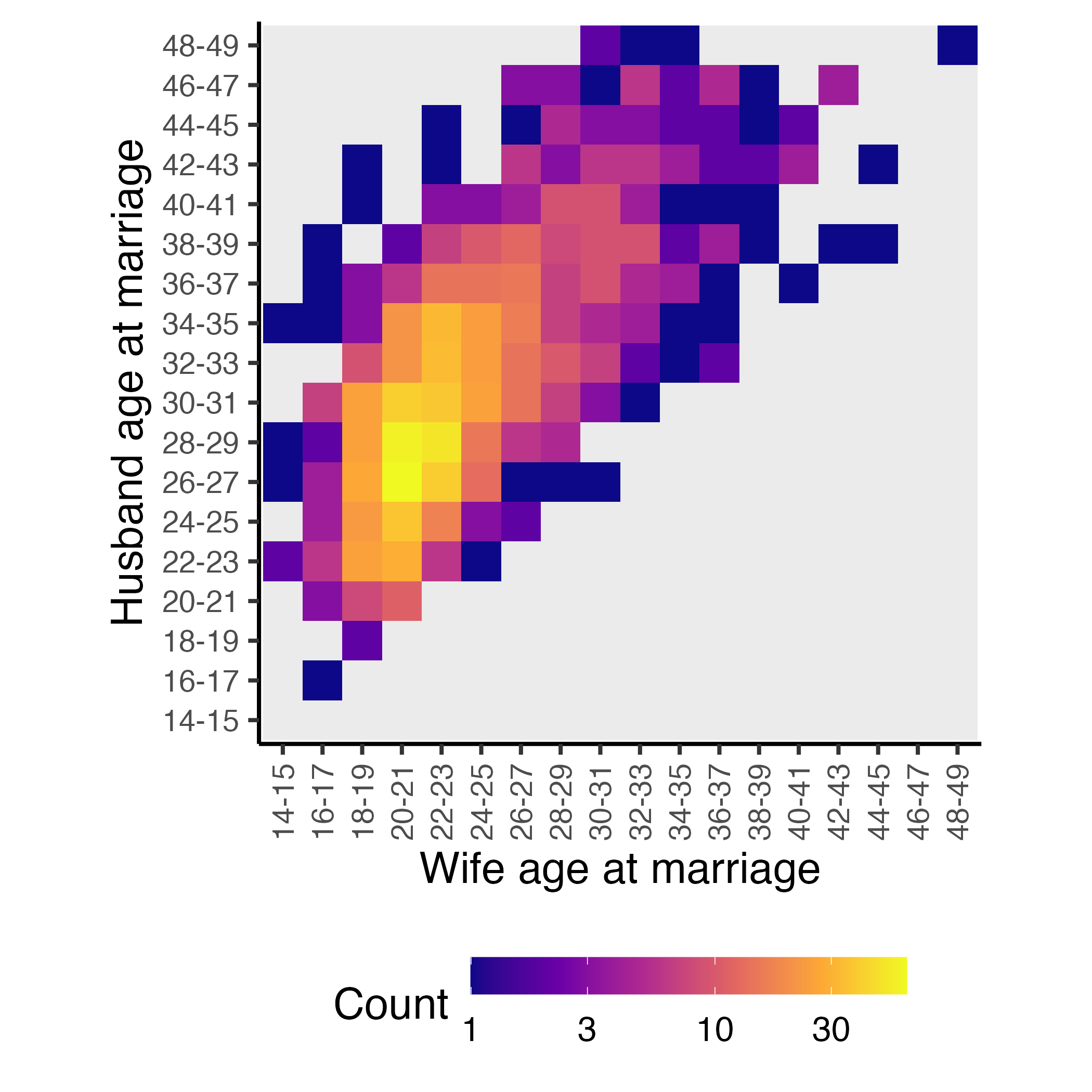}
    }
    \end{center}
    \caption{Age-at-Marriage Matching Matrix by Marriage Cohort}\label{fg:matching_matrix_cohort}
    \footnotesize
    \textit{Note:} Each panel shows the age-at-marriage matching matrix $M = (\mu_{ij})$ for a marriage cohort, with the husband's age-at-marriage bin on the vertical axis and the wife's on the horizontal axis (two-year bins from 14 to 50); color indicates the number of marriages on a $\log_{10}$ scale (empty cells in gray). Mass concentrates near the diagonal but slightly above it, reflecting that husbands are typically a few years older than their wives, and the ridge shifts toward older ages across cohorts as marriage is delayed.
\end{figure}

Figure \ref{fg:matching_matrix_cohort} plots the age-at-marriage matching matrices by cohort. In every cohort the mass concentrates along a ridge just above the diagonal---husbands a few years older than their wives---and this ridge migrates toward older ages over time, mirroring the secular rise in the age at marriage. We summarize the strength of the diagonal concentration in $M$ with the two measures introduced above.

\begin{figure}[htbp]
    \begin{center}
    \includegraphics[width = 0.9\textwidth]{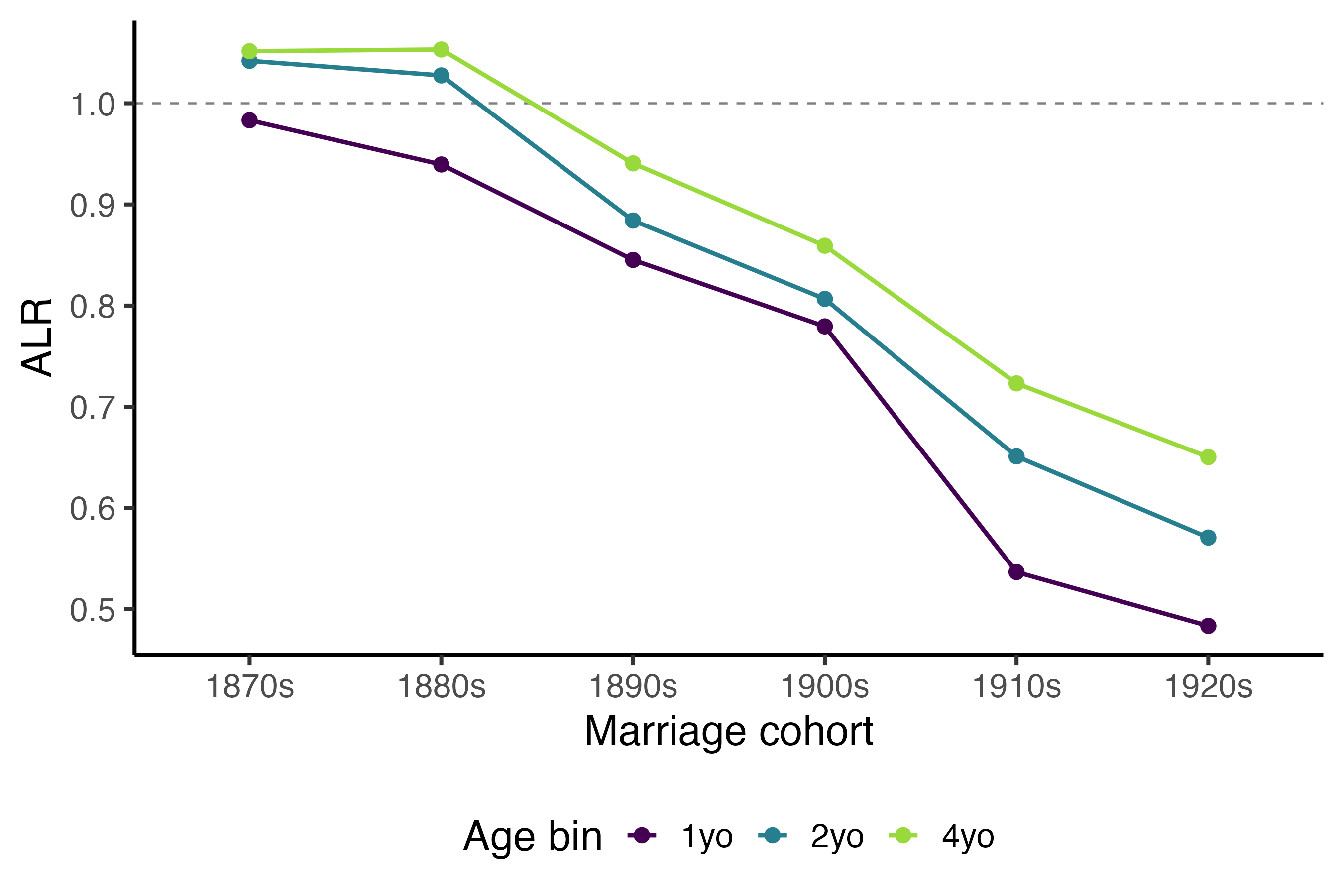}
    \end{center}
    \caption{Multi-Type ALR Transition across Marriage Cohorts}\label{fg:alr_transition}
    \footnotesize
    \textit{Note:} Each line plots the aggregate likelihood ratio (ALR, Equation \eqref{eq:alr}) of the age-at-marriage matching matrix by cohort, at one-, two-, and four-year age-bin widths. $\mathrm{ALR} > 1$ indicates positive assortative matching by age. The ALR rises mechanically with the bin width and declines across cohorts; because it is diagonal-based, this decline is confounded with the widening husband--wife age gap.
\end{figure}

\begin{figure}[htbp]
    \begin{center}
    \includegraphics[width = 0.9\textwidth]{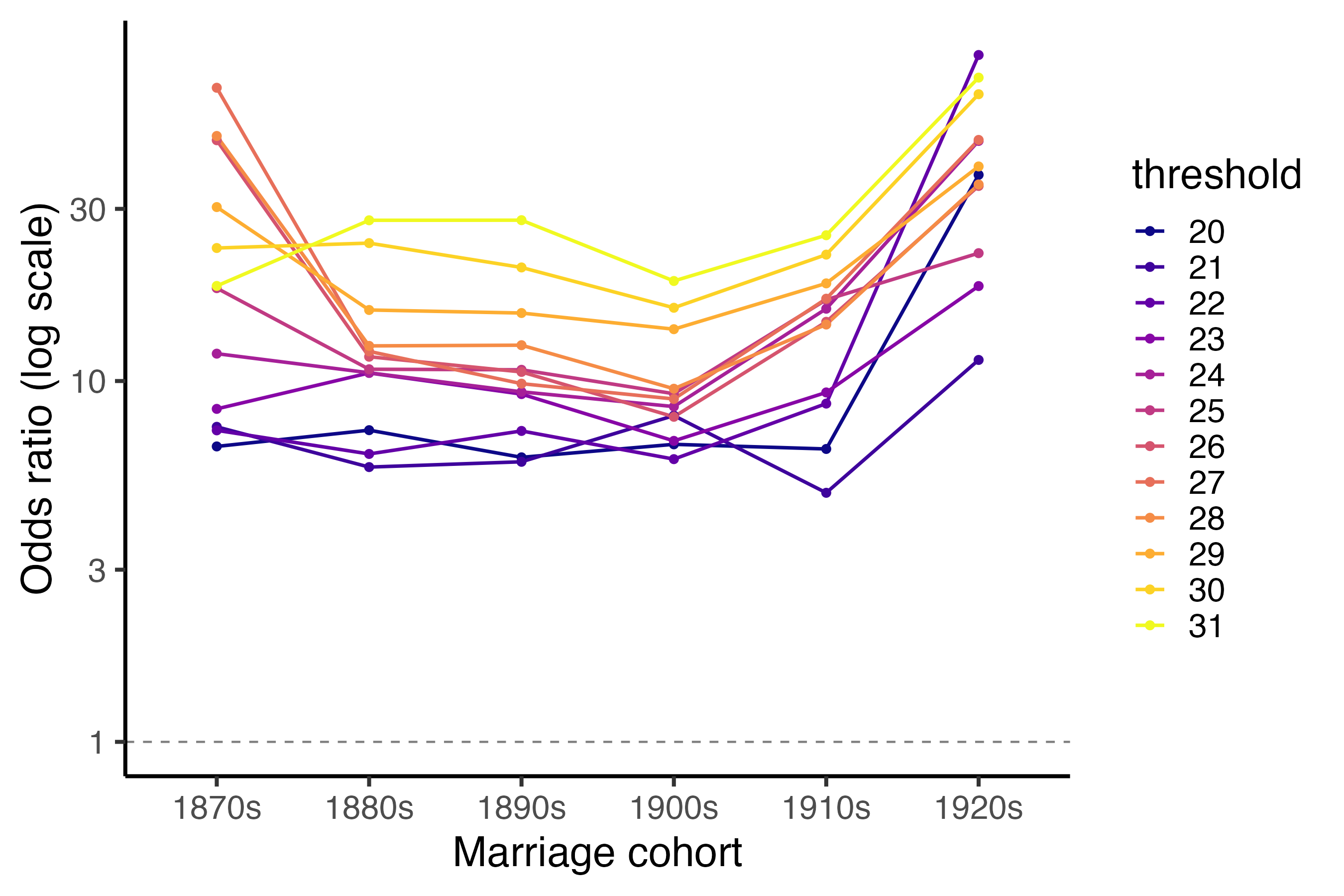}
    \end{center}
    \caption{Two-Type Odds Ratio Transition across Marriage Cohorts}\label{fg:two_type_or_transition}
    \footnotesize
    \textit{Note:} Each line plots the two-type odds ratio (Equation \eqref{eq:tor}, log scale) by cohort for an age threshold $\tau$ (young $<\tau$ vs old $\ge\tau$), for $\tau \in \{20,\dots,31\}$ (the benchmark $\tau = 30$ corresponds to the $\le 29$ vs $\ge 30$ split). $\mathrm{OR} > 1$ indicates positive assortative matching by age. Unlike the ALR, the odds ratio is invariant to the age marginals, so the strongly positive and roughly flat profile indicates stable age sorting that is not driven by the secular shift in the age distributions.
\end{figure}

Figures \ref{fg:alr_transition} and \ref{fg:two_type_or_transition} trace the two measures across cohorts. The ALR (Figure \ref{fg:alr_transition}) declines overall---from about $1.0$ in the 1870s--1880s to about $0.5$--$0.7$ in the 1920s, depending on the bin width---seemingly indicating weakening sorting. But the ALR rewards mass on the strict diagonal (the same age bin for both spouses); as the age gap widens, fewer couples fall in the same bin even if sorting is unchanged, so the decline conflates weaker sorting with the widening gap. The two-type odds ratio (Figure \ref{fg:two_type_or_transition}), which is invariant to the marginals, tells a more credible story: $\log\mathrm{OR}$ remains strongly positive and roughly flat across cohorts (for the benchmark threshold $\tau = 30$, $\mathrm{OR} \approx 16$--$62$), and this pattern is robust across all thresholds $\tau \in \{20,\dots,31\}$. We conclude that positive age assortative matching among the elite is strong and stable---if anything strengthening late in the period---across marriage cohorts, and that the apparent ALR decline is an artifact of the widening husband--wife age gap rather than a genuine erosion of sorting.

\subsection{Spousal Age Gap at Marriage and Elite Family Formation}\label{sec:marriage_timing}

Having documented who marries whom, we next examine how marriage timing is associated with family formation among the elite. Historical demography has long emphasized that fertility was regulated not only through behavior within marriage, but also through the timing of marriage itself. In pre-transition societies, fertility depended strongly on age at marriage and the proportion who married, implying that delayed marriage mechanically reduced completed fertility by shortening the reproductive period. An important feature of the historical fertility transition was the gradual shift from controlling fertility through marriage behavior to controlling fertility within marriage \citep{guinnane2011}.

For elite households, marriage timing may have been particularly important because family size affected the supply of potential heirs. When biological heirs were scarce, Japanese elite households could resort to adoption---often through the adoption of a son-in-law---to maintain the continuity of the household. We therefore examine whether marriage timing and spousal age differences are systematically associated with completed family size and adoption among elite households.

We reconstruct each couple's age at marriage from the birth year of the eldest biological child, as in Section \ref{sec:age_assortativeness}, approximating the marriage year as one year earlier. Let $HusbandAge_i$ and $WifeAge_i$ denote the husband's and wife's ages at marriage, respectively, and let $AgeGap_i = HusbandAge_i - WifeAge_i$ denote the spousal age difference. The main specification is
\begin{equation}\label{eq:fertility}
  Y_i = \alpha + \beta_1\, WifeAge_i + \beta_2\, AgeGap_i + \lambda_{\kappa} + X_i'\gamma + \varepsilon_i ,
\end{equation}
where $Y_i$ is a family outcome---the total number of biological children, biological sons, biological daughters, or an indicator for having an adopted son---$\lambda_{\kappa}$ is a marriage-cohort fixed effect, with $\kappa$ indexing marriage cohorts, and $X_i$ collects husband characteristics (social group, education, occupation, and origin-prefecture fixed effects). Because $(WifeAge, AgeGap)$ is a one-to-one reparametrization of the two spouses' ages, both coefficients are identified: $\beta_1$ reflects the wife's remaining childbearing window, and $\beta_2$ the husband's relative age holding the wife's age fixed. To allow non-linear wife-age effects, we also bin the wife's age at marriage ($<18$, $18$--$20$, $21$--$23$, $24$--$26$, $27+$), with the youngest bin as the reference. The analysis is descriptive: the sample is restricted to couples with at least one datable biological child, and the estimates are associations rather than causal effects.

Table \ref{tb:age_gap_fertility_main} reports the main estimates. A one-year increase in the wife's age at marriage is associated with approximately $0.16$ fewer biological children, with reductions in both sons and daughters. It is also associated with a higher probability of having an adopted son. Holding the wife's age fixed, a larger husband--wife age gap is associated with slightly fewer biological children and a somewhat higher probability of having an adopted son.

\begin{figure}[htbp]
    \begin{center}
    \includegraphics[width = 0.9\textwidth]{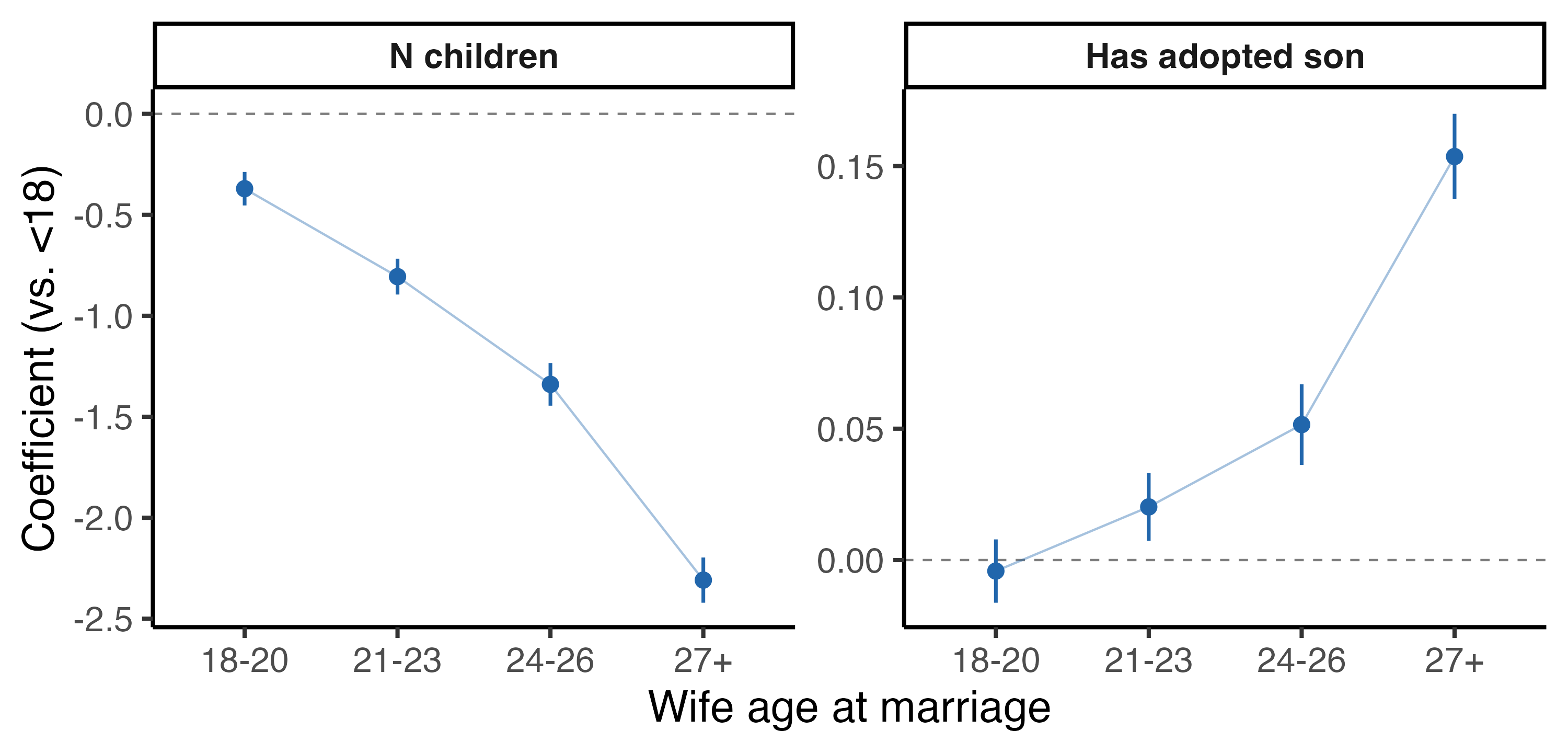}
    \end{center}
    \caption{Wife Age at Marriage and Family Formation}\label{fg:age_gap_fertility_wife_age_bin}
    \footnotesize
    \textit{Note:} Coefficients (with 95\% confidence intervals) on wife age-at-marriage bins, estimated from Equation \eqref{eq:fertility} with the bins replacing $WifeAge_i$, for two outcomes: the total number of biological children (left) and an indicator for having an adopted son (right). The omitted reference category is wives marrying before age 18. All specifications include the age gap, marriage-cohort fixed effects, and husband characteristics.
\end{figure}

Figure \ref{fg:age_gap_fertility_wife_age_bin} further shows that the relationship is markedly non-linear. Relative to wives marrying before age 18, the number of biological children falls by approximately $0.4$, $0.8$, $1.3$, and $2.3$ for the $18$--$20$, $21$--$23$, $24$--$26$, and $27+$ age groups, respectively. The probability of having an adopted son also rises with the wife's age at marriage and is about $15$ percentage points higher among wives marrying at age 27 or older.

The negative association between wife's age at marriage and completed fertility is consistent with a central finding in the historical demography literature: fertility in pre-transition societies was strongly regulated through marriage timing. Later marriage shortened the period available for childbearing and therefore reduced the number of biological children. Research on family succession in Japan further shows that adopting a son provided an alternative means of securing an heir when biological succession was unavailable \citep{kurosu_adoption_1995, moriguchi_comparative_2010, kurosu2013adoption, mehrotra_adoptive_2013, kumon_adoption_2025, kumanomido_elite_2026}. Our estimates connect these two strands of literature by examining how marriage-age structure is associated with both biological fertility and the use of adoption. Because marriage year is inferred from the eldest biological child, these estimates should be read as descriptive associations for couples with observed biological fertility rather than as population-wide causal effects.

\begin{table}[!htbp]
  \begin{center}
      \caption{Spousal Age Gap at Marriage and Elite Family Formation}
      \label{tb:age_gap_fertility_main}
      
\begin{tabular}{lllll}
\toprule
 & N children & N sons & N daughters & Has adopted son\\
\midrule
Wife age at marriage & -0.156*** & -0.079*** & -0.078*** & 0.012***\\
 & (0.003) & (0.002) & (0.002) & (0.000)\\
Age gap (husband - wife) & -0.041*** & -0.021*** & -0.020*** & 0.004***\\
 & (0.003) & (0.002) & (0.002) & (0.000)\\
Num.Obs. & 25459 & 25459 & 25459 & 25459\\
R2 & 0.164 & 0.092 & 0.096 & 0.077\\
\bottomrule
\end{tabular}
  \end{center}\footnotesize
  \textit{Note}: OLS estimates of Equation \eqref{eq:fertility}; each column is a family outcome. All specifications include marriage-cohort fixed effects and husband characteristics (social group, education, occupation, and origin-prefecture fixed effects), with missing categories kept as an explicit ``Unknown'' level; a linear-probability model is used for the adopted-son indicator. The sample is elite marriages with at least one datable biological child (1870s--1920s cohorts). Standard errors in parentheses; ${}^{*}\,p<0.1$, ${}^{**}\,p<0.05$, ${}^{***}\,p<0.01$.
\end{table}

\clearpage

\section{Conclusion}\label{sec:conclusion}

This paper introduces and analyzes the Japanese Personnel Inquiry Records (PIR), a biographical dataset covering distinguished individuals in prewar Japan. The dataset includes detailed biographical information for roughly the top 0.1\% of the population during Japan's transition from a feudal to a modern society. Furthermore, we reconstruct intergenerational linkages between elite fathers and their sons and daughters. For future research, we can utilize this dataset as the primary data source for examining the transmission of elite status across generations.

Using this dataset, we first document the composition and family structure of listed elites. The descriptive evidence shows both continuity and change during Japan's transition from a feudal to a modern society: commoners became increasingly represented among listed elites, higher education became more visible but was not a universal route to elite status, and sons from elite families were overrepresented among listed elites. These patterns indicate that modernization broadened some paths to elite status, while family-based mechanisms of succession remained salient.

We then examine three dimensions of elite formation. First, the geographical analysis shows that elite formation combined metropolitan mobility with locally rooted persistence: some listed individuals moved toward major urban centers, while many remained active in their birth prefectures. Second, the marriage analysis shows stable positive age assortative matching across marriage cohorts, even as husband--wife age gaps widened over time. Third, the family-formation analysis shows that later wife age and larger husband--wife age gaps at marriage were associated with fewer biological children and greater use of adoption. Taken together, these findings show that new opportunities for mobility coexisted with persistent local and family-based mechanisms of elite formation and reproduction.

Beyond these substantive findings, this paper is intended to serve as a practical guide for researchers who use the PIR data. By documenting the structure of the original records, the construction of individual- and family-level variables, and the preprocessing steps needed to link individuals and family members across editions, the paper and accompanying code provide a transparent basis for future empirical work. We hope that this resource will facilitate further research on a wide range of topics, including elite formation, social and intergenerational mobility, marriage and family strategies, regional development, and the long-run consequences of institutional change.

\bibliographystyle{ecca}
\bibliography{jpir_project}

\newpage
\appendix

\section{Online Appendix}

\subsection{Data Source}\label{sec:PIRDataSource}

As Japanese society grew increasingly complex in the early twentieth century, it became more difficult for individuals to know one another's backgrounds, personal histories, or family ties, even within business and social networks. The Personnel Inquiry Office (\textit{Jinji Koshinjo}) was established in 1902 to meet this need. The aim was to promote social transparency and stability by helping individuals make informed decisions in contexts such as hiring, marriage, or business partnerships.

To address this mission, the office began compiling the Personnel Inquiry Records (PIR), including not only occupational trajectories but also information on family structure, marriage ties, and social standing.

PIR sourced information from multiple directories, such as the government's official personnel list (\textit{Shokuinroku}), the Who's Who (\textit{Shinshiroku}) published by \textit{Kojunsha}, bank and corporate directories (\textit{Teikoku Ginkou Kaisha Youroku}) published by \textit{Teikoku Koshinjo}, bank and corporate directories (\textit{Ginkou Kaisha Youroku}) published by \textit{Tokyo Koshinjo}, and the member list of corporate executives (\textit{Zenkoku Shokaisha Yakuinroku}) published by \textit{Shogyo koshinjo}. Based on these sources, the editors supplemented information, including family registry documents and private investigations. As a result, the dataset includes detailed records on family structure, occupational history, and educational background.

\subsection{Data Processing}

\paragraph{Construction of Elite Father-Son Pairs}

The first three editions of PIR, published in 1903, 1915, and 1928, include family information on children, such as first name, sex, birth year, and relationship to listed individuals. By extensive name and birth year matching, we identify whether the sons themselves are also listed in any of the PIR listed above. A challenge of string matching in this process is that it requires the sons' full names (surname + first name), but this dataset lacks the sons' surnames. Since sons have the same surname as their fathers, we needed to retrieve sons' surnames from their fathers. Furthermore, PIR does not list fathers' surnames and first names separately but provides only their full names in a single string. Therefore, we applied the following method to construct the sons' full names:\footnote{\cite{feigenbaum_multiple_2018, long2013intergenerational} provide methods for constructing data to identify father-son relationships using full names, birth years, and regions of origin. However, we conduct father-son matching without using regional information, as fathers and sons possibly have different regional information in PIR.}

\begin{enumerate}
    \item[step 1] Standardization of \textit{Kanji} characters: To resolve the variation in \textit{Kanji} characters in names, we convert old \textit{Kanji} to modern \textit{Kanji} for all elites' names and their children's names listed in PIR.
    \item[step 2] Data extraction: we obtain the first names and birth years of all children from PIR published in 1903, 1915, and 1928.\footnote{We exclude illegitimate sons.}
    \item[step 3] Construct sons' full name: we extract the first one, two, and three characters from the father's full name and combine each with the child's first name. By this data processing, we obtain three patterns of the child's full names for each child.
\end{enumerate}

By these steps, we identify 71,008 unique father-son pairs (including adopted sons) and 63,071 unique father-daughter pairs, excluding children who lack information on first name and birth year. The method for matching father-son pairs across the five editions of PIR is as follows\footnote{We exclude father-daughter pairs for identifying whether the elite sons are listed in the latter PIR editions.}:

\begin{enumerate}
    \item[step 4] Identify whether sons are listed in PIR: Using the three patterns of sons' full names and sons' birth years, we identify whether the sons are listed in PIR in any of the five editions of PIR. 
    \item[step 5] Identify individuals who succeeded their father's first name: Among elite families, some inherited family headship and succeeded to the same name as their fathers. In this case, sons' first names are possibly different from those of their childhood; instead, they have the same name as their father. To address this problem, we identify individuals in PIR who have the same full names as their fathers and whose age is the same as that of the sons.
\end{enumerate}

\paragraph{Attrition}

Table \ref{tb:data_observation_attrition_rate_across_version} reports the attrition rate of observations with missing or invalid names and birth years during cleaning. In edition 1, the attrition rate is 13.7\%. From edition 4 onward, the attrition rates are minimal--below 2.5\%--indicating that nearly all observations are retained.

\begin{table}[!htbp]
  \begin{center}
      \caption{Raw and Cleaned Data Sample Size and Attrition Rate}
      \label{tb:data_observation_attrition_rate_across_version} 
       
\begin{tabular}[t]{r|rrr}
\toprule
Edition & Raw Data Obs & Cleaned Obs & Attrition Rate\\
\midrule
1 & 3353 & 2892 & 0.137\\
4 & 14060 & 13759 & 0.021\\
8 & 25221 & 24931 & 0.011\\
10 & 26185 & 25846 & 0.013\\
12 & 55781 & 54497 & 0.023\\
\bottomrule
\end{tabular}

  \end{center}\footnotesize
  \textit{Note}: This table shows the attrition rate during data processing. We exclude individuals who do not have names (family name and first name) or birth year.
\end{table} 

\subsection{Additional Tables and Figures}\label{sec:additional_tables_and_figures}

Table \ref{tb:summary_statistics_wife_version_covariates} reports summary statistics of family-edition level variables.
Information on fathers and mothers is limited to names, without accompanying attributes such as birth year or occupation.
For wives, birth year is consistently available across editions, and the table shows a gradual increase in the mean birth year over time.

\begin{table}[!htbp]
  \begin{center}
      \caption{Summary Statistics of Wife-edition Level Variables}
      \label{tb:summary_statistics_wife_version_covariates} 
       
\begin{tabular}[t]{llrrrrr}
\toprule
edition &   & N & mean & sd & min & max\\
\midrule
1 & birthy & 2294 & 1864.79 & 10.86 & 1804.00 & 1897.00\\
4 & birthy & 12578 & 1872.14 & 10.91 & 1830.00 & 1919.00\\
8 & birthy & 22680 & 1883.13 & 10.50 & 1835.00 & 1919.00\\
\bottomrule
\end{tabular}

  \end{center}\footnotesize
  \textit{Note}: This table presents summary statistics for the wives of listed individuals across editions. Wives without birth year information are excluded.
\end{table} 

\begin{table}[!htbp]
  \begin{center}\scriptsize
      \caption{Summary Statistics of Daughters' Birth Year by Birth Order across Editions}
      \label{tb:summary_statistics_daughter_version_covariates} 
      
\begin{tabular}[t]{llrrrrr}
\toprule
edition &   & N & mean & sd & min & max\\
\midrule
1 & birthy 1 & 1926 & 1887.59 & 11.13 & 1841.00 & 1926.00\\
 & birthy 2 & 1372 & 1891.81 & 10.41 & 1861.00 & 1926.00\\
 & birthy 3 & 872 & 1894.73 & 9.96 & 1869.00 & 1926.00\\
 & birthy 4 & 530 & 1896.43 & 9.82 & 1871.00 & 1923.00\\
 & birthy 5 & 315 & 1897.72 & 10.15 & 1871.00 & 1926.00\\
 & birthy 6 & 187 & 1898.65 & 10.25 & 1874.00 & 1925.00\\
 & birthy 7 & 112 & 1899.88 & 8.92 & 1879.00 & 1925.00\\
 & birthy 8 & 56 & 1899.54 & 9.37 & 1879.00 & 1922.00\\
 & birthy 9 & 29 & 1900.10 & 8.67 & 1879.00 & 1924.00\\
 & birthy 10 & 17 & 1901.12 & 10.61 & 1879.00 & 1925.00\\
 & birthy 11 & 8 & 1901.75 & 8.08 & 1893.00 & 1913.00\\
 & birthy 12 & 5 & 1903.20 & 8.58 & 1896.00 & 1914.00\\
 & birthy 13 & 3 & 1904.67 & 9.29 & 1897.00 & 1915.00\\
 & birthy 14 & 2 & 1907.50 & 14.85 & 1897.00 & 1918.00\\
 & birthy 15 & 2 & 1909.50 & 14.85 & 1899.00 & 1920.00\\
 & birthy 16 & 1 & 1899.00 &  & 1899.00 & 1899.00\\
 & birthy 17 & 1 & 1901.00 &  & 1901.00 & 1901.00\\
 & birthy 18 & 1 & 1907.00 &  & 1907.00 & 1907.00\\
4 & birthy 1 & 10242 & 1894.98 & 11.42 & 1841.00 & 1926.00\\
 & birthy 2 & 7219 & 1899.21 & 10.71 & 1862.00 & 1926.00\\
 & birthy 3 & 4515 & 1902.13 & 9.97 & 1868.00 & 1927.00\\
 & birthy 4 & 2522 & 1903.92 & 9.66 & 1871.00 & 1926.00\\
 & birthy 5 & 1304 & 1905.20 & 9.54 & 1871.00 & 1926.00\\
 & birthy 6 & 599 & 1905.69 & 9.84 & 1874.00 & 1926.00\\
 & birthy 7 & 269 & 1905.96 & 9.76 & 1879.00 & 1925.00\\
 & birthy 8 & 120 & 1905.58 & 10.03 & 1879.00 & 1926.00\\
 & birthy 9 & 49 & 1904.96 & 9.82 & 1879.00 & 1924.00\\
 & birthy 10 & 25 & 1905.48 & 11.27 & 1879.00 & 1925.00\\
 & birthy 11 & 12 & 1906.75 & 10.09 & 1893.00 & 1922.00\\
 & birthy 12 & 6 & 1905.50 & 9.52 & 1896.00 & 1917.00\\
 & birthy 13 & 3 & 1904.67 & 9.29 & 1897.00 & 1915.00\\
 & birthy 14 & 2 & 1907.50 & 14.85 & 1897.00 & 1918.00\\
 & birthy 15 & 2 & 1909.50 & 14.85 & 1899.00 & 1920.00\\
 & birthy 16 & 1 & 1899.00 &  & 1899.00 & 1899.00\\
 & birthy 17 & 1 & 1901.00 &  & 1901.00 & 1901.00\\
 & birthy 18 & 1 & 1907.00 &  & 1907.00 & 1907.00\\
8 & birthy 1 & 17979 & 1906.23 & 10.87 & 1841.00 & 1927.00\\
 & birthy 2 & 12209 & 1909.65 & 10.20 & 1862.00 & 1927.00\\
 & birthy 3 & 7167 & 1911.66 & 9.58 & 1868.00 & 1927.00\\
 & birthy 4 & 3674 & 1912.72 & 9.26 & 1871.00 & 1927.00\\
 & birthy 5 & 1644 & 1913.28 & 9.01 & 1871.00 & 1926.00\\
 & birthy 6 & 671 & 1913.05 & 9.02 & 1875.00 & 1926.00\\
 & birthy 7 & 274 & 1912.92 & 9.47 & 1880.00 & 1927.00\\
 & birthy 8 & 102 & 1911.09 & 9.20 & 1887.00 & 1926.00\\
 & birthy 9 & 42 & 1909.95 & 9.51 & 1889.00 & 1925.00\\
 & birthy 10 & 19 & 1910.37 & 10.33 & 1892.00 & 1925.00\\
 & birthy 11 & 10 & 1908.00 & 10.46 & 1893.00 & 1922.00\\
 & birthy 12 & 3 & 1903.67 & 11.59 & 1896.00 & 1917.00\\
 & birthy 13 & 2 & 1899.50 & 3.54 & 1897.00 & 1902.00\\
 & birthy 14 & 1 & 1897.00 &  & 1897.00 & 1897.00\\
 & birthy 15 & 1 & 1899.00 &  & 1899.00 & 1899.00\\
 & birthy 16 & 1 & 1899.00 &  & 1899.00 & 1899.00\\
 & birthy 17 & 1 & 1901.00 &  & 1901.00 & 1901.00\\
 & birthy 18 & 1 & 1907.00 &  & 1907.00 & 1907.00\\
\bottomrule
\end{tabular}

  \end{center}\footnotesize
  \textit{Note}: This table presents summary statistics for the biological daughters of listed individuals across editions. Daughters without birth year information are excluded.
\end{table} 

\begin{table}[!htbp]
  \begin{center}\scriptsize
      \caption{Summary Statistics of Biological Son-edition Level Variables}
      \label{tb:summary_statistics_son_version_covariates}
      
\begin{tabular}[t]{llrrrrr}
\toprule
edition &   & N & mean & sd & min & max\\
\midrule
1 & birthy 1 & 1963 & 1886.91 & 11.62 & 1840.00 & 1924.00\\
 & birthy 2 & 1334 & 1891.99 & 10.46 & 1859.00 & 1926.00\\
 & birthy 3 & 850 & 1895.14 & 10.15 & 1864.00 & 1926.00\\
 & birthy 4 & 503 & 1897.50 & 10.04 & 1866.00 & 1926.00\\
 & birthy 5 & 252 & 1898.55 & 9.77 & 1868.00 & 1926.00\\
 & birthy 6 & 131 & 1899.98 & 9.98 & 1870.00 & 1926.00\\
 & birthy 7 & 66 & 1901.38 & 10.33 & 1873.00 & 1924.00\\
 & birthy 8 & 31 & 1900.84 & 10.18 & 1873.00 & 1922.00\\
 & birthy 9 & 12 & 1899.33 & 12.56 & 1875.00 & 1922.00\\
 & birthy 10 & 9 & 1901.00 & 8.00 & 1890.00 & 1912.00\\
 & birthy 11 & 4 & 1907.00 & 5.35 & 1902.00 & 1913.00\\
 & birthy 12 & 0 &  &  &  & \\
 & birthy 13 & 0 &  &  &  & \\
4 & birthy 1 & 10167 & 1894.95 & 11.42 & 1833.00 & 1926.00\\
 & birthy 2 & 7159 & 1899.73 & 10.29 & 1863.00 & 1926.00\\
 & birthy 3 & 4466 & 1902.81 & 9.65 & 1865.00 & 1926.00\\
 & birthy 4 & 2375 & 1904.98 & 9.47 & 1868.00 & 1926.00\\
 & birthy 5 & 1168 & 1906.69 & 8.98 & 1878.00 & 1926.00\\
 & birthy 6 & 525 & 1907.48 & 8.72 & 1879.00 & 1926.00\\
 & birthy 7 & 207 & 1908.22 & 8.75 & 1881.00 & 1926.00\\
 & birthy 8 & 90 & 1908.17 & 8.91 & 1883.00 & 1925.00\\
 & birthy 9 & 36 & 1908.64 & 9.35 & 1886.00 & 1926.00\\
 & birthy 10 & 20 & 1908.30 & 10.20 & 1890.00 & 1925.00\\
 & birthy 11 & 5 & 1909.60 & 7.44 & 1902.00 & 1920.00\\
 & birthy 12 & 0 &  &  &  & \\
 & birthy 13 & 0 &  &  &  & \\
8 & birthy 1 & 18294 & 1906.20 & 10.75 & 1833.00 & 1927.00\\
 & birthy 2 & 12562 & 1909.94 & 9.77 & 1871.00 & 1927.00\\
 & birthy 3 & 7424 & 1912.21 & 9.02 & 1872.00 & 1927.00\\
 & birthy 4 & 3661 & 1913.62 & 8.51 & 1876.00 & 1926.00\\
 & birthy 5 & 1591 & 1914.34 & 8.20 & 1881.00 & 1926.00\\
 & birthy 6 & 645 & 1914.59 & 8.09 & 1887.00 & 1926.00\\
 & birthy 7 & 249 & 1914.74 & 7.69 & 1890.00 & 1926.00\\
 & birthy 8 & 84 & 1914.68 & 8.05 & 1891.00 & 1926.00\\
 & birthy 9 & 34 & 1914.97 & 8.33 & 1892.00 & 1926.00\\
 & birthy 10 & 12 & 1912.92 & 10.48 & 1893.00 & 1925.00\\
 & birthy 11 & 1 & 1920.00 &  & 1920.00 & 1920.00\\
 & birthy 12 & 0 &  &  &  & \\
 & birthy 13 & 0 &  &  &  & \\
\bottomrule
\end{tabular}
 
  \end{center}\footnotesize
  \textit{Note}: This table presents summary statistics for the biological sons of listed individuals across editions. Sons without birth year information are excluded.
\end{table} 

\begin{table}[!htbp]
  \begin{center}
      \caption{Summary Statistics of Adopted Son-edition Level Variables}
      \label{tb:summary_statistics_adopted_son_version_covariates} 
\begin{tabular}[t]{llrrrrr}
\toprule
edition &   & N & mean & sd & min & max\\
\midrule
1 & birthy 1 & 492 & 1882.26 & 12.86 & 1847.00 & 1921.00\\
 & birthy 2 & 100 & 1889.23 & 12.03 & 1863.00 & 1923.00\\
 & birthy 3 & 25 & 1890.84 & 10.83 & 1873.00 & 1915.00\\
 & birthy 4 & 11 & 1894.27 & 13.32 & 1875.00 & 1918.00\\
 & birthy 5 & 2 & 1891.50 & 10.61 & 1884.00 & 1899.00\\
 & birthy 6 & 1 & 1889.00 &  & 1889.00 & 1889.00\\
 & birthy 7 & 0 &  &  &  & \\
 & birthy 8 & 0 &  &  &  & \\
4 & birthy 1 & 2926 & 1887.00 & 12.21 & 1846.00 & 1924.00\\
 & birthy 2 & 656 & 1891.34 & 11.00 & 1863.00 & 1925.00\\
 & birthy 3 & 161 & 1893.19 & 10.54 & 1871.00 & 1925.00\\
 & birthy 4 & 48 & 1894.73 & 11.06 & 1875.00 & 1926.00\\
 & birthy 5 & 19 & 1896.11 & 9.52 & 1881.00 & 1914.00\\
 & birthy 6 & 8 & 1898.00 & 11.05 & 1884.00 & 1913.00\\
 & birthy 7 & 3 & 1896.00 & 4.36 & 1893.00 & 1901.00\\
 & birthy 8 & 2 & 1902.00 & 8.49 & 1896.00 & 1908.00\\
8 & birthy 1 & 4036 & 1895.99 & 12.24 & 1848.00 & 1926.00\\
 & birthy 2 & 821 & 1898.57 & 11.11 & 1863.00 & 1926.00\\
 & birthy 3 & 168 & 1899.85 & 11.03 & 1873.00 & 1926.00\\
 & birthy 4 & 45 & 1898.91 & 11.68 & 1876.00 & 1926.00\\
 & birthy 5 & 17 & 1899.53 & 10.68 & 1881.00 & 1920.00\\
 & birthy 6 & 9 & 1903.44 & 13.36 & 1884.00 & 1923.00\\
 & birthy 7 & 3 & 1906.67 & 16.26 & 1894.00 & 1925.00\\
 & birthy 8 & 1 & 1908.00 &  & 1908.00 & 1908.00\\
\bottomrule
\end{tabular}

  \end{center}\footnotesize
  \textit{Note}: This table presents summary statistics for the adopted sons of listed individuals across editions. Adopted sons without birth year information are excluded.
\end{table} 

\begin{figure}[htbp]
    \begin{center}
    \subfloat[Distribution of Number of Children per Elite by Edition]{
    \includegraphics[width = 0.70\textwidth]{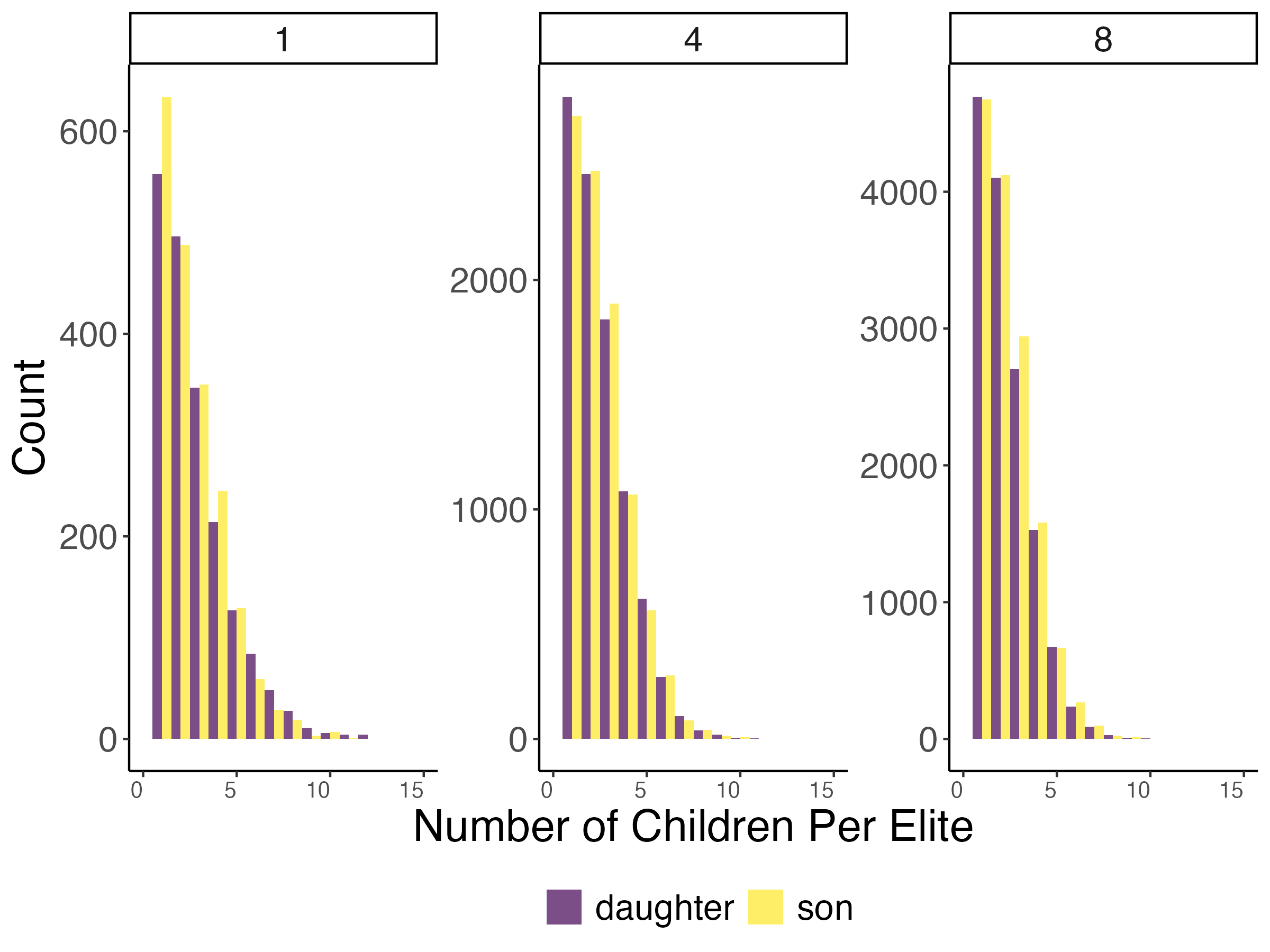}
    }\\\vspace{0.6in}
    \subfloat[Distribution of Child Birth Order by Edition]{
    \includegraphics[width = 0.45\textwidth]{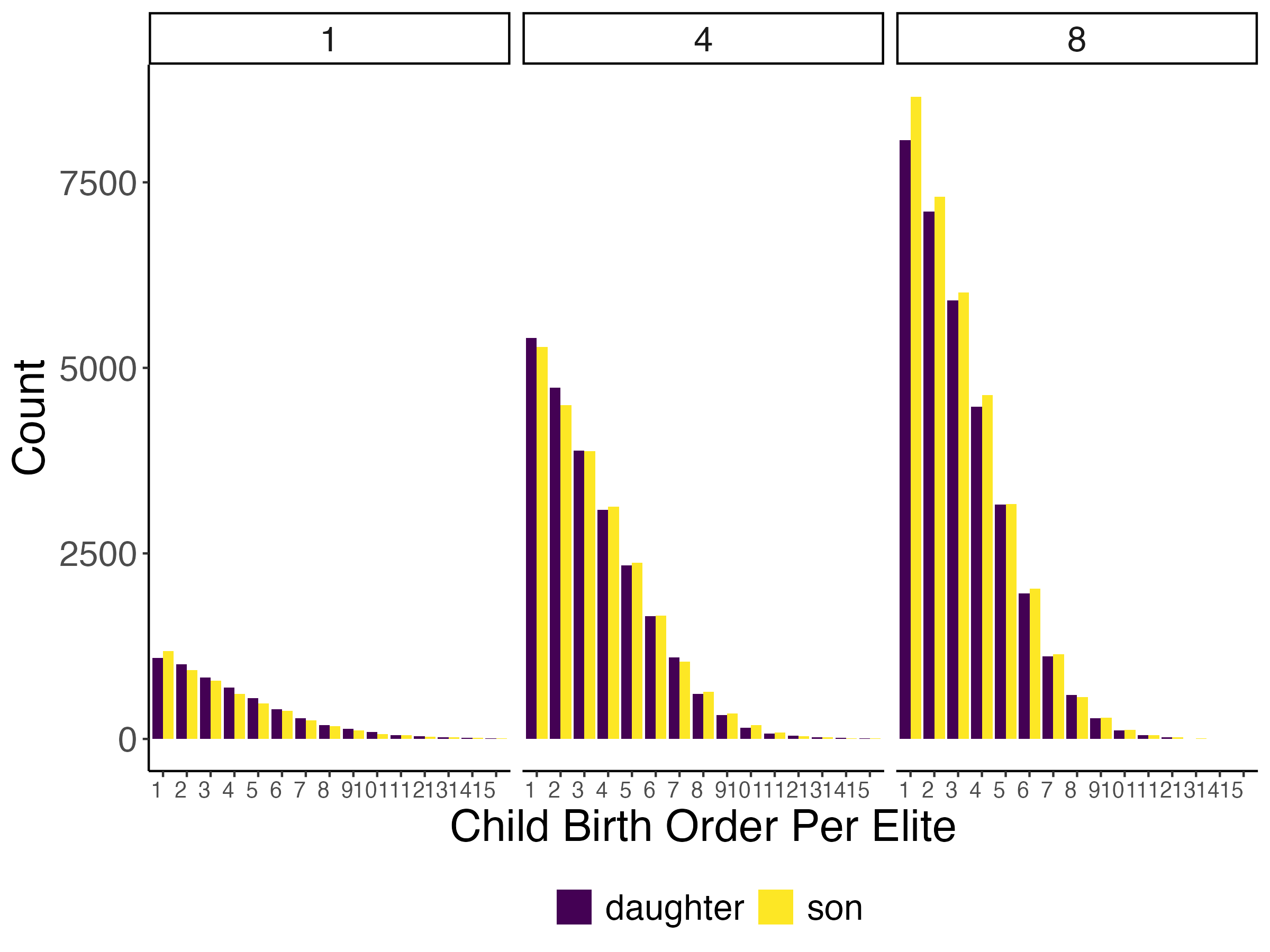}\includegraphics[width = 0.45\textwidth]{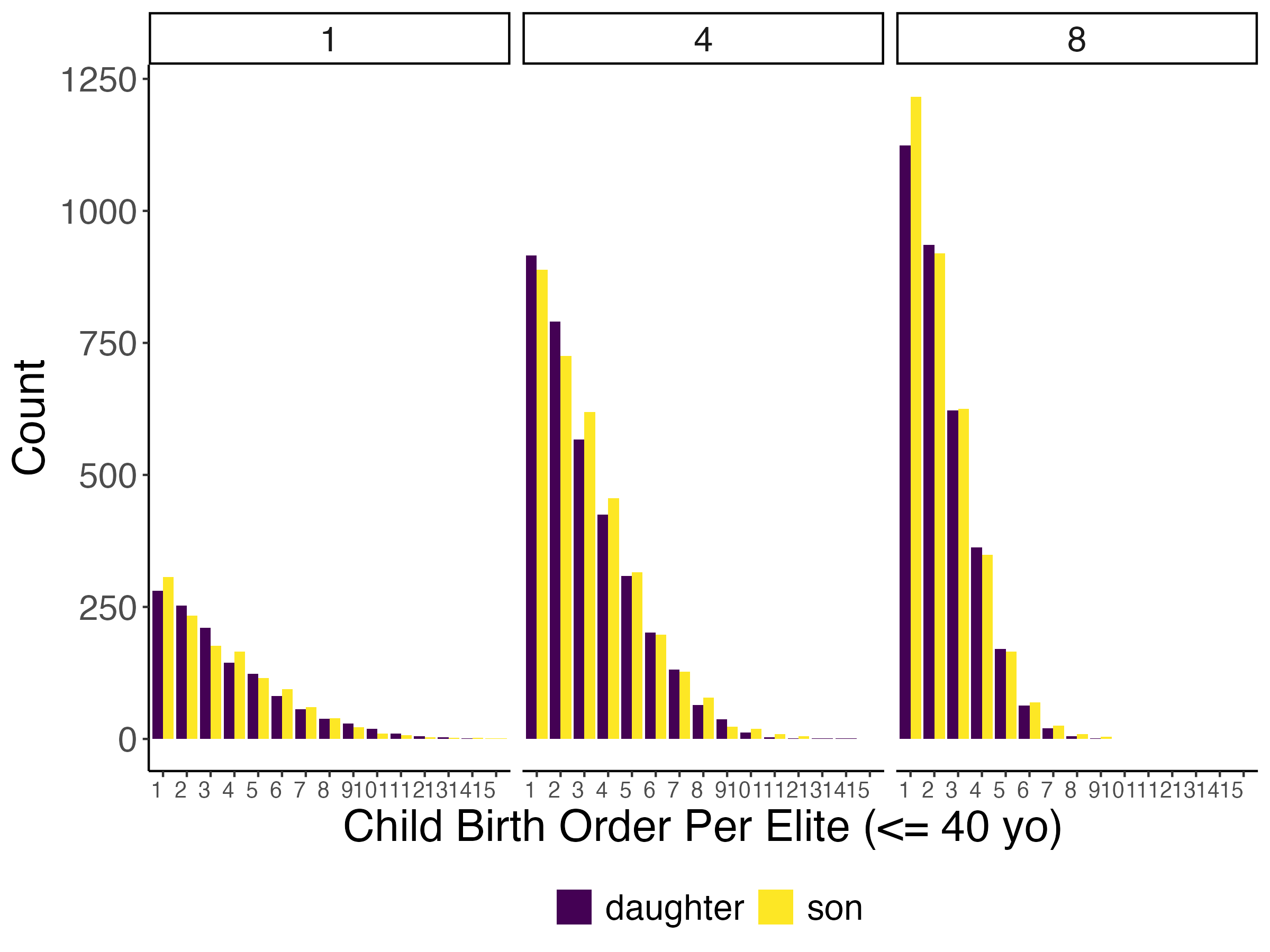}
    }
    \end{center}
    \caption{Biological Family Composition Change}\label{fg:histogram_number_of_children_type_per_birthorder_only_biological}
    \footnotesize
    \textit{Note:} These figures show the number of biological daughters and biological sons by their birth order within each family across PIR editions. In panel (a), we count the number of children at each birth order within each child type (i.e., daughters and sons). For each family, we calculate the birth order separately for each child type (e.g., first son, second daughter, etc.) and then count the number of biological children in each order within each child type across families. Panel (b, left) presents the same counts by overall biological birth order. We calculate the order of all biological children in a family based on age, then count the number of daughters and sons at each order. Panel (b, right) applies the same calculation as panel (b, left), but restricts the sample to elite fathers who were first listed in the PIR before the age of 40, to account for the possibility that daughters in older households may have already married into other families by the time of listing.
\end{figure}

\begin{table}[!htbp]
  \begin{center}
      \caption{Summary Statistics of Individual-edition Level Variables}
      \label{tb:summary_statistics_individual_version_covariates}       
\begin{tabular}[t]{llrrr}
\toprule
edition &   & N & mean & sd\\
\midrule
1 & birth year & 2892 & 1856.17 & 11.15\\
 & samurai dummy & 2892 & 0.39 & 0.49\\
 & imperial univ dummy & 2892 & 0.06 & 0.24\\
 & top 0.1 \% income earner dummy & 0 &  & \\
 & medal recipient dummy & 2892 & 0.51 & 0.50\\
4 & birth year & 13759 & 1864.56 & 10.86\\
 & samurai dummy & 13759 & 0.28 & 0.45\\
 & imperial univ dummy & 13759 & 0.11 & 0.32\\
 & top 0.1 \% income earner dummy & 0 &  & \\
 & medal recipient dummy & 13759 & 0.23 & 0.42\\
8 & birth year & 24931 & 1876.28 & 10.85\\
 & samurai dummy & 24931 & 0.17 & 0.37\\
 & imperial univ dummy & 24931 & 0.17 & 0.38\\
 & top 0.1 \% income earner dummy & 0 &  & \\
 & medal recipient dummy & 24931 & 0.22 & 0.41\\
10 & birth year & 25846 & 1881.88 & 11.29\\
 & samurai dummy & 25846 & 0.15 & 0.36\\
 & imperial univ dummy & 25846 & 0.22 & 0.42\\
 & top 0.1 \% income earner dummy & 25846 & 0.53 & 0.50\\
 & medal recipient dummy & 25846 & 0.26 & 0.44\\
12 & birth year & 54497 & 1886.35 & 10.81\\
 & samurai dummy & 54497 & 0.10 & 0.30\\
 & imperial univ dummy & 54497 & 0.20 & 0.40\\
 & top 0.1 \% income earner dummy & 54497 & 0.21 & 0.41\\
 & medal recipient dummy & 54497 & 0.18 & 0.38\\
\bottomrule
\end{tabular}

  \end{center}\footnotesize
  \textit{Note}: This table presents summary statistics for elites across PIR editions. As documented in Section \ref{data:definition}, tax information is available only in editions 10 and 12. For specific subgroups, samurai, graduates of imperial universities, top 0.1\% income earners, and medal recipients, we report the share of such elites in each edition (shown in the ``mean'' column). We exclude individuals who do not have information on birth year or whose reported age at the time of listing is below 17.
\end{table} 

\begin{figure}[htbp]
    \begin{center}
    \subfloat[Edition 1 (1903)]{
    \includegraphics[width = 0.45\textwidth]{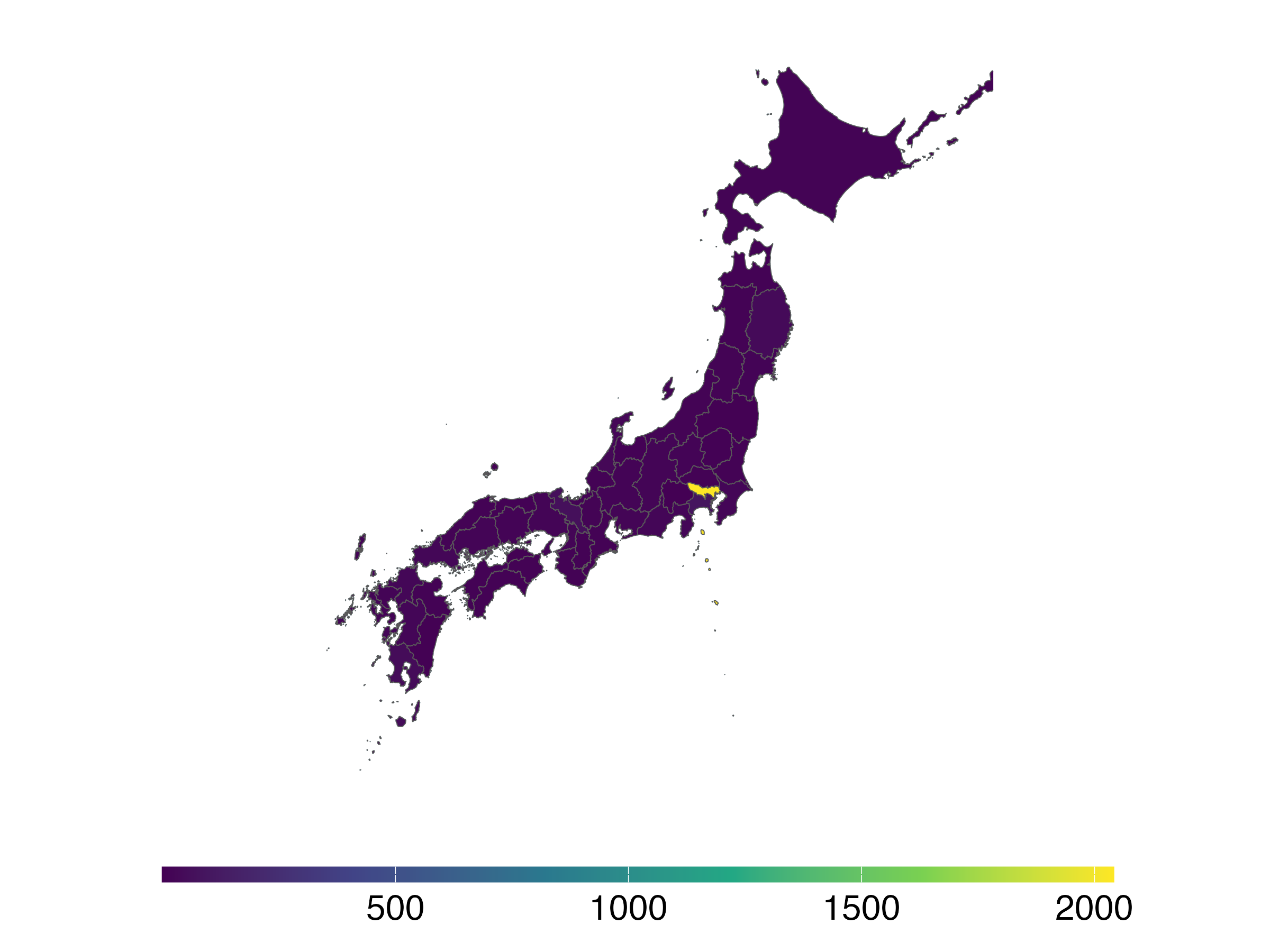}
    }
    \subfloat[Edition 4 (1915)]{
    \includegraphics[width = 0.45\textwidth]{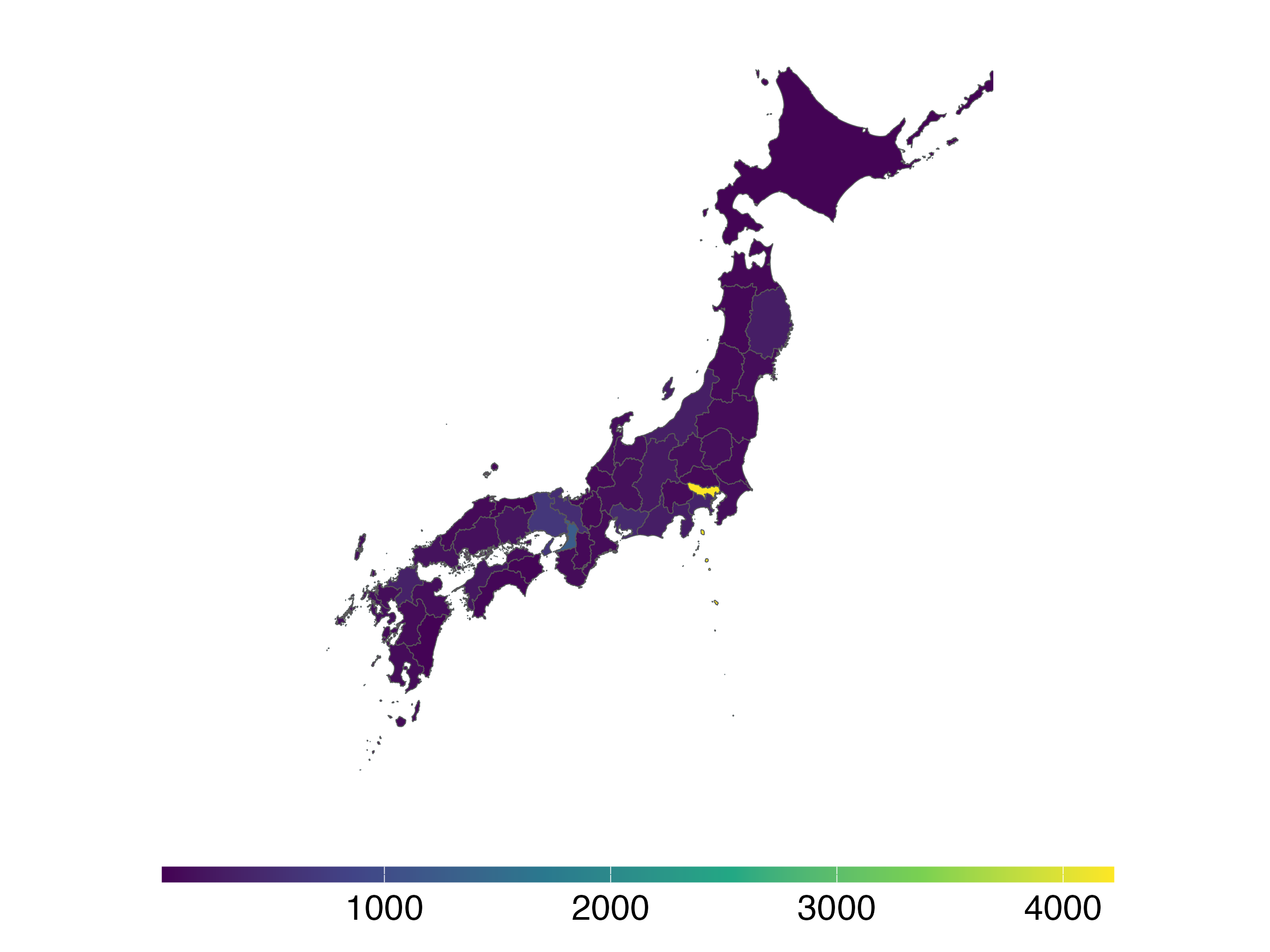}
    }\\
    \subfloat[Edition 8 (1928)]{
    \includegraphics[width = 0.45\textwidth]{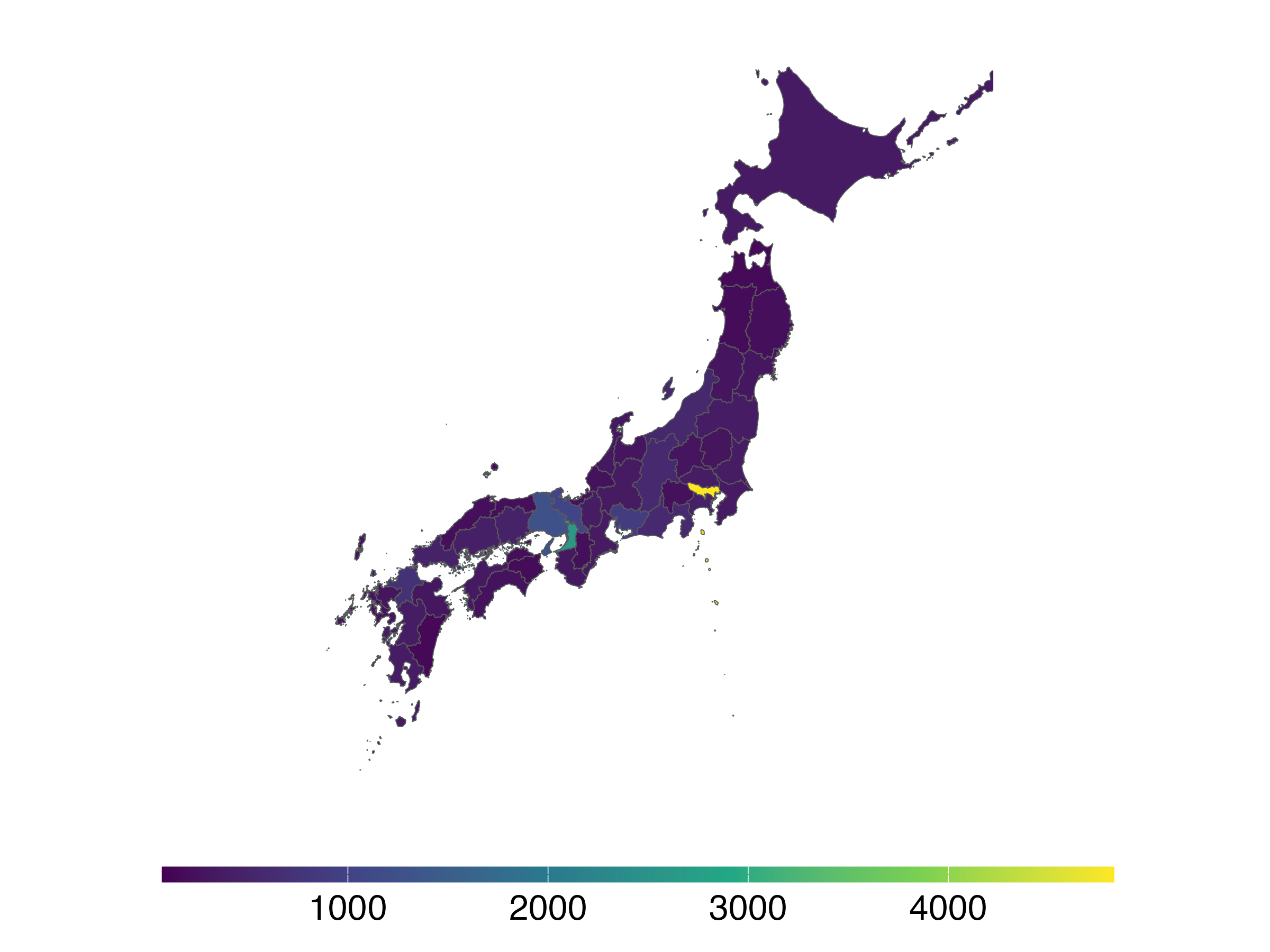}
    }
    \subfloat[Edition 10 (1934)]{
    \includegraphics[width = 0.45\textwidth]{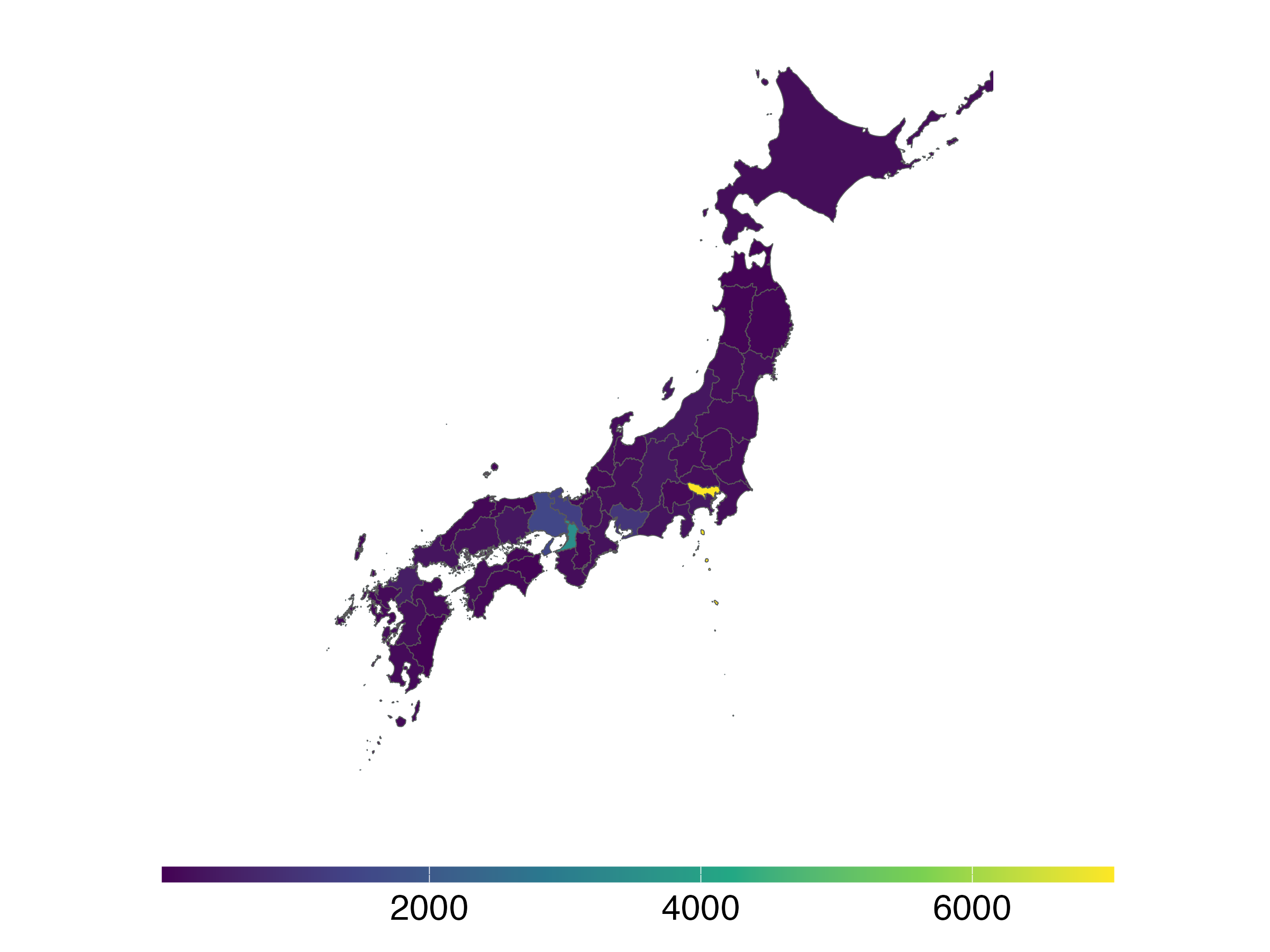}
    }\\
    \subfloat[Edition 12 (1939)]{
    \includegraphics[width = 0.45\textwidth]{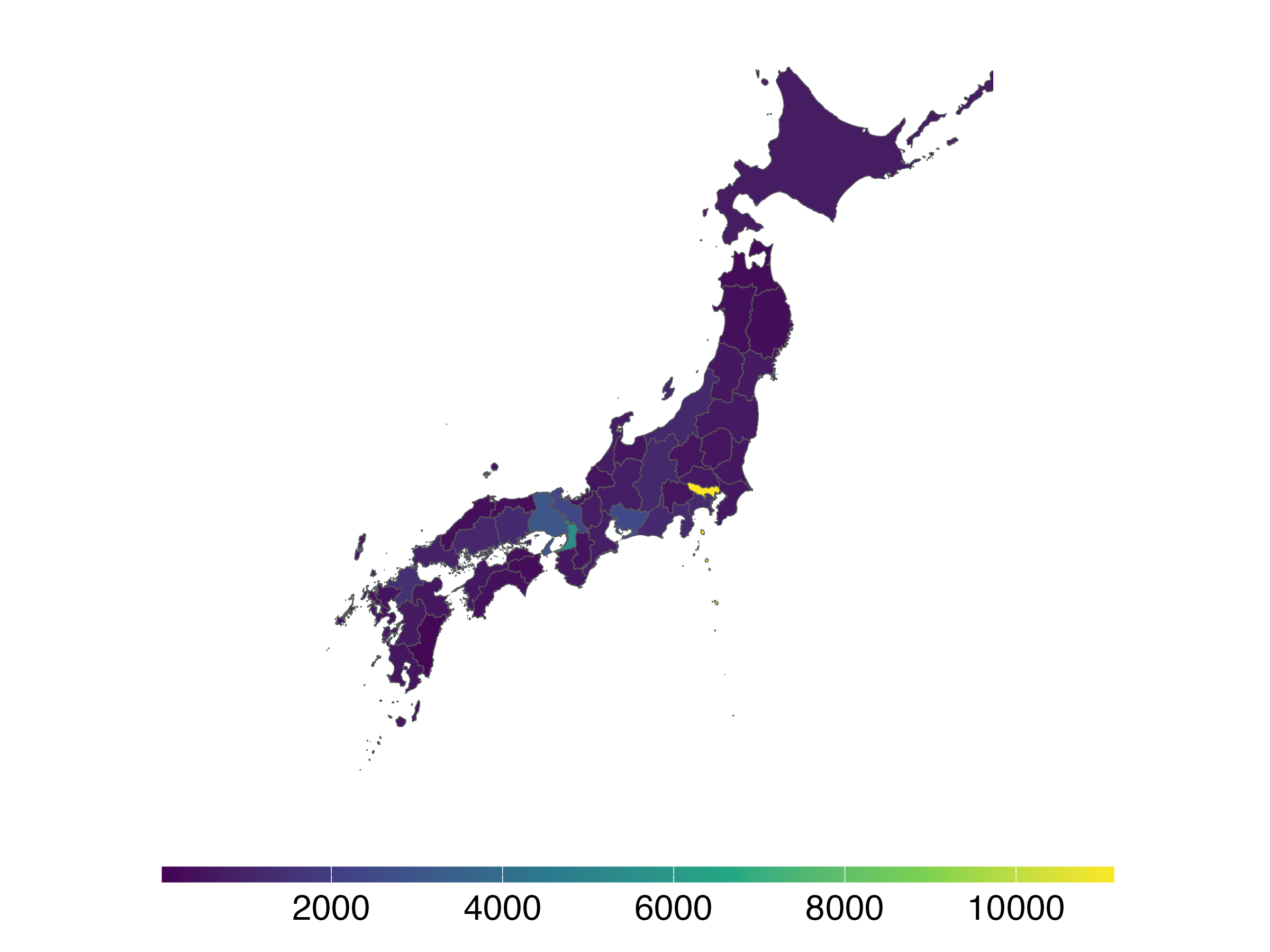}
    }
    \end{center}
    \caption{Number of Elites by Prefecture}
    \label{fg:listing_num_pref_ver12}
    \footnotesize
    \textit{Note:} In these figures, we report the number of elites by residential prefecture at the time of listing (47 prefectures). Individuals with missing residential information or who live in foreign countries at the time of listing are excluded.
\end{figure}

\begin{figure}[htbp]
    \begin{center}
    \subfloat[Edition 1 (1903)]{
    \includegraphics[width = 0.45\textwidth]{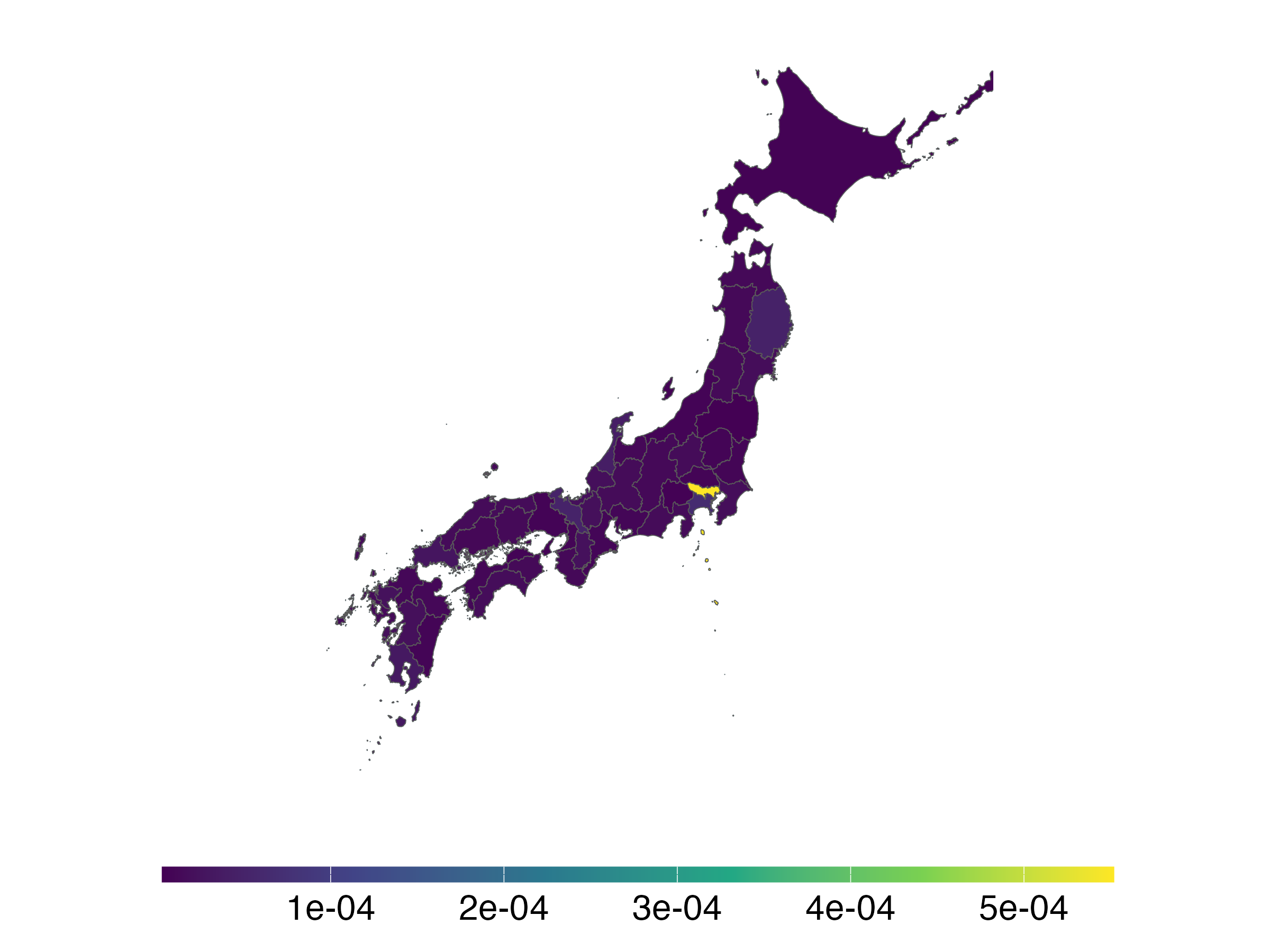}
    }
    \subfloat[Edition 4 (1915)]{
    \includegraphics[width = 0.45\textwidth]{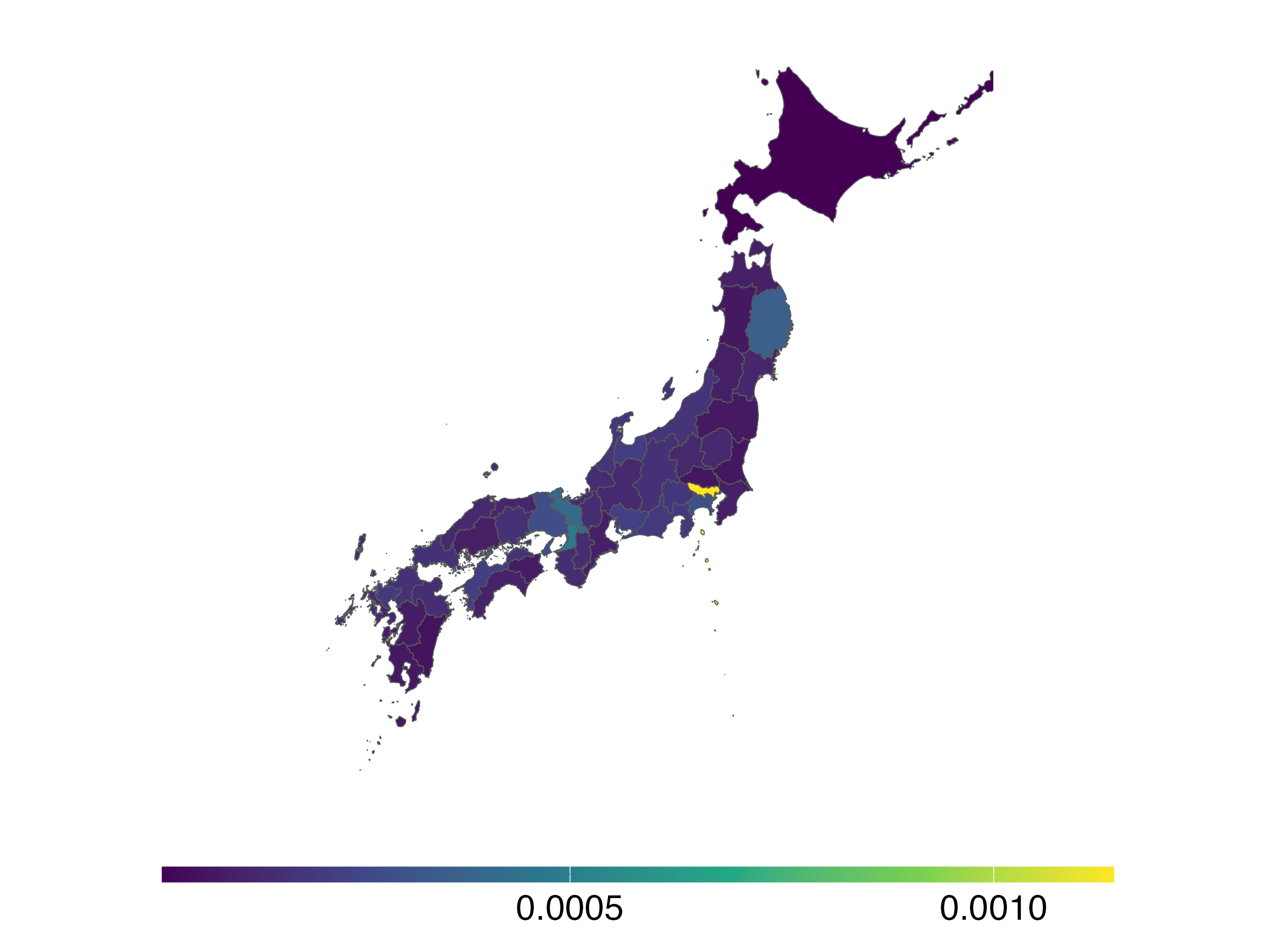}
    }\\
    \subfloat[Edition 8 (1928)]{
    \includegraphics[width = 0.45\textwidth]{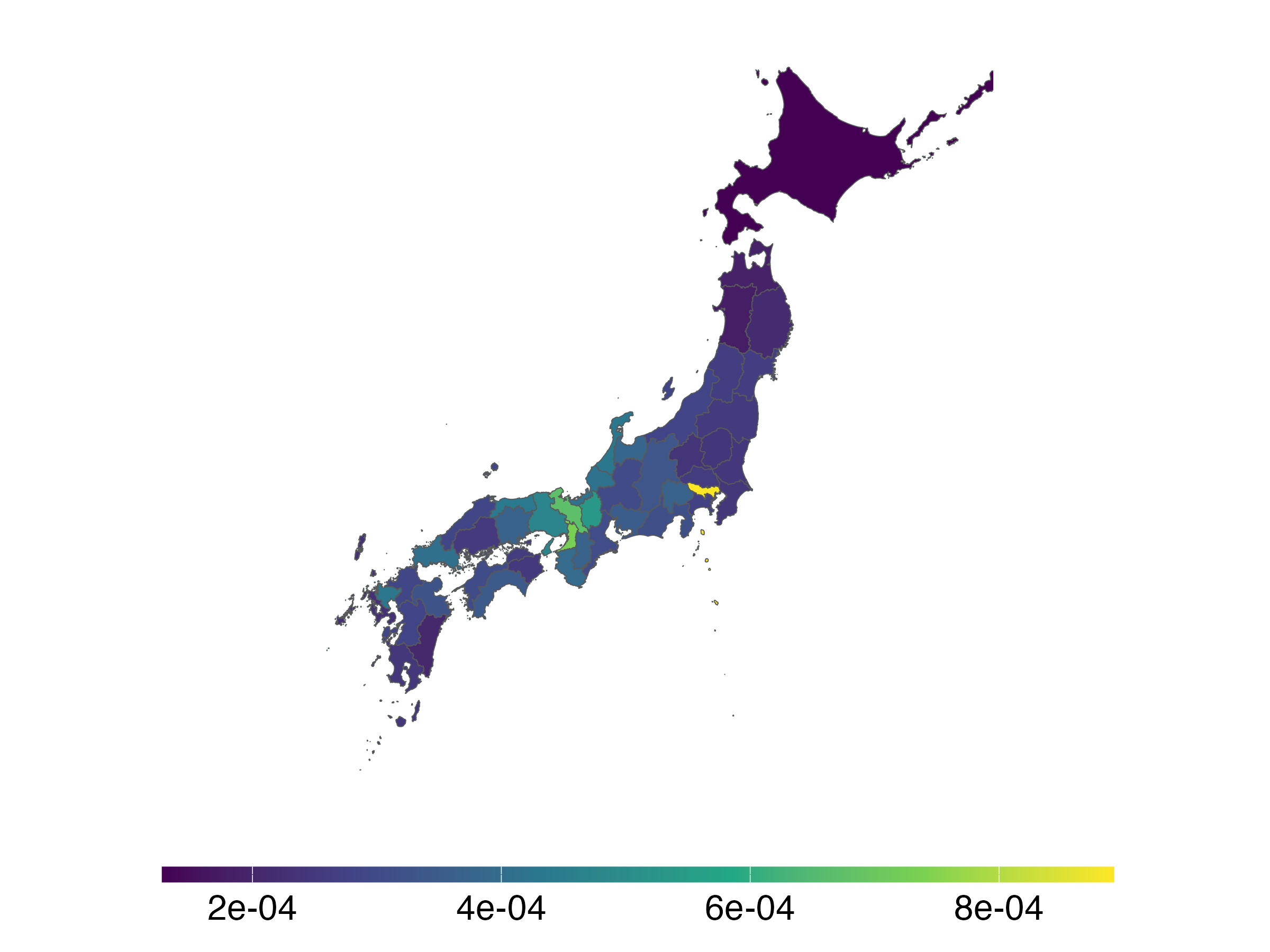}
    }
    \subfloat[Edition 10 (1934)]{
    \includegraphics[width = 0.45\textwidth]{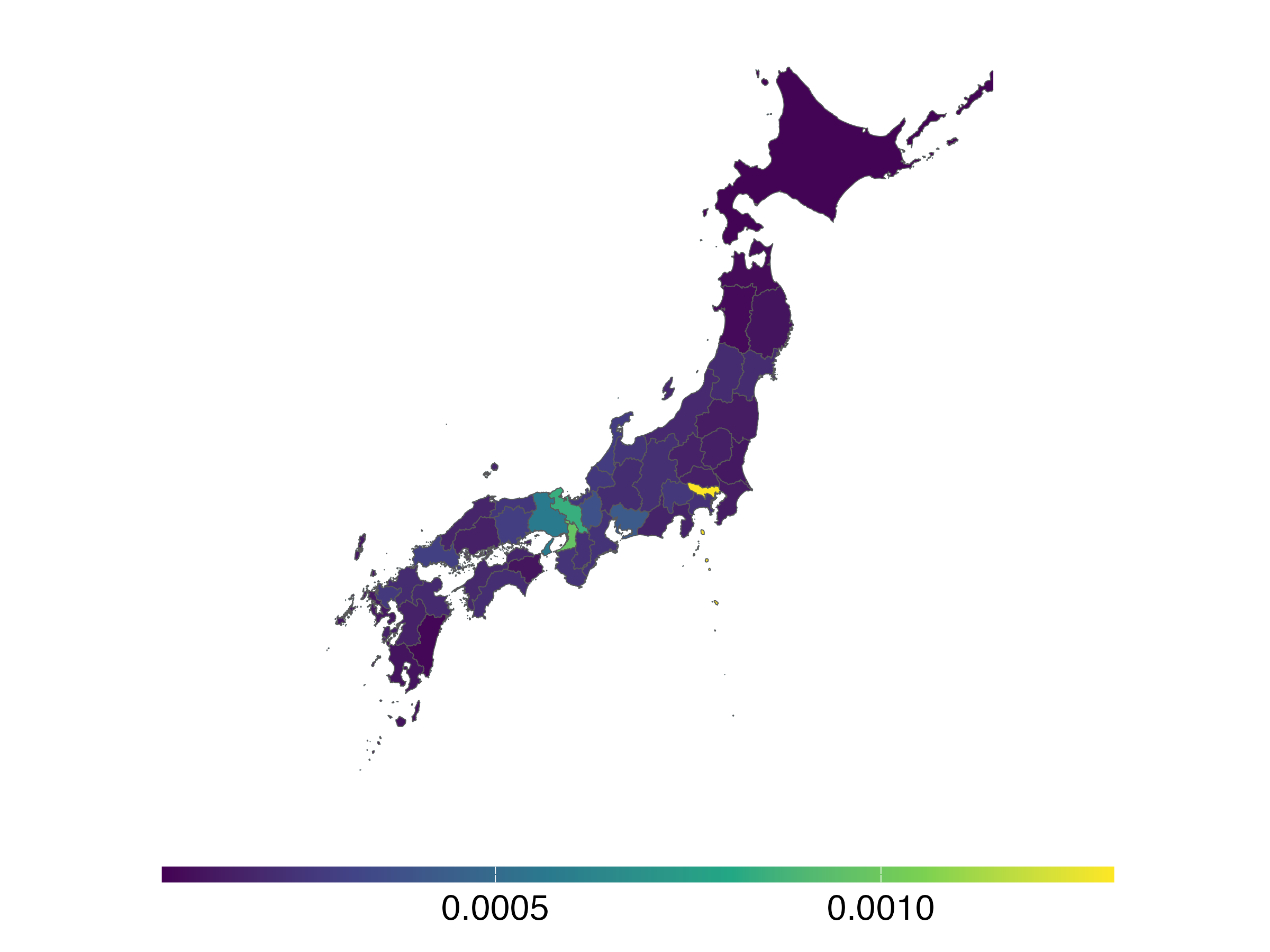}
    }\\
    \subfloat[Edition 12 (1939)]{
    \includegraphics[width = 0.45\textwidth]{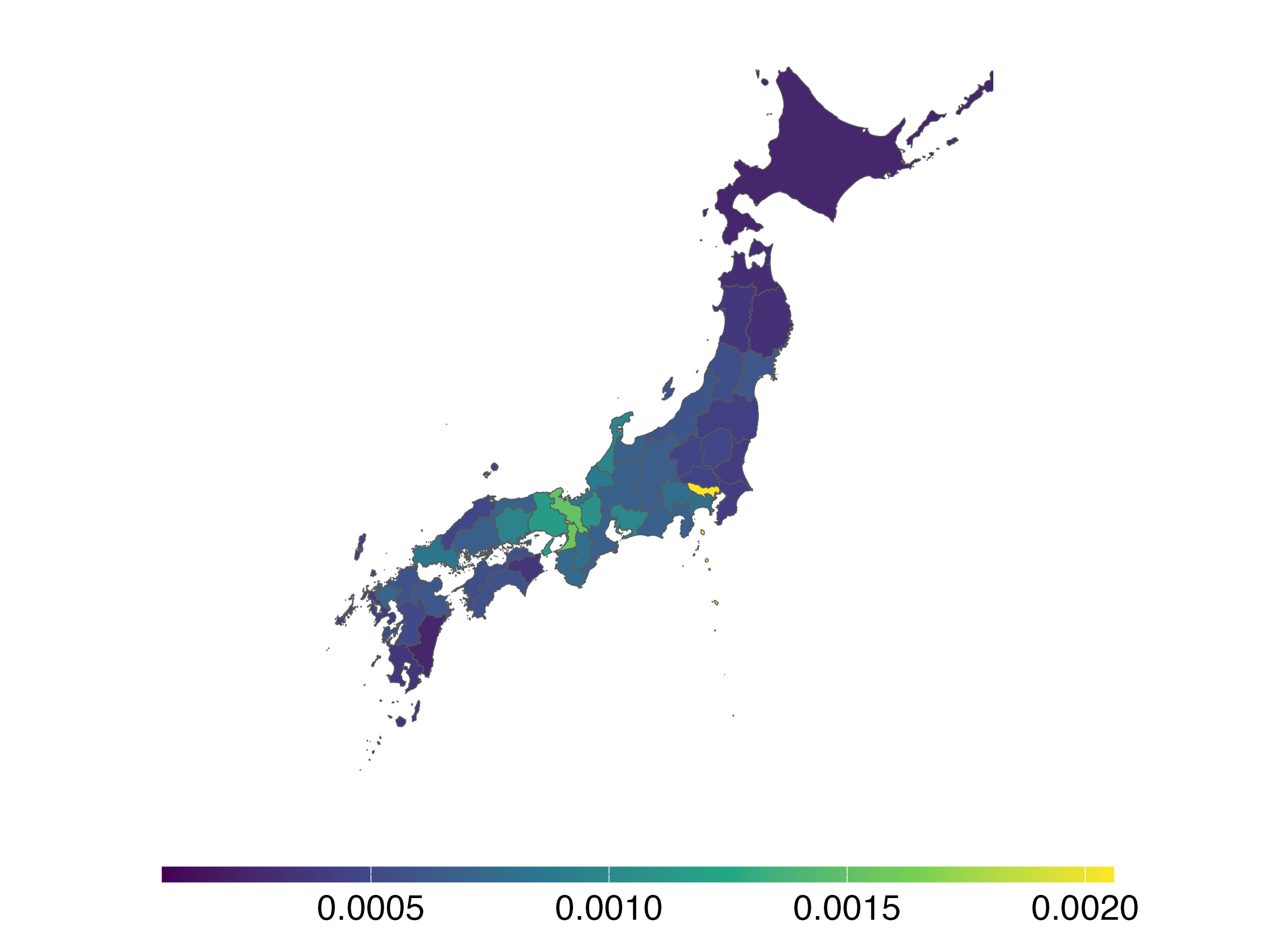}
    }
    \end{center}
    \caption{Prefecture-Level Share of Listed Elites in Population}
    \label{fg:listing_ratio_pref_ver12}
    \footnotesize
     \textit{Note:} These figures show the share of elites relative to the population across PIR editions for all 47 prefectures. As the population census was conducted every five years starting in 1920, we use the prefectural population data in 1920 for panels (a) and (b) and those in 1930 for panels (c)--(e). Individuals with missing residential information or who live in foreign countries at the time of listing are excluded.
\end{figure}

\begin{figure}[htbp]
    \begin{center}
    \includegraphics[width = 0.7\textwidth]{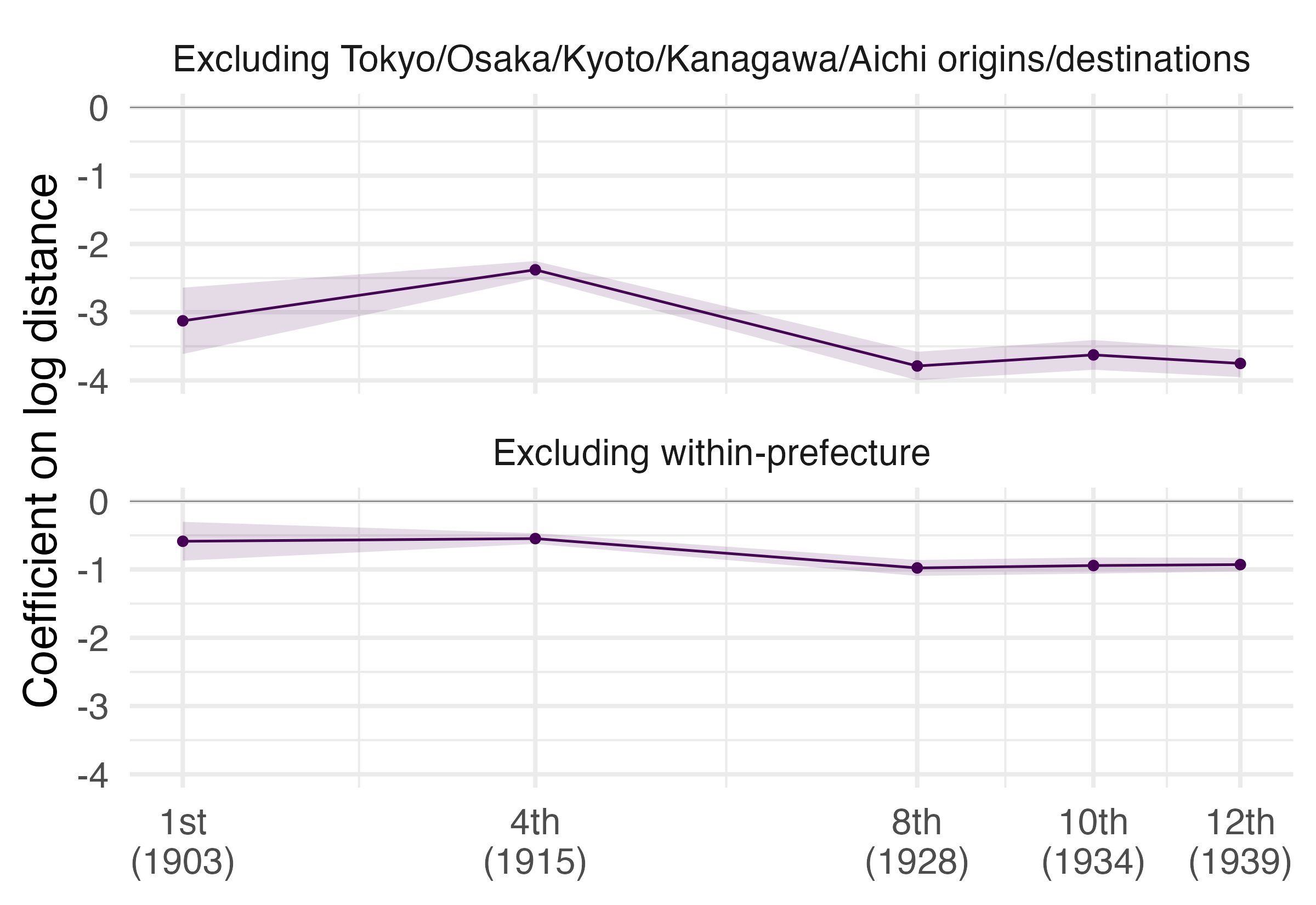}
    \end{center}
    \caption{Coefficient Plots on Distance in PPML Gravity with Subsample}\label{fig:gravity_ot_rt_fe_sub}
    \footnotesize
    \textit{Note:} This figure reports additional results for the gravity estimates. The top panel excludes observations associated with Tokyo, Osaka, Kyoto, Kanagawa, and Aichi. The bottom panel excludes within-prefecture observations. All specifications include origin-by-edition and destination-by-edition fixed effects and are estimated by Poisson pseudo-maximum likelihood. Vertical bars indicate 95\% confidence intervals. Standard errors are clustered at the origin-destination pair level.
\end{figure}

\end{document}